\documentclass[11pt]{article}
\pdfoutput=1
\usepackage{amsmath}
\usepackage{amssymb}
\usepackage{graphicx,bbm,mathrsfs}
\usepackage{nicefrac}
\usepackage{slashed}
\usepackage{bbm}
\usepackage{geometry}
\usepackage{empheq}
\usepackage{stackrel}
\usepackage{ulem}
\usepackage{jheppub}
\usepackage{setspace}
\usepackage{relsize}
\usepackage{empheq}
\usepackage{pifont}% http://ctan.org/pkg/pifont
\usepackage{mathtools}
\usepackage{enumitem}
\usepackage{dcolumn}   % needed for some tables
\usepackage{bm}        % for math
\usepackage{graphicx,mathrsfs}
\usepackage{nicefrac}
\usepackage{multirow}
\usepackage{color}
\usepackage{mathtools}
\usepackage{makecell}
\usepackage{environ} 
\usepackage{lipsum} 
\usepackage{wasysym}
\usepackage{hhline,colortbl}  
\usepackage{tikz}
\usepackage{graphicx}
\usepackage{tensor}
\usepackage{dsfont}
\usepackage{simpler-wick}

\usepackage{booktabs}
\usepackage{mdframed}
\usepackage{xcolor}
\definecolor{EqFrame}{RGB}{235,245 ,250 }

 \NewEnviron{Smaller11}{
           \scalebox{1.1}{$\BODY$} 
 } 
 \NewEnviron{Smaller08}{
           \scalebox{0.8}{$\BODY$} 
 }

\newcommand{\be}{\begin{equation}}
\newcommand{\ee}{\end{equation}}
\newcommand{\bea}{\begin{eqnarray}}
\newcommand{\eea}{\end{eqnarray}}

\renewcommand{\O}{\mathcal O}

\renewcommand{\c}{c}

\newcommand{\B}{{\cal\bar B}}

\renewcommand\labelenumi{(\roman{enumi})}
\renewcommand\theenumi\labelenumi

\newcommand{\nn}{\nonumber}

\usepackage{titlesec}

\titleformat*{\section}{\Large\bfseries}
\titleformat*{\subsection}{\large\bfseries}
\titleformat*{\subsubsection}{\large\bfseries}
\titleformat*{\paragraph}{\large\bfseries}
\titleformat*{\subparagraph}{\large\bfseries}

\makeatletter
\newcommand*{\prodsym}{%
  \DOTSB
  \mathop{
    \mathchoice
      {\rlap{\kern.3em\rotatebox[origin=c]{-90}{}}{\prod}}
      {\vcenter{\rlap{\kern.2em\rotatebox[origin=c]{-90}{}}}{\prod}}
      {\sum}{\sum}
  }\slimits@
}
\makeatother

\makeatletter
\DeclareFontFamily{OMX}{MnSymbolE}{}
\DeclareSymbolFont{MnLargeSymbols}{OMX}{MnSymbolE}{m}{n}
\SetSymbolFont{MnLargeSymbols}{bold}{OMX}{MnSymbolE}{b}{n}
\DeclareFontShape{OMX}{MnSymbolE}{m}{n}{
    <-6>  MnSymbolE5
   <6-7>  MnSymbolE6
   <7-8>  MnSymbolE7
   <8-9>  MnSymbolE8
   <9-10> MnSymbolE9
  <10-12> MnSymbolE10
  <12->   MnSymbolE12
}{}
\DeclareFontShape{OMX}{MnSymbolE}{b}{n}{
    <-6>  MnSymbolE-Bold5
   <6-7>  MnSymbolE-Bold6
   <7-8>  MnSymbolE-Bold7
   <8-9>  MnSymbolE-Bold8
   <9-10> MnSymbolE-Bold9
  <10-12> MnSymbolE-Bold10
  <12->   MnSymbolE-Bold12
}{}

\let\llangle\@undefined
\let\rrangle\@undefined
\DeclareMathDelimiter{\llangle}{\mathopen}%
                     {MnLargeSymbols}{'164}{MnLargeSymbols}{'164}
\DeclareMathDelimiter{\rrangle}{\mathclose}%
                     {MnLargeSymbols}{'171}{MnLargeSymbols}{'171}
\makeatother

\begin{document}

\vspace*{4mm}

\thispagestyle{empty}

\begin{center}

%  {\LARGE
% \sc
\begin{minipage}{20cm}
\begin{center}
\hspace{-5cm }
\huge
\sc
Running  Love Numbers   and the   \\   
\hspace{-5cm }   Effective Field Theory  of Gravity 
\end{center}
\end{minipage}
\\[30mm]

\renewcommand{\thefootnote}{\fnsymbol{footnote}}

{\large  
Sergio~Barbosa$^{\, a}$ \footnote{sergio.barbosa@aluno.ufabc.edu.br}\,, 
Philippe~Brax$^{\, b}$ \footnote{philippe.brax@ipht.fr}\,
Sylvain~Fichet$^{\, a}$ \footnote{sylvain.fichet@gmail.com}\,, 
Lucas~de~Souza$^{\, c}$ \footnote{souza.l@ufabc.edu.br}\,, 
}\\[12mm]
\end{center} 
\noindent

\indent \; ${}^a\!$ 
\textit{CCNH, Universidade Federal do ABC,} \textit{Santo Andr\'e, 09210-580 SP, Brazil}

\indent \; $^b\!$ \textit{Institut de Physique Th\'{e}orique, Universit\'e Paris-Saclay, CEA, CNRS, }

\indent \; \textit{ F-91191 Gif/Yvette Cedex, France}

\indent \; ${}^c\!$ 
\textit{CMCC, Universidade Federal do ABC,} \textit{Santo Andr\'e, 09210-580 SP, Brazil}
\\

\addtocounter{footnote}{-4}

\vspace*{8mm}
 
\begin{center}
{  \bf  Abstract }
\end{center}
\begin{minipage}{15cm}
\setstretch{0.95}
% \small

Massive states produce higher derivative corrections to Einstein gravity in the infrared,   which are encoded into operators of
the  Effective Field Theory (EFT) of gravity. 
These EFT operators  modify the geometry  and affect the tidal properties of black holes, either neutral or charged.
 A thorough  analysis of the perturbative tidal deformation problem  leads us  to introduce a tidal  Green function, which  we  use  to derive two
  universal formulae that  efficiently  provide the  constant and running Love numbers induced by the EFT. 
 We apply these formulae to determine the  tidal response of EFT-corrected non-spinning black holes induced by vector and tensor  fields, reproducing existing results where available and deriving new ones.   
We find that neutral black hole Love numbers run classically 
for $\ell\geq 3$ while charged ones run for $\ell\geq2$.  Insights from the Frobenius method and from EFT principles confirm that the Love number renormalization flow is a well-defined physical effect.  
We find that extremal black holes can have Love numbers much larger than neutral ones, up to ${\cal O}(1)$ within the EFT validity regime, 
and  that the EFT cutoff corresponds to the exponential suppression of the Schwinger effect. 
We  discuss the possibility of probing an Abelian dark  sector through gravitational waves, considering a scenario in which dark-charged extremal black holes exist in the present-day Universe.

    \vspace{0.5cm}
\end{minipage}

\newpage
\setcounter{tocdepth}{2}

\tableofcontents  

\vspace{1cm}
\hrule
\vspace{1cm}

\section{Introduction \label{se:Intro}}

Gravitation and the dynamics of spacetime at distances larger than the Planck length can be described  by an Effective Field
Theory (EFT) \cite{Donoghue:1995cz,Burgess:2003jk}.  The effective Lagrangian of gravity has the structure of a derivative expansion, that includes the  expansion in spacetime curvature. 
From this modern point of view, the theory of General Relativity (GR) is seen  as the lowest order term of the {effective Lagrangian of gravity}, here denoted  ${\cal L}_{\rm eff}$. 
The first leading terms  
 are schematically
\be
{\cal L}_{\rm eff}=
{\cal L}_{\rm GR}+
{\cal L}_{\partial^4}+
{\cal L}_{\partial^6}
+\O(\partial^8)\;.
\label{eq:GREFT_LO}
\ee
 In this work we study some of the consequences of   
 the four and six-derivatives terms in
\eqref{eq:GREFT_LO}.

Matter fields are generally present in \eqref{eq:GREFT_LO}, both in
${\cal L}_{\rm GR}=\frac{1}{2\kappa^2}R+{\cal L}_{\rm matter}$ and in the higher derivative terms.
Notably, in $d=4$ dimensions,
 ${\cal L}_{\partial^4}$ vanishes in the absence of matter. The leading EFT corrections to pure gravity occur instead from  ${\cal L}_{\partial^6}$.

The EFTs considered in this work are valid at large enough distances i.e. in the infrared, below some typical mass scale. 
This  cutoff scale is the Planck scale for the   EFT that emerges from the UV completion of quantum gravity,  but the cutoff  can be much lower  for other EFTs. 
In particular, the  infrared EFT of our macroscopic world arises at energies below the  neutrino mass, for which only gravity and electromagnetism remain dynamical.\,\footnote{We specifically refer to the EFT of the real gravitational world  as the GREFT, in analogy with the SMEFT that denotes the EFT Lagrangian of the Standard Model (SM) of Particle Physics.}

One may ask whether the  ${\cal L}_{\partial^6}$ term in \eqref{eq:GREFT_LO} could  be zero in the vacuum. 
In fact, in the hypothesis that the UV completion of quantum gravity is a superstring theory, the  EFT emerging below the string scale does predict vanishing curvature-cubed terms \cite{Jack:1988sw,Gross:1986iv,Gross:1986mw,Kikuchi:1986rk,Becker:2006dvp} --- the leading corrections are instead quartic in curvature.  This feature is a consequence of  the supersymmetry of the UV completion. 
At lower energies, whenever spacetime gets compactified from $d=10$ to $d=4$, or when massive particles are integrated out, nonzero contributions to ${\cal L}_{\partial^6}$ generically appear  in the infrared EFT.

The physical objects  studied in this work  are black holes. 
At distances scales larger than the Planck length, black holes are classical objects described by general relativity.  
In the framework of GR in $d=4$ dimensions, {asymptotically flat} black holes are perfectly rigid objects in the sense that their response to static tidal deformations vanishes \cite{1972ApJ...175..243P,Martel:2005ir,Fang:2005qq,Damour:2009va,Binnington:2009bb,Kol:2011vg,Landry:2015cva,Landry:2015zfa,Porto:2016pyg,Poisson:2020mdi,LeTiec:2020spy,LeTiec:2020bos,Chia:2020yla,Goldberger:2020fot,Hui:2020xxx,Charalambous:2021mea,Ivanov:2022qqt,Pereniguez:2021xcj,Rai:2024lho,Landry:2014jka,Charalambous:2021kcz,Hui:2021vcv,Hui:2022vbh,Charalambous:2022rre}, even at the nonlinear level \cite{Poisson:2020vap,Poisson:2021yau,DeLuca:2023mio,Riva:2023rcm,Hadad:2024lsf,Iteanu:2024dvx,Combaluzier-Szteinsznaider:2024sgb,Kehagias:2024rtz,Gounis:2024hcm}.\,\footnote{This can be explained from approximate symmetries arising in the near-horizon region. See \cite{Charalambous:2021kcz,Hui:2021vcv,Hui:2022vbh,Charalambous:2022rre,Sharma:2024hlz,Combaluzier-Szteinsznaider:2024sgb,Kehagias:2024rtz,Gounis:2024hcm} for progress on this aspect. On the other hand,  dynamical  Love numbers are found to be non-zero \cite{Mandal:2023hqa}, and 
 static  Love numbers of
asymptotically (anti)-de Sitter black holes do not vanish either, see e.g. \cite{Emparan:2017qxd, Nair:2024mya}. }

In the real world, however, spacetime at subPlanckian energies serves as the stage for quantum phenomena. 
Quantum field theory (QFT) predicts that, even when there is no matter around, a black hole is surrounded by bubbles of virtual particles. 
In Minkowski space, this causes the black hole to Hawking decay, while at classical level the black hole would be eternal. 
This is an instance where the existence of the quantum vacuum causes a major change in the classical properties of black holes.  

\begin{figure}[!t]
     \centering
    \includegraphics[width=0.8\linewidth,trim={0.5cm 5.5cm 1cm 5.5cm},clip]{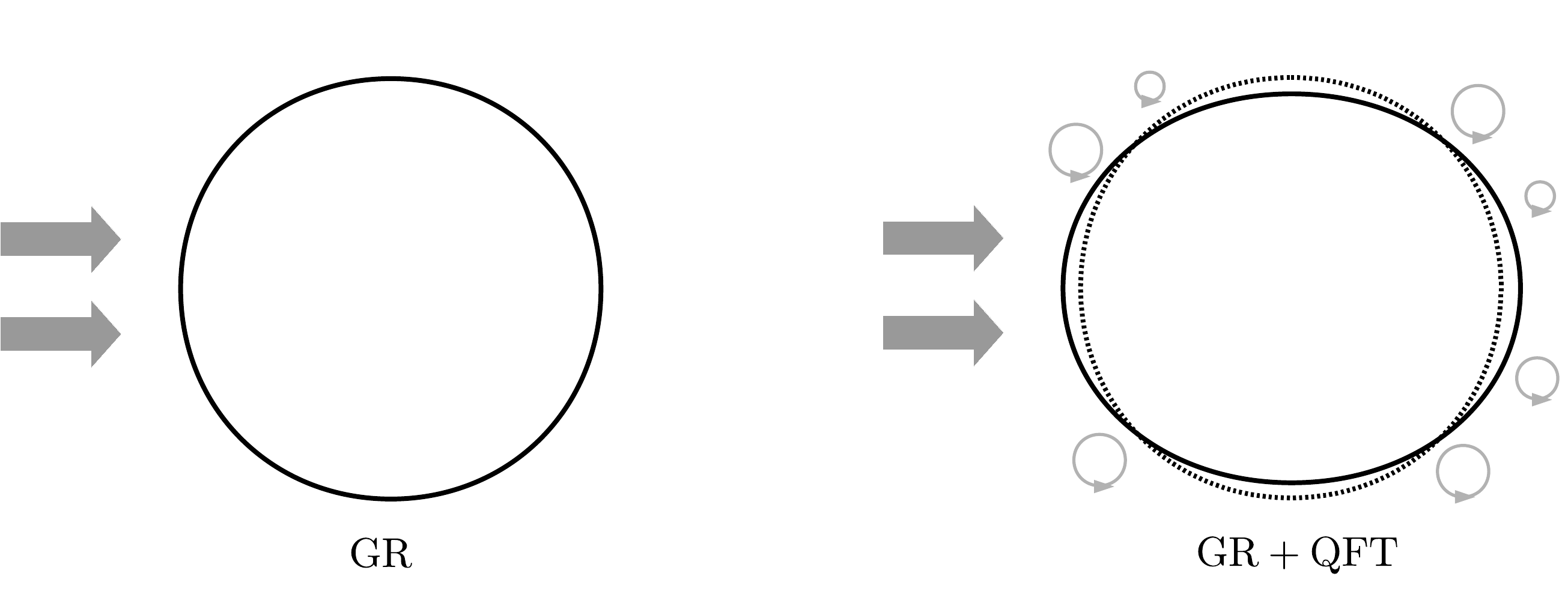}
     \label{fig:BH_deformation}
     \caption{In 4D general relativity, black holes in  classical vacuum are rigid: they do not respond to tidal fields. When spacetime is filled with fluctuations from the quantum vacuum, black holes become tidally deformable.}
 \end{figure}

In analogy with the Hawking phenomenon, we may wonder if the QFT vacuum may crucially affect other properties of black holes, such as their absolute rigidity. 
 Can black holes tidally deform in the presence of the quantum vacuum? 
The answer is positive, as shown in Refs.\,\cite{Cardoso:2018ptl,DeLuca:2022tkm,Charalambous:2022rre}. 
We  sum up the above statements as in Fig. \ref{fig:BH_deformation}, where the arrows represent the perturbation from an external tidal field. 
When the black hole radius is much larger than the Compton wavelength of the matter particles, i.e. $r_h\gg \frac{1}{m}$, the loops can be shrunk to points and their effect on the black hole are encapsulated by the infrared EFT described in \eqref{eq:GREFT_LO}.

It might seem at first view that the computation of the Love numbers in gravitational EFTs is plagued by both technical and conceptual difficulties, as discussed in \cite{DeLuca:2022tkm,Charalambous:2022rre}.
One of the aim of this work is to unravel these apparent challenges. 
In a nutshell, we show that tidal Love numbers in gravitational EFTs are well-defined and  easy to compute. Namely, they are extracted from a certain overlap function that has the schematic form
\be
\psi_{0}D^{(1)} \psi_{0}\;,
\ee
where $\psi_0$ is the tidal field solution that is regular on the horizon and $D^{(1)}$ is the correction of the tidal wave operator  produced by the EFT.\,\footnote{Other aspects of black hole physics affected by the EFT of gravity   include  gravitational waves \cite{
Endlich:2017tqa,Brandhuber:2019qpg,AccettulliHuber:2020dal,Cayuso:2023xbc,Cardoso:2019mqo,McManus:2019ulj,deRham:2020ejn,Cano:2020cao,Sennett:2019bpc,Silva:2022srr,Maenaut:2024oci}, 
quasinormal modes  \cite{Cano:2023jbk,Miguel:2023rzp,Melville:2024zjq, Cano:2024ezp,Cano:2024wzo} 
and UV conjectures on extremal black holes \cite{Kats:2006xp,
Cheung:2014ega,
Endlich:2017tqa,
Cheung:2018cwt,
Hamada:2018dde,
Loges:2019jzs,Goon:2019faz,Jones:2019nev, Chen:2019qvr, Bellazzini:2019xts,Loges:2020trf,Arkani-Hamed:2021ajd,Cao:2022iqh,DeLuca:2022tkm, Bittar:2024xuc, Knorr:2024yiu}. 
More generally, gravitational EFTs are constrained by infrared consistency conditions based on causality and unitarity, see e.g. \cite{ 
Camanho:2014apa,
 Goon:2016une, 
Bellazzini:2015cra, Cheung:2016wjt,Hamada:2018dde, Chen:2019qvr, 
Bellazzini:2019xts, 
Arkani-Hamed:2020blm,Alberte:2020jsk,Alberte:2020bdz,
Arkani-Hamed:2021ajd,
Bern:2021ppb, 
deRham:2021bll,
 Caron-Huot:2022ugt,
Caron-Huot:2022jli,
Hamada:2023cyt,
Bellazzini:2023nqj,
Bittar:2024xuc, 
Eichhorn:2024wba,
Knorr:2024yiu}.
 }

Here is the plan of our work. 
Section\,\ref{se:EFT_basics} reviews the basic principles and practical rules of effective field  theory. Section\,\ref{se:EFT_R3} determines the operator basis for the gravitational EFT with and without matter. The EFT-corrected field equations are also computed,
that serve to compute the corrections to black hole geometries. 
Section\,\ref{se:OneLoopEFT} contains a computation of the contributions to the order-$\partial^6$ operators  from loops of massive particles with spin $0$, $\frac{1}{2}$, $1$.  
 Section\,\ref{se:TLNs} reviews the notion of tidal deformability and the worldline EFT. It contains a general analysis of the perturbative tidal problem, introduces the tidal Green function and provides the universal formulae to compute the Love numbers with and without running.  
 Section\,\ref{se:TLN_results} applies the formalism to non-spinning black holes. The EFT-corrected black hole geometries are computed. We then derive the vector and tensor tidal equations of motion and obtain the Love numbers of the neutral and charged black holes. 
 Section\,\ref{se:bounds} discusses the search for light particles. Exclusion bounds are derived in the specific case of black holes being near-extremal under a dark $U(1)$. 
  Section\,\ref{se:Conclusion} summarizes our study. 
The Appendix contains technical details on variational computations (\ref{app:Var_L4}), on the heat kernel coefficients (\ref{app:HK}),  on the tidal Green function (\ref{app:Green}),
on EFT-corrected black hole geometries (\ref{se:bh_calculation}),
on the spherical harmonics (\ref{sec:spherical_harmonics}),  and on the tidal equations of motion (\ref{se:source}).

\section{Effective Field Theory in a Nutshell}

\label{se:EFT_basics}

We present the notion of low-energy EFT from the viewpoint of the quantum effective action  in section \ref{se:Effective_action}. 
The practical rules governing EFT Lagrangians are then exposed in section \ref{se:EFT_rules}. 

\subsection{Effective Action and Effective Field Theory}

\label{se:Effective_action}

Consider a gravitational theory with metric $g_{\mu\nu}$ and containing a massive matter field $\Phi_h$ with mass $m$. Our interest may lie in finding classical solutions for $g_{\mu\nu}$, or in computing graviton scattering amplitudes. 
In either case, all the information needed is contained in the gravitational partition function.

For technical convenience, we couple the metric to a non-dynamical, abstract source $T^{\mu\nu}_\ell$. The  $T^{\mu\nu}_\ell$ source can be thought as a simplified version for the stress tensor of light fields. Its  only role is to source  the metric, it will disappear  upon Legendre transform. 

The partition function is 
\be
Z[T^{\mu\nu}_\ell] = \int {\cal D}g_{\mu\nu}{\cal D}\Phi_h e^{iS[\Phi,R,\nabla]+i\int d^dx g_{\mu\nu}T_\ell^{\mu\nu}}\,. 
\ee
We can perform the  field integral over $\Phi_h$. This defines a ``partial'' quantum effective action $\Gamma_h[R,\nabla]$, with  
\be
Z[T^{\mu\nu}_\ell] =\int {\cal D}g_{\mu\nu}  e^{i\Gamma_h[R,\nabla]+i\int d^dx g_{\mu\nu}T_\ell^{\mu\nu}}\,.
\ee
The $\Gamma_h[R,\nabla]$ action depends only on the metric, but still encodes all the information about the $\Phi_h$ field. 

Consider then the long distance regime for which the  distance scales encoded in  $T_\ell^{\mu\nu}$ are much larger than the Compton wavelength of the heavy field, $\frac{1}{m}$.
Equivalently, in the context of scattering amplitudes,  we consider the low-energy regime for which  the external momenta flowing through the $T^{\mu\nu}_\ell$ sources  are much smaller than $m$. 
In this limit, the  quantum effective action $\Gamma_h$ can be organized as an expansion in powers of derivatives over $m$ and becomes a local functional.

This is conveniently expressed as an effective Lagrangian ${\cal L }_{\rm eff}$  
\be
\Gamma_h[g_{\mu\nu}] \equiv \int d^dx \sqrt{-g}\, {\cal L}_{\rm eff}[g_{\mu\nu}]\;, 
\ee
where  ${\cal L }_{\rm eff}$  is made out of monomials of the Riemann tensor {$R_{\mu \nu \rho \sigma}$ and its contractions $R_{\mu \nu}$ and $R$, denoted collectively by ``Riem'',} and its covariant derivatives, suppressed by powers of $m$. Schematically we have a series of the form
\be {\cal L}_{\rm eff}[g_{\mu\nu}]\sim \sum_{a,b} \frac{\nabla^{2a}({\rm Riem})^b}{m^{2a+2b-4}}\,. \ee
In practice, ${\cal L }_{\rm eff}$   is typically truncated at some order of the derivative expansion $\frac{\partial}{m}$, that counts both  
$\nabla$'s and curvatures. 
This defines an infrared effective field theory (EFT)  that encodes all the effects of the $\Phi_h$ field at energies below $m$, within the accuracy of the truncation of ${\cal L}_{\rm eff}$.

The derivative expansion applies at each order of the loop expansion of $\Gamma_h$, $\Gamma_h=\Gamma^{(0)}_h+\Gamma^{(1)}_h+\ldots$.  Hence the effective Lagrangian can be organized with respect to this loop expansion: ${\cal L}_{\rm eff}={\cal L}^{(0)}_{\rm eff}+{\cal L}^{(1)}_{\rm eff}+\ldots .$ The ${\cal L}^{(0)}_{\rm eff}$ term arises from the tree diagrams involving  $\Phi_h$ encoded in $\Gamma^{(0)}_h$. 
The ${\cal L}^{(1)}_{\rm eff}$ arises from the one-loop diagrams involving $\Phi_h$  encoded in $\Gamma^{(1)}_h$, etc. 

In this paper, we work  at the one-loop level in Section \ref{se:OneLoopEFT}. The finer details of EFT at loop level can be found in \cite{Manohar:1996cq,Manohar:2018aog}.

\subsection{EFT in Practice}

\label{se:EFT_rules}

Effective field theory is a powerful tool because the properties of the effective Lagrangian ${\cal L}_{\rm eff}$ are described by a few simple rules. Using them we can efficiently write down effective Lagrangians --- without necessarily knowing the UV completion of the EFT. An effective Lagrangian is naturally structured as follows.

\subsubsection{Derivative Expansion}
The infinite amount of effective operators is organized in terms of the number of derivatives, $\partial^{2n}$. The derivatives are accompanied with an inverse mass scale $\Lambda$, such that the combination $\frac{\partial}{\Lambda}$ can be thought as an expansion parameter.   That $\frac{\partial}{\Lambda}$ provides an expansion scheme becomes transparent when applied to physical observables.   In  black hole perturbation theory, for instance, each derivative contributes as $\partial\sim \frac{1}{r_h}$, hence $\frac{1}{r_h\Lambda}$ is an expansion parameter. 

Importantly there is a {\it finite} number of independent operators at a given order $n$.

\subsubsection{Validity Range}

In the practical use of  EFT, the derivative series is always
truncated at a finite order $n$, such that all the effects of the UV physics are encapsulated into a finite number of parameters. Truncating the derivative expansion is possible only in a definite domain of distances. For example the black hole perturbation theory becomes inaccurate when $r_h$ approaches $1/\Lambda$. $\Lambda$ is referred to as the cutoff scale of the EFT.

\subsubsection{Symmetries}
The set of effective operators satisfies the symmetries of the UV theory. The description of EFT from the gravitational partition function presented in section \ref{se:Effective_action} makes it clear that the effective operators can be written covariantly, hence manifestly retaining the symmetries of the UV theory. In gravitational EFTs, the UV theory has diffeomorphism invariance, the symmetry of GR. Hence the effective operators can be written in terms of 
curvature tensors and covariant derivatives.

The fact that each effective operator satisfies diffeomorphism invariance by itself has an important consequence. It implies that, just like the stress-energy tensor is conserved, the divergence of the variation of each of the effective operators must vanish. This property is used in section \ref{se:EFE}.

\subsubsection{Field Equations}
The field equations derived from the effective Lagrangian at  order $\partial^{2n}$ can be used to reduce the set of effective operators at order $\partial^{2n+2}$. This is allowed both at the level
of perturbation theory and at the level
of the scattering amplitudes of QFT.  

At the level of perturbative calculations, one solves differential equations order by order. 
The perturbation at  order $n$ is used to obtain the  perturbation at order $n+1$. 
By construction, this structure implies that  we can apply the field equations deformed at order $n$ in the set of operators of order $n+1$.

For scattering amplitudes this is stated more generally in terms of  field redefinition. Applying the equation of motion amounts to a particular case of a field redefinition. By construction, the $S$-matrix is invariant under field redefinitions due to the LSZ reduction, see e.g. \cite{Manohar:2018aog}. It follows that we are allowed to use the equation of motion directly in ${\cal L}_{\rm eff}$.

\section{The EFT of Gravity at $\partial^6$ Order }
\label{se:EFT_R3}

We apply the general rules exposed in section \ref{se:EFT_rules} to the effective Lagrangian up to six derivatives. 
We assume $d=4$ Minkowski space, hence operators that are total derivatives do not contribute to the action due to the divergence theorem.  
We distinguish the cases with no matter (in Sec.\,\ref{se:puregravEFT}) and with electromagnetism (in Sec.\,\ref{se:EMEFT}).  We work out the former in details  and simply review the latter.

\subsection{The EFT of Pure Gravity at  $\partial^6$ Order}

\label{se:puregravEFT}

In the absence of matter, the effective Lagrangian reads  
\be
{\cal L}_{\rm eff}=\frac{1}{2}\hat R+  {\cal L}_{4} + {\cal L}_{6}+\O(\partial^{8})\;,
\label{eq:L_eff_gen}
\ee
where ${\cal L}_n\equiv {\cal L}_{\partial^{2n}}$ contains the  terms of $\partial^{2n}$ order. We use $\hat R\equiv \frac{1}{\kappa^2} R$, with $\kappa^2 = 8 \pi G=\frac{1}{M_{\rm Planck}^2}$.

At the order of four derivatives, $ {\cal L}_{4}$ contains  the curvature square operators $R^2$, $R_{\mu\nu}R^{\mu\nu}$, $R_{\mu\nu \rho \sigma}R^{\mu\nu\rho \sigma}$ as well as  $\square R$. At the order of six derivatives, $ {\cal L}_{6}$ contains curvature cubed operators such as $R^3$,
%$(R_{\mu\nu}^{~~~\rho \sigma})^3$
curvature squared with one d'Alembertian such as $R \square R$, and the curvature with two d'Alembertians, $\square^2 R$.

\subsubsection{Reducing the Set of Operators}

A first simplification from the $d=4$ assumption is  that the Gauss-Bonnet combination
\be
GB=(R_{\mu\nu\rho\sigma})^2- 4(R_{\mu\nu})^2+R^2 
\ee
is a topological invariant, hence it cannot contribute to the field equations. For our purposes we can thus use $GB=0$ to eliminate the $(R_{\mu\nu\rho\sigma})^2$ term.\,\footnote{In higher dimensions this combination still vanishes at quadratic order in the fluctuations. Hence this property can still be useful depending on the application, see \cite{Bittar:2024xuc}.}

We then use the field equations, with the assumption of flat empty space. 
At leading order one would have $R_{\mu\nu}=0$. Here, however, we must take into account the correction of $\partial^4$ order. 
The EFT-corrected field equation takes the form\,\footnote{The explicit variations are given in App.\,\ref{app:Var_L4}.} 
\be
\hat R_{\mu\nu}-\frac{1}{2}\hat R g_{\mu\nu}= -\frac{2}{\sqrt{-g}}\frac{\delta(\sqrt{-g}{\cal L}_4)}{\delta g^{\mu\nu}} + \O(\partial^{6})\;. 
\label{eq:EFE_del4}
\ee
Substituting in the curvature square terms produce contributions to ${\cal L}_6$. These contributions necessarily involve at least one scalar curvature or Ricci tensor, essentially because the field equation has only two indexes. 

We also use the field equation to reduce ${\cal L}_6$. At that order only the leading order of \eqref{eq:EFE_del4} contributes  since its correction produces higher order $\O(\partial^8)$ terms that are neglected.   This implies that all the terms of ${\cal L}_6$ with at least one $R_{\mu\nu}$ or $R$ are eliminated. In particular all the contributions produced by using the field equation in ${\cal L}_4$ are eliminated. 

The remaining operators are of the form Riem$^3$ and Riem$\square$Riem. 

\subsubsection{The Curvature Cubed Operators}

A rigorous counting of the Riem$^3$ structures can be done as follows.\,\footnote{ 
We focus on parity-even operators through this work. Parity-odd operators are similarly treated by building invariants involving $\tilde R^{\mu\nu\rho\sigma}=\varepsilon^{\mu\nu}_{~~\alpha\beta}R^{\alpha\beta\rho\sigma }$, with $\epsilon^{\mu\nu}_{~~\alpha\beta}$ the Levi-Civita tensor, see e.g. \cite{Endlich:2016jgc}. 
}
Using the elementary symmetries of the Riemann tensor, we can count all the inequivalent ways of the two blocks  of each Riemann tensor  (i.e. the index groups $(1,2)$ and $(3,4)$) to be  connected with each others. 
The structure of each operator can be characterized by its cycles of contractions. We identify five different ${\rm Riem}^3$ structures,

\begin{subequations}
\begin{align}
{\cal O}_{3\times 2} &= R_{\mu\nu}^{~~\,\rho\sigma} R_{\rho\sigma}^{~~\,\alpha\beta}R_{\alpha\beta}^{~~\,\mu\nu} 
\, \\
{\cal O}_{2+4}&= R_{\mu\nu}^{~~\,\rho\sigma} R_{\rho~\,\sigma}^{~\,\alpha~\,\beta} R_{\alpha \beta}^{~~\,\mu\nu} 
\, \\
{\cal O}_{2\times 3} &= R_{\mu~\,\nu}^{~\,\rho~\,\sigma} 
R^{~\,\alpha~\,\beta}_{\rho~\,\sigma}
R^{~\,\mu~\,\nu}_{\alpha~\,\beta}
 \\
{\cal O}_{6a} &=R_{\mu\nu}^{~~\,\rho\sigma} R_{~\,\alpha~\,\beta}^{\mu~\, \nu} 
R_{\rho~\, \sigma }^{~\, \alpha~\,\beta}
 \\
{\cal O}_{6b}& =R_{\mu~\,\nu}^{~\,\rho~\,\sigma} 
R^{~\,\alpha~\,\beta}_{\rho~\,\sigma}
R^{~\,\mu~\,\nu}_{\beta~\,\alpha}\;
\end{align}
\end{subequations}
where the labels denote the block contraction cycles. 
For example,  ${\cal O}_{2\times 3}$  
 features a 3-cycle $\wick{\c1 \mu\rho - \rho\alpha - \alpha \c1 \mu  }$.   
The ${\cal O}_{3\times 2}$ has three 2-cycles, ${\cal O}_{2+4}$ has a 2-cycle and a 4-cycle, ${\cal O}_{2\times 3}$ has two 3-cycles, and there are two inequivalent operators ${\cal O}_{6a,b}$ with a single 6-cycle. 

We have not used the first Bianchi identity so far. Applying it to each operator gives  three relations 
 \be
 {\cal O}_{3\times 2} = 2 {\cal O}_{2+4}\,,\quad \quad 
 {\cal O}_{2+4} = 2{\cal O}_{6a} \,,\quad \quad 
 {\cal O}_{6a} = {\cal O}_{2\times 3}- {\cal O}_{6b}\;.
 \ee
The first Bianchi identity applied to  ${\cal O}_{2\times 3}$ or  ${\cal O}_{6b}$ gives again the last relation. 
We conclude there are two independent ${\rm Riem}^3$ operators.

\subsubsection{The $\partial^6$ Basis and a Non-Reduced Set}

\label{se:O6basis}

Two  Riem$\square$Riem terms are possible, $R_{\mu\nu\rho\sigma}\square R^{\mu\nu\rho\sigma}$ and $\nabla^\mu R_{\mu\nu\rho\sigma}\nabla_{\lambda} R^{\lambda\nu\rho\sigma}$. Using the contracted second Bianchi identity $\nabla^\mu R_{\mu\nu\rho\sigma}= \nabla_\rho R_{\sigma\nu} - \nabla_\sigma R_{\rho\nu}$, the latter  is expressed in terms of the Ricci tensor, and using  the field equation \eqref{eq:EFE_del4} we conclude it contributes only at order $\O(\partial^8)$. 
Similarly, the former is reduced  to 
\be
R_{\mu\nu\rho\sigma}\square R^{\mu\nu\rho\sigma} = 
-{\cal O}_{3\times 2} -4 {\cal O}_{2\times 3}\;,
\ee
using integration by parts. Finally, the antisymmetric identity  \cite{Edgar:2001vv}
\be
R_{[\mu\nu}^{~~\,\rho\sigma} R_{\rho\sigma}^{~~\,\alpha\beta}R_{\alpha\beta]}^{~~\,\mu\nu}=0 
\ee
holds in $d\leq4$. It implies 
\be
{\cal O}_{3\times 2}=2{\cal O}_{2\times 3}\;.
\ee
In summary, the  EFT of gravity  at $\partial^6$ order in $d=4$ Minkowski space with no matter can be  described by a single operator that we choose to be $  (R_{\mu\nu}^{~~\,\rho\sigma})^3 \equiv {\cal O}_{3\times 2}   $.

 It is however  very useful to also use a set of \textit{non-reduced} operators. They can be used to perform consistency checks of the results, and to assess whether a given quantity is physical. 
We use the set 
\begin{align}
{\cal L}_{\rm eff} = \frac{1}{2}\hat R + \alpha_1 R_{\mu\nu\rho\sigma}\square R^{\mu\nu\rho\sigma} 
+\alpha_2 ( R_{\mu\nu}^{~~\,\rho\sigma})^3   +\alpha_3 (R_{\mu~\,\nu}^{~\,\rho~\,\sigma})^3 
= \frac{1}{2}\hat R +\alpha ( R_{\mu\nu}^{~~\,\rho\sigma})^3\;, 
\label{eq:EFT_lag1}
\end{align}
where $(R_{\mu~\,\nu}^{~\,\rho~\,\sigma})^3 \equiv R_{\mu~\,\nu}^{~\,\rho~\,\sigma} 
R^{~\,\alpha~\,\beta}_{\rho~\,\sigma}
R^{~\,\mu~\,\nu}_{\alpha~\,\beta}={\cal O}_{2\times 3}$.  
The relation between the coefficients is 
\be
\alpha=-3\alpha_1+\alpha_2+\frac{1}{2}\alpha_3\,. \label{eq:alpha_comb}
\ee
Any physical quantity in $d=4$ Minkowsky with no matter must receives corrections only via the combination \eqref{eq:alpha_comb}. This will be put to use in Section \ref{se:TLN_results}. 

\subsection{The Einstein-Maxwell EFT at $\partial^4$ Order}

 \label{se:EMEFT}

In the presence of matter, the leading effective operators are at 
$\partial^4$ order.
We consider gravity coupled with electromagnetism, i.e. the Einstein-Maxwell EFT. 
This is the relevant EFT in the case of charged black holes. 

The effective Lagrangian contains the Maxwell action and the respective high-order terms involving the gauge invariant field strength $F_{\mu \nu}$, such as $R F^2$, $R^{\mu \nu}F_{\mu \rho}F\indices{_\nu ^\rho}$, $R^{\mu \nu \rho \sigma}F_{\mu \nu}F_{\rho \sigma}$, $F^4$, etc.
 Our focus is again on parity-even operators. 
The reduction follows similar steps as in \ref{se:puregravEFT}, where both the Einstein and Maxwell field equations 
\be
\nabla_\mu F^{\mu\nu}=\O(\partial^4)\,,\quad \hat R_{\mu\nu}-\frac{1}{2}g_{\mu\nu}\hat R=T_{\mu\nu}^{\rm e.m.} + \O(\partial^4)\,,\quad 
 T_{\mu \nu}^{\rm e.m.} = -F_{\mu \rho}F\indices{^\rho _\nu } - \frac{1}{4}g_{\mu \nu}F_{\rho \sigma}F^{\rho \sigma} 
\ee
are used. The reduction is well-known, see e.g. \cite{Cheung:2014ega,Bittar:2024xuc}. One obtains a  basis of three operators, chosen to be 
\begin{align}
    \mathcal{L}_{\mathrm{eff}} = \frac{1}{2}\hat{R} - \frac{1}{4}F_{\mu \nu}F^{\mu \nu}  + \gamma_1 R^{\mu \nu \rho \sigma}F_{\mu \nu}F_{\rho \sigma} + \gamma_2 (F_{\mu \nu}F^{\mu \nu})^2 +  \gamma_3  (F_{\mu \nu}\tilde F^{\mu \nu})^2    \;.
\label{eq:EinsteinMawxellEFT}
\end{align}
It turns out that the $(F_{\mu \nu}\tilde F^{\mu \nu})^2 $ operator does not deform the black hole metrics studied in this work. Hence we preemptively set $\gamma_3\equiv 0$.  This is equivalent to say that the $F_{\mu\nu} F^{\nu\rho}F_{\rho\sigma}F^{\sigma\mu}$ operator contributes as half of $(F_{\mu \nu}F^{\mu \nu})^2 $, as noted in \cite{Kats:2006xp}, due to the relation $(F_{\mu \nu}\tilde F^{\mu \nu})^2 = 4 F_{\mu\nu} F^{\nu\rho}F_{\rho\sigma}F^{\sigma\mu} -2 (F_{\mu \nu}F^{\mu \nu})^2 $.

\subsection{EFT-corrected Field Equations  }
\label{se:EFE}

\subsubsection{EFT of Pure Gravity}

The  ${\cal L}_6$ Lagrangian corrects the vacuum field equations, 
\be
\hat R_{\mu\nu}-\frac{1}{2}\hat R g_{\mu\nu}=  \Tilde{T}_{\mu \nu,6} + \O(\partial^{8}) \,,\quad \quad \Tilde{T}_{\mu \nu, 6} = -\frac{2}{\sqrt{-g}}\frac{\delta(\sqrt{-g}{\cal L}_6)}{\delta g^{\mu\nu}}\,,
\label{eq:EFE_del6}
\ee
where we have introduced the effective stress tensor $\Tilde{T}_{\mu \nu,6}$.
% \be
% \Tilde{T}_{\mu \nu} = -\frac{2}{\sqrt{-g}}\frac{\delta(\sqrt{-g}{\cal L}_6)}{\delta g^{\mu\nu}}\;.
% \ee
The explicit computation of  the variations gives
\begin{align}
    \Tilde{T}^{}_{\mu \nu, 6} &= \alpha_1\Big[-g_{\mu \nu}(\nabla_\rho R_{\gamma \sigma \lambda \delta})(\nabla^\rho R^{\gamma \sigma \lambda \delta}) -8 \nabla^\rho \nabla^\sigma \square R_{\mu \rho \nu \sigma} +2 (\nabla_\mu R\indices{^\lambda _\sigma _\alpha _\beta})(\nabla_\nu R\indices{_\lambda ^\sigma ^\alpha ^\beta})  \nn \\ 
    & \; \; \; \; \; \; \; \; \; \;+ 4(\nabla_\rho R\indices{_\mu ^\lambda ^\delta ^\sigma})(\nabla^\rho R_{\nu \lambda \delta \sigma})-8 \nabla_\rho(R\indices{_\mu ^\lambda ^\delta ^\sigma}\nabla_{\nu} R\indices{^\rho _\lambda _\delta _\sigma}) +8 \nabla_\rho(R^{\rho \lambda \delta \sigma}\nabla_{\mu} R_{\nu \lambda \delta \sigma}) \nn \\ &\; \; \; \; \; \; \; \; \; \;-2 \square(R\indices{_\mu ^\alpha ^\beta ^\gamma}R\indices{_\nu _\alpha _\beta _\gamma}) \Big] \nn \\ &+ \alpha_2 \Big[g_{\mu \nu} R\indices{^\delta ^\zeta _\alpha _\beta}R_{\delta \zeta \rho \sigma}R^{\rho \sigma \alpha \beta}   -12\nabla_\alpha \nabla_\beta (R\indices{^\alpha _\mu _\rho _\sigma}R\indices{^\rho ^\sigma ^\beta _\nu})  -6 R^{\delta \gamma \rho \sigma}R_{\delta \gamma \mu \lambda}R\indices{_\rho _\sigma _\nu ^\lambda} \Big] \nn\\&+ \alpha_3 \Big[g_{\mu \nu} R\indices{_\gamma ^\lambda _\rho ^\sigma}  R^{\gamma \alpha \rho \beta} R_{\lambda \alpha \sigma \beta} -6\nabla_\alpha \nabla_\beta (R\indices{_\mu _\lambda _\nu _\sigma}R\indices{^\alpha ^\lambda ^\beta ^\sigma} - R\indices{^\alpha _\lambda _\mu _\sigma}R\indices{_\nu ^\lambda ^\beta ^\sigma})\nn\\&\; \; \; \; \; \; \; \; \; \;-6 R\indices{^\lambda_\alpha ^\sigma _\beta}R\indices{_\mu _\lambda _\sigma _\rho}R\indices{_\nu ^\alpha ^\beta ^\rho} \Big]\;
\end{align}
in which the $\mu,\nu$ have to be symmetrized. 
The expansion and identities necessary to obtain these formulae are collected in App.\,\ref{app:Var_L4}.

As explained in section \ref{se:EFT_rules},  the divergence of the effective stress tensor vanishes as a result of the diffeomorphism invariance of GR, 
\begin{equation}
    \nabla_{\mu}\Tilde{T}^{\mu \nu}_6 = 0\;.
\end{equation}

\subsubsection{Einstein-Maxwell EFT}

The ${\cal L}_4$ Lagrangian corrects both the Einstein and Maxwell equations, 
\begin{align}
    \hat{R}_{\mu \nu} - \frac{1}{2}\hat{R}g_{\mu \nu} &= T_{\mu \nu}^{{\rm e.m.}} + \Tilde{T}_{\mu \nu,4}\\
    \nabla_\nu F^{\mu \nu} &= \Tilde{J}^\mu\;
\end{align}
with the effective stress tensor $\Tilde{T}_{\mu \nu,4}$ and the effective Maxwell source $\Tilde{J}^\mu$. We find
\begin{align}
    \Tilde{T}_{\mu \nu,4} &= 
    \gamma_1 \Big(g_{\mu \nu}R^{\kappa \lambda \rho \sigma}F_{\kappa \lambda}F_{\rho \sigma} - 6 F\indices{_\alpha _(_\nu}F^{\beta \gamma}R\indices{^\alpha _\mu_) _\beta _\gamma} - 4 \nabla_\beta \nabla_\alpha(F\indices{^\alpha _( _\mu}F\indices{^\beta _\nu_)}) \Big)\nonumber \\ &+ \gamma_2 \Big(g_{\mu \nu}(F^2)^2 - 8 F^2 F\indices{_\mu ^\sigma}F\indices{_\nu _\sigma} \Big)\;,
\end{align}
and
\begin{align}
    \Tilde{J}^\mu &= 4 \gamma_1 \nabla_\nu (R^{\alpha \beta \mu \nu}F_{\alpha \beta}) + 8 \gamma_2 \nabla_\nu(F_{\rho \sigma}F^{\rho \sigma}F^{\mu \nu})\;.
\end{align}
These sources satisfy respectively the conservation equations  $\nabla_{\mu}\Tilde{T}^{\mu \nu}_4 = 0$ and  $\nabla_\mu \Tilde{J}^\mu = 0$.

\section{Gravitational EFT at One-Loop from Heavy Particles}
\label{se:OneLoopEFT}

We present an explicit computation of the order-$\partial^6$
gravitational EFT produced by loops of massive neutral particles.
This computation exemplifies some EFT aspects discussed in Section  \ref{se:EFT_basics} and also provides  the main contribution to the GREFT. 

Consider a UV Lagrangian ${\cal L}_{\rm eff, UV}$ including massive fields with spin $0,\frac{1}{2},1$,
\be
{\cal L}_{\rm eff, UV}=\frac{1}{2}\hat R+ {\cal L}_{\rm matter}+   {\cal L}_{4, {\rm UV}}+   {\cal L}_{6, {\rm UV}}+\O(\partial^{8})\;.
\ee
It  contains local higher dimensional operators with the properties discussed in Section \ref{se:EFT_R3},  that are not involved in the present loop computation. The matter particles are described by the following matter Lagrangians.

\paragraph{Spin $0$.}
The Lagrangian is
\be
{\cal L}_{0}= -\frac{1}{2}(\partial_\mu\Phi)^2 -\frac{1}{2}m^2\Phi^2 \,.
\label{eq:Lag0}
\ee
A non-minimal coupling to the scalar curvature $\Phi^2 R$ does not contribute due to the field equation \eqref{eq:EFE_del6}. 

\paragraph{Spin $\frac{1}{2}$.}
The Lagrangian is
\be
{\cal L}_{1/2}=-\frac{1}{2}\bar\Psi (\slashed D-m)\Psi  \,,
\label{eq:Lag_half}
\ee
where $\Psi$ is  a Dirac spinor. We have $\slashed D = \gamma^\mu D_\mu$ with $\gamma^\mu$ the $n\times n$ Dirac matrices in $d$ dimensions, with $n=2^{[d/2]}$ the dimension of spinor space \cite{Strathdee:1986jr,Hoover:2005uf}.

\paragraph{Spin $1$.}

The Lagrangian, including a $R_\xi$-type gauge fixing, is
\be
{\cal L}_1+ {\cal L}^{\rm  gf}_1
= -\frac{1}{4}  (F^{\mu\nu})^2  - \frac{1}{2\xi}(\nabla_\mu A^\mu)^2 \,.
\label{eq:Lag1}
\ee
The Lagrangian produces a $A_\mu A_\nu  R^{\mu\nu}$ term that vanishes due to the field equation \eqref{eq:EFE_del6}.  
In the following, we choose the Feynman gauge $\xi=1$.

\subsection{Integrating Out Massive Particles at One-Loop }

The massive particles contribute to graviton  interactions through one-loop diagrams. At energy scales below the particle mass, these contributions are encoded into the order-$\partial^6$ effective Lagrangian  given in Eq.\,\eqref{eq:EFT_lag1}.

All the effects from the loops of the  particle are encoded into the one-loop effective action.\,\footnote{For even spacetime dimensions, some of the loop diagrams  contain UV divergences. In $d=4$ these divergences renormalize the operators in ${\cal L}_4$.  These operators do not contribute in the context of our study since they can be eliminated, see Section \ref{se:EFT_basics}. } 
An efficient way to extract this information is to use the well-known expansion of the effective action into  heat kernel coefficients.  See  \cite{Gilkey_original, vandeVen:1984zk} for seminal papers and \cite{Vassilevich:2003xt} for a review. Other useful references are \cite{Fradkin:1983jc,Metsaev:1987ju,Hoover:2005uf}. Our main technical references are \cite{Vassilevich:2003xt,Hoover:2005uf}.

\subsubsection{Expanding the One-Loop Effective Action}

The one-loop effective action induced by the matter fields takes the form 
\be
\Gamma^{(1)}_{\rm mat} = (-)^F \frac{i}{2}{\rm Tr}\log\left[\left( -\square +m^2 +X \right)_{ij}\right] \,, 
\ee
with $\square =g_{\mu\nu}D^\mu D^\nu $ the Laplacian built from background-covariant derivatives and $F=0,1$ for bosonic and fermionic fields, respectively.   
The covariant derivatives give rise to a background-dependent field strength $\Omega_{\mu\nu}=[D_\mu,D_\nu]$, encoding both gauge and curvature connections. It takes the general form
\be
\Omega_{\mu\nu} = -\frac{i}{2}R_{\mu\nu}^{~~~~\rho\sigma}J_{\rho\sigma}\,,
\ee
where $J_{\rho\sigma}$ are the generators of the spin representation of the quantum fluctuation. 
$X$ is the ``field-dependent mass matrix'' of the quantum fluctuations, it is a local background-dependent quantity. 
The effective field strength $\Omega_{\mu\nu}$ and the effective mass $X$ are, together with the curvature tensor, the building blocks of the heat kernel coefficients. 
Using the heat kernel method, $\Gamma^{(1)}_{\rm mat} $ is expanded as
\begin{align}
& \Gamma^{(1)}_{\rm mat} =  (-)^F \frac{1}{2}
\frac{1}{(4\pi )^{\frac{d}{2}}}
\int_{\cal M} d^{d}x \sqrt{ |g|} \sum^\infty_{r=0} \frac{ \Gamma(r-\frac{d}{2})   }{m^{2r-d}}{\rm tr}\,b_{2r}(x)\,, 
\label{eq:Gam1_b}
\end{align}
with $\rm tr$ the trace over internal (non-spacetime) indexes. The local quantities $b_{2r}$ are referred to as the heat kernel coefficients. Terms with $2r\leq d$ with even $d$  have divergences that renormalize  ${\cal L}^{(0)}_{\rm eff}$.
In contrast, the terms with negative powers of masses in \eqref{eq:Gam1_b}  are finite. They correspond to an expansion for large $m$ and give rise to 
the one-loop contribution to the effective Lagrangian ${\cal L}^{(1)}_{\rm eff}$,
\be
 {\cal L}^{(1)}_{\rm eff} = (-)^F
\frac{1}{2}
\frac{1}{(4\pi )^{\frac{d}{2}}}
\sum^\infty_{r=[d/2]+1} \frac{ \Gamma(r-\frac{d}{2})   }{m^{2r-d}} {\rm tr}\, b_{2r}(x)\,.
\label{eq:Leff_oneloop}
\ee
Only the first heat kernel coefficients are explicitly known, we use up to $b_6$.

\paragraph{Spin $0$.}
The one-loop effective action following from the Lagrangian \eqref{eq:Lag0} is 
\be
\Gamma^{(1)}_0= \frac{i}{2} {\rm Tr}\log\left[\left( -\square +m^2   \right)\right]\,.  
\ee
The geometric invariants are $X=0$, $\Omega_{\mu\nu}=0$.

\paragraph{Spin $1/2$.}
The one-loop effective action following from the Lagrangian \eqref{eq:Lag_half} is 
\be
\Gamma^{(1)}_{1/2}= - \frac{i}{4}{\rm Tr}\log\left[\left( -\square +m^2  \right)\right]\,.  
\ee
The geometric  invariants are $X=0$ and 
\be  \Omega_{\mu\nu}= \frac{1}{4}\gamma^\rho \gamma^{\sigma}R_{\rho\sigma\mu\nu} \,. \ee

\paragraph{Spin $1$.}
For the massive spin $1$ particle, the contributions from the ghosts and the Goldstone boson must be included. In the Feynman gauge, these degrees of freedom are degenerate and do not mix. The ghosts contribute as $-2$  times a scalar adjoint. Similarly,  the Goldstone contributes as $+1$ the scalar term. As a result, the one-loop effective action following from the Lagrangian \eqref{eq:Lag1} is 
\be
\Gamma^{(1)}_{1}=  \frac{i}{2} {\rm Tr}\log\left[\left( (-\square +m^2)\delta^\mu_{~\,\nu}    \right)\right] -  \frac{i}{2} {\rm Tr}\log\left[\left( -\square +m^2   \right)\right]\,,  
\ee
where the last term is the ghost + Goldstone contribution. 
The geometric  invariants of the vector fluctuation are $X^\mu_{~\,\nu}=0$ and
\be
    (\Omega_{\mu\nu})^{\rho}_{~~\sigma}= - R^\rho_{~~\sigma\mu\nu} \,.
\ee

\label{se:EFT_1loop}

\subsection{The One-Loop Order-$\partial^6$ Effective Lagrangian }

In $d=4$ flat empty space, the $b_6$ heat kernel coefficient reduces to 
\begin{align}
 b_6 = &  \frac{1}{360}\bigg(
4 \Omega_{\mu\nu}\square \Omega^{\mu\nu}
+6 R_{\mu\nu\rho\sigma}\Omega^{\mu\nu}\Omega^{\rho\sigma}
-12 \Omega_{\mu\nu}\Omega^{\nu\rho}\Omega^{~~\mu}_{\rho}
\bigg)   
\\ \nn & 
+\frac{1}{7!}\bigg(
 3 R_{\mu\nu\sigma\lambda} \square R^{\mu\nu\sigma\lambda} 
+ \frac{44}{9} (R^{\mu\nu}_{~~\alpha\beta})^3   + \frac{80}{9} (R_{\mu~~\rho~~}^{~~\nu~~\sigma})^3 
\bigg) I \;.
\label{eq:b6simplified}
\end{align}
Putting together the ingredients from the previous subsections, we obtain the coefficients $\Delta\alpha_{i,s}$ of the $\Delta{\cal L}_{6,s}$ effective Lagrangian  \eqref{eq:EFT_lag1},
\be
 {\cal L}_{6, {\rm IR}}= {\cal L}_{6, {\rm UV}}+\Delta {\cal L}_{6,s}\,,
\ee
\be
\Delta {\cal L}_{6,s} = 
 \Delta\alpha_{1,s} R_{\mu\nu\rho\sigma}\square R^{\mu\nu\rho\sigma} + 
 \Delta\alpha_{2,s} (R_{\mu\nu}^{~~\,\rho\sigma})^3 +
 \Delta\alpha_{3,s}(R_{\mu~\,\nu}^{~\,\rho~\,\sigma})^3\,,
\ee
with 
\begin{subequations}
\begin{align}
\left(\Delta\alpha_{1,0},\Delta\alpha_{2,0},\Delta\alpha_{3,0}\right) & =\frac{1}{32\pi^2m^2}\left(\frac{1}{1680},\frac{11}{11340},\frac{1}{567} \right) \,,
\label{eq:R3_1loop_scalar}
\\
\left(\Delta\alpha_{1,\frac{1}{2}},\Delta\alpha_{2,\frac{1}{2}},\Delta\alpha_{3,\frac{1}{2}}\right)&=\frac{1}{32\pi^2m^2}\left(\frac{1}{315},\frac{101}{22680},\frac{109}{11340}\right)\,,
\label{eq:R3_1loop_fermion}
\\
%%%%%%%%%%%%%%%%%%%%%%%%%%%
%Vector Contribution:
% \left(\Delta\alpha_{1,3},\Delta\alpha_{2,3},\Delta\alpha_{3,3}\right)=\frac{1}{32\pi^2m^2}\left(-\frac{11}{1260},-\frac{29}{2268},-\frac{149}{5670} \right) \\
%%%%%%%%%%%%%%%%%%%%%%%%%%%%
\left(\Delta\alpha_{1,1},\Delta\alpha_{2,1},\Delta\alpha_{3,1}\right)& =\frac{1}{32\pi^2m^2}\left(-\frac{47}{5040},-\frac{13}{945},-\frac{53}{1890} \right) \,.
\label{eq:R3_1loop_vector}
\end{align}
\end{subequations}
Reducing the operators using the relations in Section \ref{se:O6basis}, we obtain 
\be
 {\cal L}_{6, {\rm IR}}= {\cal L}_{6, {\rm UV}}+\Delta \alpha_s (R^{\mu\nu}_{~~\alpha\beta})^3 
\ee
with 
\be
\left(\Delta\alpha_{0},\Delta\alpha_{\frac{1}{2}},\Delta\alpha_{1}\right)=\frac{1}{483840\pi^2}\left(1,-4,3\right)\,. 
\label{eq:R3_loop}
\ee
In the spin-$1$ case, we used that $\Delta\alpha_{i,1}=\Delta\alpha_{i,V}-\Delta\alpha_{i,0}$. The scalar and vector are real, and the fermion is Dirac type. The result from a Majorana fermion is obtained by dividing   the $\Delta\alpha_{i,\frac{1}{2}}$ by two. 

The final result  \eqref{eq:R3_loop} is remarkably simple
and vanishes in supersymmetric theories. It matches the one found in \cite{Goon:2016mil}.

\subsection{The Gravitational EFT of the Real World}

\label{se:GREFT}

The gravitational EFT that arises just below the string scale has vanishing order-$\partial^6$ operators, i.e., $\alpha=0$ due to supersymmetry.\,\cite{Jack:1988sw,Gross:1986iv,Gross:1986mw,Kikuchi:1986rk,Becker:2006dvp}. 
In contrast, \eqref{eq:R3_loop} makes clear that integrating out massive particles features generically order-$\partial^6$ operators suppressed by $\frac{1}{m^2}$. 
This is the case of  the gravitational EFT of the real world, that arises at scales larger than the Compton wavelength of Standard Model particles.

Gravity in the real world is observed at scales down to $\O(10 )\,\mu$m, see e.g. \cite{Smullin:2005iv}. 
The lightest known particle masses are the neutrinos, with $m_\nu=\O(0.1)$eV.
Hence the massive neutrinos produce the leading SM contribution to the GREFT at length scales larger than the $\mu $m. 
 The exact GREFT coefficients  from one Majorana neutrino are   \be \alpha_{i,\frac{1}{2},{\rm GREFT}}=\frac{1}{2}\Delta \alpha_{i,\frac{1}{2}}\,,  \ee
which are given in \eqref{eq:R3_1loop_fermion}.

\section{Love Numbers in the EFT of Gravity 
\label{se:TLNs}} 

\subsection{The Tidal  Deformability of Black Holes}

From the viewpoint of an observer  localized at $r\gg r_h$,  a black hole, like any other spatially localized  object, can be described by a  wordline effective field theory in pure Minkowski space. 
The leading term of the worldline EFT describes the black hole as a pointlike object, to which  external fields couple. 
The higher order derivative terms of this EFT encode information about the black hole  shape and its response to external fields.\,\footnote{
The worldine EFT also has application in other fields such as atomic physics \cite{Burgess:2017mhz} or superradiance \cite{Endlich:2016jgc}. The  electromagnetic polarizability of composite objects such as neutral ions \cite{Feinberg:1968zz} or neutral strings \cite{Fichet:2016clq}, which amounts to  deformability  under a vector tidal field,  
is also similarly described via EFTs. 
} 

 \subsubsection{The Worldline Effective Field Theory in a Nutshell}

General details about the wordline EFT applied to black holes can be found in e.g.\,\cite{Goldberger:2004jt, Goldberger:2005cd,Porto:2016pyg,Hui:2020xxx,Ivanov:2022hlo}. 
Our focus here is on the deformability of black holes under the effect of an external field. This is encoded into the \textit{quadratic} terms of the worldline effective action. 

For a brief conceptual review it is enough to focus on a scalar external field $\Phi$ and choose the black hole rest frame.  
The worldline effective action reads
\begin{align}
S_{\rm WL}[\Phi]=S_{\rm kin}[\Phi] + \int d\tau e \left( {\cal L}_{\rm point} + {\cal L}_{\rm quad}[\Phi] \right)\,, \\ 
S_{\rm kin}[\Phi]=-\frac{1}{2}\int d^dx \partial_\mu \Phi\partial^\mu \Phi\,,\quad\quad 
{\cal L}_{\rm point} = \frac{1}{2} e^{-2} \partial_\tau x_\mu \partial_\tau x^\mu-\frac{1}{2}m^2\,,\\ 
{\cal L}_{\rm quad}[\Phi] = \sum_{\ell=1}^\infty 
\frac{\lambda_\ell}{2\ell!} {\cal O}_\ell\,,\quad\quad 
{\cal O}_\ell = \left(\partial_{(i_1} \partial_{i_2}\ldots \partial_{i_\ell)_{\rm T}}\Phi \right)^2\,,
\end{align}
where $\tau $ is a coordinate that parametrizes the worldline of the black hole. 
The leading term ${\cal L}_{\rm point}$ is simply the Polyakov point particle Lagrangian. 
The ${\cal L}_{\rm quad}$ effective Lagrangian 
encodes the information about the finite shape of the black hole.  
The quadratic operators ${\cal O}_\ell$  are defined such that  each of them transform in a different irreducible representation of the spatial rotation group, and thus describes the deformability of the black hole in a given spherical harmonic $\ell$ (see App.\,\ref{sec:spherical_harmonics}). 

The action can easily be covariantized. Hence the $\lambda_\ell$ coefficients represent a covariant description of the black hole deformability  under the effect of the $\Phi$ tidal field.

\subsubsection{Matching the Worldline EFT to the EFT of Gravity }

  \begin{figure}[t]
      \centering
\includegraphics[width=0.8\linewidth,trim={1.5cm 6cm 2cm 5cm},clip]{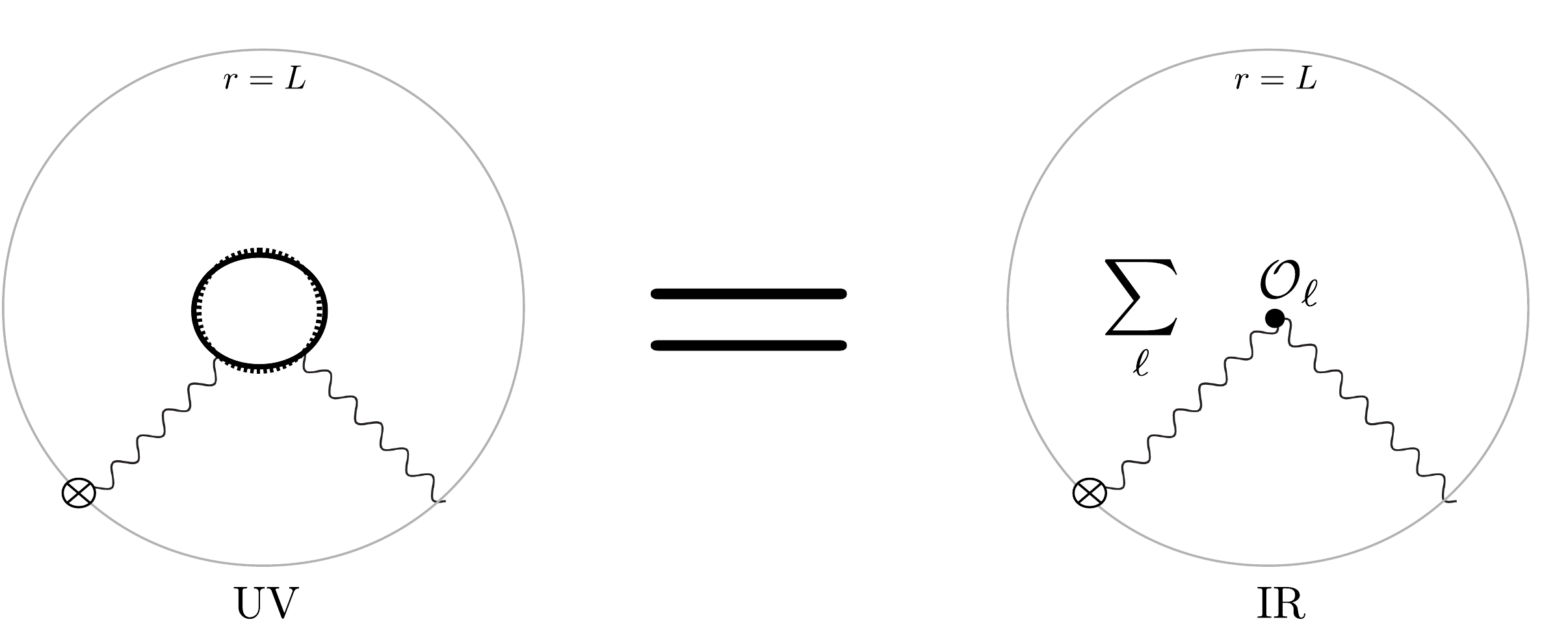}
\caption{The matching of the black holes static response of the gravitational theory (UV) to the worldine EFT (IR).
\label{fig:matching_WL_EFT} }
  \end{figure}

The worldline EFT can be used as the infrared EFT of the black hole solutions computed in GR. 
As usual in effective field theory, we are able to derive the infrared EFT from a given
 ultraviolet theory. 
We can thus derive the $\lambda_\ell$ coefficients from the EFT of gravity. 

Since the ${\cal O}_\ell$ operators are quadratic, a convenient setup   to probe them is to assume that the tidal field is sourced in a given harmonic. The strength of the response, computed perturbatively on both worldline EFT and gravity sides, gives then access to $\lambda_\ell$. 

This is a familiar matching procedure between an IR EFT and its UV completion. 
The matching relies on physical observables, hence the physical quantities in the worldline EFT must depend on the combination of operators \eqref{eq:alpha_comb}, i.e. on the $\alpha$ coefficient. 
The matching of the observables is performed on a sphere with radius $r|_{\rm matching}\equiv L$, as described in Fig.\,\ref{fig:matching_WL_EFT}.  This aspect is important when the coefficients feature a renormalization flow, in which case $\lambda_\ell = \lambda_\ell(L)$. 

The coefficients characterizing the response on the gravity side are generically referred to as {\it Love numbers}, denoted $k_\ell$.\,\footnote{{Historically, the Love numbers refer to the even-parity sector of the tensorial tidal response in the Newtonian limit} \cite{Love}. Here we extend the naming to all types of tidal fields {and parity sectors}. 
}
Their correspondence to the $\lambda_\ell$ coefficients involves a normalization factor $N_\ell$ such that 
\be
\lambda_\ell \equiv N_\ell k_\ell\,.  
\label{eq:matching_coef}
\ee
The $N_\ell$ were computed systematically in \cite{Hui:2020xxx} for each tidal fields, some of them will be specified in the next sections.

\subsection{Structure and Properties}

Consider a generic tidal field  $\Psi$ living on a non-spinning black hole background that may be either neutral or charged. In the charged case, the $r_h$ coordinate denotes  the outer horizon. 

 We denote the exact wave operator as ${\cal D}$. The corresponding tidal equation of motion is 
\be
{\cal D}_r \Psi(r) =0\,. \label{eq:EOM} \ee
The equation of motion is a second order linear differential equation with  the following elementary properties: 
\begin{enumerate}[label=(\roman*)]
    \item ${\cal D}_r \Psi(r) =0$ reduces to the Minkowski space equation of motion in the limit  $\frac{r_h}{r}\to 0$. \label{Prop:flat}
    \item ${\cal D}_r \Psi(r) =0$ has a regular singularity at the outer horizon $r=r_h$. 
    \label{Prop:sing}
\end{enumerate}
The  singular point is of the  regular kind, as writing ${\cal D}_r \Psi(r) \propto \Psi''(r)+p(r)\Psi'(r)+q(r)\Psi(r)$  we have $q(r),p(r)\sim \frac{1}{r-r_h}$ near the singularity. Since the  equation of motion has a  regular singularity, we know that it admits only two independent solutions and that we can apply the Frobenius (i.e. generalized power series) method \cite{gray2008linear}.

\subsubsection{Near-horizon Behavior}

We consider the near-horizon region, $r\sim r_h$. Applying the Frobenius method, the indicial equation obtained using  $\Psi(r)\sim (r -r_h)^q$  gives simply $q^2=0$. It has degenerate roots, which readily implies  the following property:
 \be
 \shortstack[l]{
\textrm{
In the vicinity of the horizon, there exists one regular solution,} \\ \textrm{ while the other diverges logarithmically.}} 
\label{Prop:solutions}
 \ee
 This fact  depends only on Prop.\,\ref{Prop:sing}, i.e. on the singularity structure of the tidal equation of motion, that is not altered in the presence of the EFT corrections. Prop.~\eqref{Prop:solutions} ensures  that the Love numbers are well-defined in the EFT framework, because their computation relies on the solution that is regular at the horizon.

Using Prop.~\eqref{Prop:solutions}, the general solution near the horizon  can be written as  
\be \Psi_{\ell}(r)|_{r\sim r_h}=c_{\ell} \psi_{\ell}(r)+ \tilde c_{\ell} \tilde\psi_{\ell}(r)\,, \ee  
where  $\psi$ is the regular solution on the horizon: 
\be
\psi_{\ell}(r\to r_h)={\rm finite}\,,\quad\quad  \tilde \psi_{\ell}(r\to r_h)\sim \log(r-r_h)\,. 
\ee
Isolating  the regular solution $\psi_{\ell}$ is key for the computation of Love numbers.

\subsubsection{Asymptotic Behavior and Scale Anomaly}
\label{se:large_r}

We consider the limit $r\gg r_h$. 
Naively setting $r_h\to 0$ in the equation of motion, Prop. \ref{Prop:flat} implies that the solutions to \eqref{eq:EOM}  have the form 
$
\Psi_\ell(r)  \underset{ r_h=0}{\sim} \hat a_{\ell} r^{\ell+1}+ \frac{\hat b_{\ell}}{r^{\ell}}\,,  
$
which is simply  the Minkowski space solutions. This property is independent of the presence of the EFT corrections, which decrease faster than the leading behavior for large $r$. 

There is however an important subtlety tied to taking the $r_h \to 0$ limit, which rigorously holds only for $r=\infty$. To analyze more finely the asymptotic behavior, we perform a variable change $r\propto \frac{1}{z} $ that maps infinity to $z=0$. The tidal equation of motion in these coordinates presents again a regular singular point. The associated indicial equation from $\Psi\sim z^q$ gives two roots $q_1=\ell$, $q_2=-\ell -1$. It follows from the Frobenius method that whenever $\ell$ takes its physical values $\ell\in \mathbb{N}$, we have $q_1-q_2\in \mathbb{N}^*$ so that the solution translated back to the $r$ coordinates 
takes the general form
\be
\Psi_\ell(r) =  r^{\ell+1}\sum^{\infty}_{n=0} \hat a_{n,\ell} r^{-n}+  \frac{1}{r^\ell}\left(1+B \log\left(\frac{r_h}{r}\right)\right)\sum^{\infty}_{n=0} \hat d_{n,\ell} r^{-n}\,,
\label{eq:Psi_asymptotic}
\ee
where the coefficient $B$ may be zero. The fact that only the $r^{-\ell}$ term in \eqref{eq:Psi_asymptotic} can have a log factor is an intrinsic property of the solution dictated by the Frobenius method.\,\footnote{
The log term is associated to the solution with lower indicial root, which is here $q_2=-\ell-1$, in which it appears as a factor of the solution  with highest indicial root, here  $q_1=\ell$. 
In \eqref{eq:Psi_asymptotic} the log term has been factored into the $r^{-\ell}$ term, however in terms of independent solutions it accompanies the $r^{\ell+1}$ term. 
}

The key difference with the solution at $r_h=0$ is the appearance of the logarithm of $r$, in which we have included the only scale available in the problem, $r_h$,  in order to make the argument dimensionless. We see that sending $r_h$ to $0$ at fixed $r$ is ill-defined unless the $B$ coefficient goes to zero with $r_h$ as a polynomial. This is why one  misses this term if one just requires $r_h=0$. 
The  asymptotic form \eqref{eq:Psi_asymptotic} is valid also in the presence of the EFT corrections, which decrease faster than the leading asymptotic terms and thus do not modify the singularity structure  at $z=0$.

\paragraph{Some notations.}
We introduce the $r$-dependent coefficients
\be
 \hat b_{n,\ell} \equiv   \left(1+B \log\left(\frac{r_h}{r}\right)\right)\hat d_{n,\ell}\,. \label{eq:hatb_def}
\ee
The $\hat b_{n,\ell}$ have a logarithmic dependence in $r$, but 
it is understood that the solutions are ultimately matched to the worldline EFT at a given length scale $r\sim L$. Hence 
the $\hat b_{n,\ell}$ can in all aspects be treated as constants in the following.

The leading coefficients in the solution \eqref{eq:Psi_asymptotic}
are denoted \be \hat a_{0,\ell}\equiv 
\hat a_{\ell}\,,\quad\quad
\hat b_{0,\ell}\equiv 
\hat b_{\ell}\,.
\ee
It is also convenient to define a {transfer} matrix $\cal T$ that relates the solutions at infinity  to those near the horizon, 
\be
\begin{pmatrix}
\hat a_{\ell} \\
\hat b_{\ell}
\end{pmatrix} = 
{\cal T}
\begin{pmatrix}
c_{\ell} \\
\tilde c_{\ell}
\end{pmatrix}\,.
\ee

\subsubsection{Resolving Ambiguities in Monomial Identifications}

In the scope of computing the Love numbers, we need to identify certain monomials in the solution \eqref{eq:Psi_asymptotic}. In the absence of the scale anomaly ($B=0$), an ambiguity appears for integer $\ell$, because the terms with $\hat a_{\ell+1+p,\ell}$, $p\in \mathbb{N}$  overlap with the $\hat b_{n,\ell}$ ones. The $\hat a_{\ell+p,\ell}$ terms are identified in the worldline EFT as graviton corrections to the source, that should be subtracted \cite{Kol:2011vg}. 
A proposed shortcut in the literature is to isolate the wanted monomials  for $\ell$ analytically continued to  $\mathbb{R}$ \cite{LeTiec:2020bos,Charalambous:2021mea}.

Our computation leading to \eqref{eq:Psi_asymptotic} shows that whenever $B\neq 0$, no ambiguity is possible between the monomials due to the presence of the logarithm. The Frobenius method dictates that the log necessarily accompanies the response, not the source. 
On the worldline EFT side, this means that 
no logarithms are expected to appear in the corrections to the source. 
This resolves, for $B\neq 0$, the puzzle of the ambiguous determination of the monomials.

\subsection{The Love Numbers}
\label{se:TLN_def}

The Love number for a given $\ell$ is obtained by considering the asymptotic behavior of the solution that is regular on the horizon. The structure is the same as in \eqref{eq:Psi_asymptotic}, but the coefficients are different, we denote them with no hat. Our interest lies in the leading terms 
\be \Psi_{\ell}(r) \bigg|_{\rm regular}  \underset{r\gg r_h}{\sim}  a_{\ell} \,r^{\ell+1}+ \frac{b_{\ell}}{r^{\ell}}\,,
  \quad \quad
 \begin{pmatrix}
 a_{\ell} \\
b_{\ell}
 \end{pmatrix} = 
 {\cal T}
 \begin{pmatrix}
 c_{\ell} \\
 0
 \end{pmatrix}\,.
\ee  
The Love numbers are defined by the ratio 
\be
k_\ell =  \frac{1}{r_h^{2\ell+1}} \frac{b_{\ell}}{a_{\ell}}= \frac{1}{r_h^{2\ell+1}}\frac{{\cal T}_{2,1}}{{\cal T}_{1,1}}\,.
\ee
In this definition, the $r_h$ power is simply introduced by dimensional analysis to make $k_\ell$ dimensionless. 
These numbers   measure the static response of the black holes geometry under a tidal source with amplitude $a_{\ell}$ at infinity. 

In GR with $d=4$, the {transfer} matrix happens to be ${\cal T}_{\rm GR}=\mathds{1}_2$. As a result, the regularity at the horizon implies $\tilde c_{\ell,h}=b_{\ell} =0$ for all $\ell$, so that the Love numbers vanish.  It is, however, a particular case that can be explained from symmetry arguments \cite{Charalambous:2021kcz,Hui:2021vcv,Hui:2022vbh,Charalambous:2022rre,Combaluzier-Szteinsznaider:2024sgb,Kehagias:2024rtz,Gounis:2024hcm}.

\subsection{Scale Anomaly and Running Love Numbers }

\label{se:RG_discussion}

We have seen via the Frobenius method in section 
\eqref{se:TLN_def} that the Love numbers may depend on $\log(\frac{r}{r_h})$.
 The matching to a corresponding coefficient of the worldline EFT $\lambda(L)$ for a response defined at the matching scale $r=L$
 gives  
\be
\lambda(L)\equiv - N_\ell B d_{0,\ell} \log\left(\frac{L}{r_h} \right)+{\rm cst}\,, \label{eq:lambdaLdef}
\ee
where the irrelevant extra constant is present from the gravity side and can be also introduced arbitrarily in the definition of the wordline coefficient.

Applying $-L \frac{d}{dL}$ to \eqref{eq:lambdaLdef} gives the beta function of the wordline EFT coefficient. 
\be
\beta_\lambda \equiv - \frac{d}{d\log L} \lambda(L) = N_\ell B d_{0,\ell} \,. 
\ee
The minus sign is introduced to match the usual definition of the beta function from QFT. 

Formally, this beta function describes how the wordline EFT coefficient changes if we look at the theory at different scales $L$. 
It also  describes a physical phenomenon. Starting from an observed value $\lambda_0$ at a scale $r_0$, the beta function controls how the observable effect changes at a scale $r_1$ via the renormalization flow $\lambda(r_1)=\lambda(r_0)- N_\ell B d_{0,\ell}\log\left(\frac{r_1}{r_0}\right)$, where $\lambda(r_0)=\lambda_0$.\,\footnote{The renormalization flow of Love numbers was first pointed out in \cite{Kol:2011vg} for non-physical (half-integer) values of $\ell$. The running due to the $R_{\mu\nu}^3$ operator was pointed out in \cite{Charalambous:2022rre,DeLuca:2022tkm}.
A running of the dynamical Love numbers is also discussed in \cite{Mandal:2023hqa}.
The running of dimensionless quantities at the classical level might seem surprising at first view, but this phenomenon
also occurs in certain holographic quantities, see e.g. \cite{Contino:2002kc,Randall:2001gb,Goldberger:2002hb,Fichet:2019owx,Fichet:2021xfn}. 
}

\subsection{Static Response Perturbation Theory: General Structure \label{se:perturbation_theory}}

In the validity regime of the EFT, the contributions of the effective operators from ${\cal L}_{\rm eff}$
to the Love numbers can be treated perturbatively.
In this section and the next we consider a single perturbation with coefficient $\alpha$. The generalization to various perturbations used in section \ref{se:TLN_results} is trivial. We also omit the $\ell$ index of all the $\psi$'s and $\Psi$'s for simplicity.

We write the  wave operator and the tidal field from \eqref{eq:EOM} as
\be
{\cal D}= {\cal D}^{(0)}+ \alpha {\cal D}^{(1)}  +\O(\alpha^2) \,,\quad \quad  \Psi=\Psi^{(0)}+\alpha \Psi^{(1)}+ \O(\alpha^2)\,.
\label{eq:PT_WO}
\ee
Plugging these expansions into the equation of motion provides the equation for the tidal field perturbation
\be
{\cal D}^{(0)} \Psi^{(1)} =-{\cal D}^{(1)} \Psi^{(0)}  +\O(\alpha^2) 
\label{eq:EOM_Psi1}
\,.
\ee
Schematically, the  {transfer} matrix from horizon to infinity will be corrected as
\be{\cal T}={\cal T}^{(0)}+\alpha {\cal T}^{(1)}+ \O(\alpha^2)\,. 
\ee
Hence the asymptotic coefficients from the regular solution are generically corrected, $a=a^{(0)}+\alpha a^{(1)}+\O(\alpha^2)$, $b=b^{(0)}+\alpha b^{(1)}+\O(\alpha^2)$. In turn the Love number is corrected as 
\be
k=k^{(0)}+\alpha\left(\frac{a^{(1)}}{b^{(0)}}-\frac{b^{(1)}}{(a^{(0)})^2}\right)+\O(\alpha^2)\,. \label{eq:TLN_correction_rough}
\ee
Below we determine in more details the content of \eqref{eq:TLN_correction_rough}. 

\subsubsection{The perturbed solutions}

\label{se:PT_sol}

The two solutions to ${\cal D}_r \Psi=0$ can be written in perturbative form, 
\begin{align}
\Psi = c_1\psi_1 + c_2 \psi_2\,,\quad\quad   \psi_i= \psi^{(0)}_i+\alpha \psi^{(1)}_i + \O(\alpha^2) 
\label{eq:psi1def}
\end{align}
with 
\be
\Psi^{(0)}= c_1\psi^{(0)}_1 + c_2 \psi^{(0)}_2\,,\quad \quad
\Psi^{(1)}= c_1\psi^{(1)}_1 + c_2 \psi^{(1)}_2\,.
\ee
We have the freedom  to choose $\psi^{(0)}_1$, $\psi^{(0)}_2$ to be respectively the regular and divergent solutions at leading order, i.e.
\be
\psi^{(0)}_1 \equiv  \psi^{(0)}\,,\quad \quad \psi^{(0)}_2 \equiv  \tilde \psi^{(0)} \label{eq:sol_gen_PT}
\ee
with 
${\cal D}^{(0)}_r\psi^{(0)}_i(r)=0$. This convenient choice will make some simplifications manifest.  

Using Prop.\,\ref{Prop:solutions}, we know that we can parametrize the asymptotic behaviors as 
\be
\tilde \psi^{(0)}(r)\underset{r\to r_h}{\sim}  C \log\left(r-r_h\right)\,,\quad \quad 
\psi_i^{(1)}(r) \underset{r\to r_h}{\sim} C_i \log\left(r-r_h\right)\,.
\ee
with $C\neq 0$ by definition of $\tilde \psi$, while $C_1$, $C_2$ may or not be zero.  
Plugging in the general solution \eqref{eq:sol_gen_PT}, regularity requires 
$(C+\alpha C_2)c_2=-\alpha c_1 C_1$. The regular perturbed solution is thus 
\be
\Psi  \big|_{\rm regular} = (C+\alpha C_2)\psi^{(0)}- \alpha C_1 \tilde \psi^{(0)} +\alpha C \psi^{(1)}_1
+ \O(\alpha^2)\,. 
\label{eq:Psi_PT_1}
\ee

We can see that a simplification occurs: the $\psi^{(1)}_2$ perturbation,  which in our definition  is associated with the singular solution $\tilde \psi^{(0)}$, does not appear in the combination \eqref{eq:Psi_PT_1}. In fact, the only piece of information remaining about $\psi^{(1)}_2$ is the $C_2$ coefficient.

We then expand for large $r$ using 
\be
\psi^{(0)} \underset{r\gg r_h}{\sim}  a^{(0)}_{\ell} \,r^{\ell+1}
+ \frac{b^{(0)}_{\ell}}{r^{\ell}}\,, \quad \quad
\tilde \psi^{(0)} \underset{r\gg r_h}{\sim}  \tilde  a^{(0)}_{\ell} \,r^{\ell+1}+ \frac{\tilde b^{(0)}_{\ell}}{r^{\ell}}\,.
\label{eq:psi0_asymptotics}
\ee
The $\psi^{(1)}_1$ solution has technically a more complicated asymptotic behavior, since it is the solution of \eqref{eq:EOM_Psi1}. On the other hand, 
we do know that the \textit{exact} solution to the perturbed equation of motion, given by $\psi^{(0)}_1+\alpha \psi^{(1)}_1 + \O(\alpha^2)  $,
must have the same asymptotics as in  \eqref{eq:psi0_asymptotics}. We can thus write 
\be
\psi_1^{(1)} \underset{r\gg r_h}{\sim}  a^{(1)}_{\ell} \,r^{\ell+1}
+ \frac{b^{(1)}_{\ell}}{r^{\ell}} 
+ \O(\alpha) \label{eq:psi1_asymptotics}
\,,
\ee
where it is understood that some mild  $r$ dependence might be hidden in the $\O(\alpha) $ and has to be considered as artifact of perturbation theory.

Combining  \eqref{eq:psi0_asymptotics} and \eqref{eq:psi1_asymptotics} we find the most general correction to the Love numbers at first order in $\alpha$,
\begin{align}
k=k^{(0)}+\alpha k^{(1)}+ \O(\alpha^2)\,, \quad 
k^{(1)}=k^{(0)}\left( \frac{a_1^{(1)}}{a^{(0)}}-\frac{C_1 \tilde a^{(0)}}{C a^{(0)} } \right) + \left( \frac{b_1^{(1)}}{a^{(0)}}-\frac{C_1 \tilde b^{(0)}}{C a^{(0)} } \right)\,.
\label{eq:TLN_PT}
\end{align}

\subsubsection{Discussion}

We can see that the dependence in $C_2$ has vanished from the final expression \eqref{eq:TLN_PT}, because it contributes only at higher order. Technically, this implies that the $\psi^{(1)}_2$ perturbation to the tidal field can be simply ignored  when solving the perturbed equation of motion \eqref{eq:EOM_Psi1}.

In GR with $d=4$, we have $k^{(0)}=0$. We see  from \eqref{eq:TLN_PT} that in this particular case, the EFT operators  induce nonzero Love numbers. Two contributions can be distinguished. The one from $C_1$ occurs if the   $\psi^{(1)}_1$ perturbation diverges at the horizon. This effect happens because the regularity condition gets corrected. The effect from $b_1^{(1)}$ occurs instead because of the behavior of $\psi^{(1)}_1$ at large $r$.

The computation of each coefficients in \eqref{eq:TLN_PT} can become a daunting task  at large $\ell$   if one tries to compute $\psi_1^{(1)}$ exactly and then take limits. In the following section we present a better approach.

\subsection{Static Response Perturbation Theory: Two Simple Formulae}

We solve the perturbed  tidal equation of motion \eqref{eq:EOM_Psi1}, reproduced here for convenience: ${\cal D}^{(0)} \Psi^{(1)} =-{\cal D}^{(1)} \Psi^{(0)}$. 
We are more precisely interested in the  perturbed solution that is regular at the horizon, $\Psi\big|_{\rm regular}$, whose structure has been determined in \eqref{eq:Psi_PT_1}. We may notice that the leading order  is  regular since we have $\Psi^{(0)}\propto \psi^{(0)}$ by construction. Hence the perturbation $\Psi^{(1)}$  is separately regular at the horizon.  We can thus write
\begin{align}
\Psi\big|_{\rm regular}& = \Psi^{(0)}\big|_{\rm regular}+ \alpha \Psi^{(1)}\big|_{\rm regular}\equiv C\left( \psi^{(0)} + \alpha\psi^{(1)}  \right)\,,
\end{align}
where we defined  the normalized regular perturbation
\be\psi^{(1)}  =\psi_1^{(1)} -\frac{C_1}{C}\tilde \psi^{(0)}\,.\ee
This $\psi^{(1)}$ perturbation is the key quantity to focus on. 
We have discarded the contribution from $C_2$ since we have already established it does not contribute to the Love number.

\subsubsection{Solving via Green Function}

An efficient way  to derive the regular tidal perturbation $\psi^{(1)}$ is presented here. 
Since both the leading terms and the perturbation are regular, we can write the perturbed equation of motion  as 
\be {\cal D}^{(0)} \psi^{(1)} =-{\cal D}^{(1)} \psi^{(0)}
\label{eq:EOM_pert_gen}
\,. \ee
We assume that ${\cal D}^{(0)} $ has the canonical form \be {\cal D}_r^{(0)} = \frac{d^2}{d r_\star^2} - V(r)\,. \ee
We have introduced the tortoise coordinate that satisfies $dr_\star = \frac{d r}{f(r)}$. The function  $f(r)$ is the blackening factor coming from the black hole geometry that is written explicitly in next section.

To solve equation \eqref{eq:EOM_pert_gen}, we introduce the Green function that inverts the ${\cal D}^{(0)} $ operator,
\be
{\cal D}_r^{(0)} G(r,r') = \delta(r_\star-r_\star') \,,\label{eq:Green_def}
\ee
where $\delta(r_\star-r_\star')= f(r)\delta(r-r')$.  The r.h.s. in \eqref{eq:Green_def} can in principle  be determined by the variation of the  action of the tidal field. It can also be deduced directly from consistency with the Wronskian of ${\cal D}_r^{(0)}$ \cite{Fichet:2023dju}.

A Green function is fixed upon specifying the boundary conditions. 
We require regularity at the horizon and at infinity. 
The solving follows standard ODE  techniques, see e.g. \cite{Fichet:2019owx}. The complete solution is given in App.~\ref{app:Green}. 
We obtain the tidal  Green function 
\be
G(r,r')= \frac{\psi^{(0)}(r_<)  \psi_+^{(0)}(r_>)}{N_{\psi,\psi_+}}\,, 
\label{eq:Green}
\ee
where $r_>={\rm max}(r,r') $, $r_<={\rm min}(r,r') $. Here $ \psi_+^{(0)}$ is the leading order solution that is regular  at infinity. 
$N_{\psi,\psi_+}$ is a normalization factor computed from the Wronskian using
$W(\psi_1,\psi_2)=\psi_1 \psi'_2- \psi_1' \psi_2=\frac{N_{\psi_1, \psi_2 }}{ f(r)} $ (see App.~\ref{app:Green}).
It makes the Green function invariant under rescaling of any of the solutions.\,\footnote{An analogous Green function has been independently found in \cite{Barura:2024uog} in a different context.}

The general  solution to the equation of motion is 
\begin{align}
\psi^{(1)}(r)&=\int^{\infty}_{r_h} dr_\star' G(r,r') {\cal D}^{(1)}\psi^{(0)}(r') 
\label{eq:psi1solgen}
\\
&= \frac{1}{N_{\psi,\psi_+}} \left[
\psi_+^{(0)}(r)\int^{r}_{r_h} dr_\star'  \psi^{(0)}(r')  {\cal D}^{(1)}\psi^{(0)}(r')
+
\psi^{(0)}(r)\int^{\infty}_{r} dr_\star' \psi_+^{(0)}(r') {\cal D}^{(1)}\psi^{(0)}(r')
\right]\,. \nn
\end{align}
In the second line, we have {replaced} the explicit expression for the tidal Green function \eqref{eq:Green}.

\subsubsection{Extracting the Love Numbers }

The perturbative computation of $\psi^{(1)}$  contains all the information needed to compute the Love numbers. Since $\psi^{(1)}$ is the regular perturbation, we know from \eqref{eq:Psi_PT_1} that 
it encodes the combination of coefficients 
\be
\bar a^{(1)}\equiv a_1^{(1)}-\frac{C_1}{C}\tilde a^{(0)}\,,\quad\quad
\bar b^{(1)}\equiv b_1^{(1)}-\frac{C_1}{C}\tilde b^{(0)}\,.
\ee
These are the combinations that appear in the Love number,  \eqref{eq:TLN_PT}.
Let us extract $\bar a^{(1)}$ and $\bar b^{(1)}$ from $\psi^{(1)}$.

We take the large $r$ limit of \eqref{eq:psi1solgen}.
 Using the asymptotics from \eqref{eq:psi0_asymptotics}, we know that the condition of regularity at infinity implies that 
 \be
 \psi_+^{(0)} \underset{r\gg r_h}{\sim}  \frac{b^{(0)}_{+,\ell}}{r^{\ell}}\,, 
 \label{eq:psiPM_asymptotics}
 \ee
 where $a^{(0)}_{+,\ell}=0$ due to regularity. 
We also notice that in terms of powers of $r$, the ${\cal D}^{(1)}$ operator contributes in \eqref{eq:psi1solgen} as $\sim r^{-3}$, which ensures that the integral in the second term is finite since the integrand behaves asymptotically as $r^{-2}$  {in the worst case}.

The first term in \eqref{eq:psi1solgen} behaves asymptotically as
\be
 \frac{b^{(0)}_{+,\ell}}{r^{\ell}}\int^{r}_{r_h} dr_\star'  \psi^{(0)}(r')  {\cal D}^{(1)}\psi^{(0)}(r') \,.
 \label{eq:overlap1_gen}
\ee
We see that the integral is proportional to $r^{-\ell}$, hence the constant piece of the integral contributes to 
the $\bar b^{(1)}$ coefficient in the asymptotic expansion of $\psi_1^{(1)}$ defined in  \eqref{eq:psi1_asymptotics}. 
The $r$-dependent pieces of the integral instead contribute as subleading  corrections  to the source and response.

The second term in \eqref{eq:psi1solgen} behaves asymptotically as
\be
\left(
a^{(0)}_{\ell} \,r^{\ell+1}+ \frac{ b^{(0)}_{\ell}}{r^{\ell}}
\right)
\int^{\infty}_{r} dr_\star' \frac{b^{(0)}_{+,\ell}}{(r')^{\ell}} {\cal D}^{(1)}\left(
a^{(0)}_{\ell} \,(r')^{\ell+1}
+ \frac{b^{(0)}_{\ell}}{(r')^{\ell}}
\right)\,.
\label{eq:overlap2_gen}
\ee
Since ${\cal D}^{(1)} \sim r^{-3}$, there is no constant piece in the integral, hence this term does not contribute to either the 
$\bar a^{(1)}$ or $\bar b^{(1)}$ coefficients in the asymptotic expansion of $\psi_1^{(1)}$.  Again, the $r$-dependent pieces of the integral contribute as subleading corrections to the source and response. 

In summary, our perturbative computation dictates that $\bar a^{(1)}=0$ while the $b^{(1)}$ term is controlled by the constant piece  of a simple overlap integral, Eq.\,\eqref{eq:overlap1_gen}. 
We further know by explicit computation that  for integer $\ell$,  $\psi^{(0)}$ simply is a Laurent series of order $r^{\ell+1}$. Hence the integrand in \eqref{eq:overlap1_gen} is a Laurent series. 
A distinction has to be made depending on whether or not the integral produces a logarithm.

In the absence of a logarithm, the integral produces a Laurent series. We find by putting the pieces together that the $k^{(1)}$ term of the Love number is given by
\global\mdfdefinestyle{EqFrame}{ linecolor=white,linewidth=3pt,
backgroundcolor=EqFrame,
leftmargin=0cm,rightmargin=0cm }
\begin{mdframed}[style=EqFrame] 
\be
k_\ell^{(1)}= \frac{1}{N_{\psi,\psi_+}} 
\frac{b^{(0)}_{+,\ell}}{a^{(0)}_\ell}  \frac{1}{ 2\pi i}\oint \frac{dr}{r} \int^{r}_{r_h} dr_\star'  \psi^{(0)}(r')  {\cal D}^{(1)}\psi^{(0)}(r')\,.
\label{eq:k1_full}
\ee
 \end{mdframed}
The contour integral selects the constant term of the Laurent series.

If the integral contains a logarithm, the analysis of section \ref{se:RG_discussion} applies. The integral produces a $\log(\frac{r_h}{r})$, and the matching to the worldline EFT done at a given matching scale $r\equiv L$ produces the structure of a renormalization flow for the worldline EFT coefficient, with $\beta_{\lambda_\ell} =N_\ell \beta_{k_\ell}= B d_{0,\ell}$. 
The beta function is conveniently extracted from the integrand using
\begin{mdframed}[style=EqFrame] 
\be
\beta_{k_\ell}= -\frac{1}{N_{\psi,\psi_+}} 
\frac{b^{(0)}_{+,\ell}}{a^{(0)}_\ell}  \frac{1}{ 2\pi i}\oint  dr_\star  \psi^{(0)}(r)  {\cal D}^{(1)}\psi^{(0)}(r)\,.
\label{eq:B_full}
\ee
 \end{mdframed}
The above formulae are general. They are manifestly invariant under rescaling of either the $\psi$ or $\psi_+$ solutions. 
We recall that $dr_\star =\frac{dr}{f(r)}$.

In the particular case of four dimensions, we know that ${\cal T}=\mathds{1}_2$ at leading order,
as already  discussed in section \ref{se:TLNs}. This implies that  $\psi_+^{(0)}\equiv \tilde \psi^{(0)}$, hence we simply have $b_{+,\ell}=\tilde b_{\ell}$. 

\subsubsection*{EFT Consistency}

The constant Love numbers computed by  \eqref{eq:k1_full} and the beta functions computed by \eqref{eq:B_full} are physically observable. This implies that  they must be proportional to the 
combination $\alpha=-3\alpha_1+\alpha_2+\frac{1}{2}\alpha_3$ found in \eqref{eq:alpha_comb}. 
We will find that this  is indeed the case in our explicit calculations of  Section \ref{se:TLN_results}.
In contrast, in the presence of a renormalization flow  the extra constant terms should be physically irrelevant. We will find  in an explicit computation  that these constant terms are indeed unphysical, in the sense that they are \textit{not} proportional to the physical combination \eqref{eq:alpha_comb}.

\section{Love Numbers of Non-Spinning Black Holes at $\partial^6$ Order}

\label{se:TLN_results}

 \subsection{The EFT-Corrected Non-Spinning Geometries} \label{se:BH}

A charged non-rotating black hole in GR  is described by the Reissner-Nordstr\"om metric, solution of the  coupled Einstein $G_{\mu \nu} = \kappa^2 T_{\mu \nu}^{{\rm e.m.}}$ and Maxwell $\nabla_\nu F^{\mu \nu} = 0$ equations. 
% In this section we reintroduce the gravitational constant $\kappa^2 = 8 \pi G$. 
The solution  in spherical coordinates is
\begin{align}
    ds^2 &= -f(r) dt^2 +  \frac{1}{f(r)} dr^2 + r^2 d\Omega^2 \,,\quad\quad f(r)=  \Bigg(1 - \frac{\kappa^2 M_{\displaystyle \circ}}{4 \pi r} + \frac{\kappa^2 Q_{\displaystyle \circ}^2}{32 \pi^2 r^2} \Bigg) \\
    F_{r t} &= E_r = \frac{Q_{\displaystyle \circ}}{4 \pi r^2}\;. 
\end{align}
$Q_{\displaystyle \circ}$ and $M_{\displaystyle \circ}$ are the total charge and mass of the black hole.  
For $Q_{\displaystyle \circ}=0$, we recover the neutral Schwarzschild black hole. The other independent components of $F_{\mu\nu}$ are zero. 

Let us calculate the  black hole solution in the 
EFT of gravity. We compute the  charged solution with corrections from the operators in ${\cal L}_4$ and ${\cal L}_6$ given in Section \ref{se:EFT_R3}, with respective coefficients $\gamma_{1,2}$ and $\alpha_{1,2,3}$. The neutral case is simply recovered by setting $q=0$. 
  
  %\eqref{eq:EinsteinMawxellEFT}. 
  Assuming that spacetime remains static and spherically symmetric, the metric ansatz and electric fields are written as
\begin{equation}
    ds^2 = -A(r)dt^2 + \frac{1}{B(r)}dr^2 + r^2 d\Omega^2\;,
\label{eq:metric_perturbated}
\end{equation}
where
\begin{align}
    A(r) &= 1 - \frac{\kappa^2 m}{r} + \frac{\kappa^2 q^2}{2 r^2} + A^{(1)}(r) \,,\quad \quad
    B(r) = 1 - \frac{\kappa^2 m}{r} + \frac{\kappa^2 q^2}{2 r^2} + B^{(1)}(r)\;,
\end{align}
\begin{equation}
    E_r(r) = \frac{q}{r^2} + E_r^{(1)}(r)\,, 
\label{eq:electric_perturbated}
\end{equation}
where the $A^{(1)}$, $B^{(1)}$ and $E_r^{(1)}$ functions are the ${\cal O}(\alpha_i,\gamma_i)$ 
 EFT-induced perturbations. 
Here $m$ and $q$ are just constants, however we  find  below that, to first order, they are still proportional to the mass and charge of the black hole. 

The detailed perturbative calculation of the $A^{(1)}$, $B^{(1)}$ and $E_r^{(1)}$ corrections is provided in App.~\ref{se:bh_calculation}. The corrected metric factors and electric field are found to be
\begin{align}
    A(r) &= 1 - \frac{\kappa^2 m}{r} + \frac{\kappa^2 q^2}{2 r^2} - \frac{\kappa^4 \alpha_1}{r^6} \Bigg(96 q^2 + 48 \kappa^2 m^2 - \frac{42 \kappa^4 m^3}{r} - \frac{2032 \kappa^2 m q^2}{7 r} + \frac{1300 \kappa^4 m^2 q^2}{7 r^2} \nonumber \\ &+ \frac{1052 \kappa^2 q^4}{7 r^2} - \frac{1079 \kappa^4 m q^4}{7 r^3} + \frac{289 \kappa^4 q^6}{9 r^4}\Bigg) + \frac{\kappa^6 \alpha_2}{r^7} \Bigg(10 \kappa^2 m^3 -\frac{192 m q^2}{7} - \frac{108 \kappa^2 m^2 q^2}{7 r} \nonumber \\&+ \frac{30 q^4}{r} + \frac{93 \kappa^2 m q^4}{7 r^2} - \frac{28 \kappa^2 q^6}{3 r^3} \Bigg) + \frac{\kappa^6 \alpha_3}{r^6} \Bigg(9 m^2 - \frac{17 \kappa^2 m^3}{2 r} - \frac{138 m q^2}{7 r} + \frac{156 \kappa^2 m^2 q^2}{7 r^2} \nonumber \\&+ \frac{21 q^4}{2 r^2} - \frac{507 \kappa^2 m q^4}{28 r^3} + \frac{53 \kappa^2 q^6}{12 r^4}\Bigg) - \frac{\kappa^2 \gamma_1}{r^4} \Bigg(2 q^2 + \frac{\kappa^2 q^4}{5 r^2} - \frac{\kappa^2 m q^2}{r} \Bigg) - \frac{4 \kappa^2 q^4}{5 r^6}\gamma_2\label{eq:A}
\end{align}
\begin{align}
    B(r) &= 1 - \frac{\kappa^2 m}{r} + \frac{\kappa^2 q^2}{2 r^2} - \frac{\kappa^4 \alpha_1}{r^6} \Bigg(288 q^2 + 36 \kappa^2 m^2 - \frac{30 \kappa^4 m^3}{r} - \frac{848 \kappa^2 m q^2}{r} + \frac{3818 \kappa^4 m^2 q^2}{7 r^2} \nonumber \\ &+ \frac{3784 \kappa^2 q^4}{7 r^2} - \frac{631 \kappa^4 m q^4}{r^3} + \frac{1612 \kappa^4 q^6}{9 r^4}\Bigg) + \frac{\kappa^6 \alpha_2}{r^6} \Bigg(108 m^2 -\frac{98 \kappa^2 m^3}{r} - \frac{384 m q^2}{r} \nonumber \\&+ \frac{2766 \kappa^2 m^2 q^2}{7 r^2} + \frac{336 q^4}{r^2} - \frac{471 \kappa^2 m q^4}{r^3} + \frac{431 \kappa^2 q^6}{3 r^4}  \Bigg) - \frac{\kappa^6 \alpha_3}{r^7} \Bigg(6 m q^2 - \frac{\kappa^2 m^3}{2} - \frac{6 q^4}{r} \nonumber \\&- \frac{57 \kappa^2 m^2 q^2}{14 r} + \frac{27 \kappa^2 m q^4}{4 r^2} - \frac{13 \kappa^2 q^6}{6 r^3}\Bigg) - \frac{\kappa^2 \gamma_1}{r^4} \Bigg(8 q^2 - \frac{7 \kappa^2 m q^2}{r} + \frac{16 \kappa^2 q^4}{5 r^2} \Bigg) \nonumber \\&- \frac{4 \kappa^2 q^4}{5 r^6}\gamma_2\label{eq:B}\end{align}
    \begin{align}
    E_r(r) &= \frac{q}{r^2} +\frac{\kappa^4 q \alpha_1}{r^8}\Bigg(96 q^2 - 6 \kappa^2 m^2 - \frac{1280 \kappa^2 m q^2}{7 r} + \frac{147 \kappa^2 q^4}{r^2} \Bigg) - \frac{\kappa^6 q \alpha_2}{r^8}\Bigg(54 m^2 -\frac{1248 m q^2}{7 r} \nonumber \\ &+ \frac{153 q^4}{r^2} \Bigg) + \frac{\kappa^6 q \alpha_3}{r^8}\Bigg(\frac{9 m^2}{2} - \frac{48 m q^2}{7 r} + \frac{9 q^4}{4 r^2} \Bigg)+\frac{\kappa^2 q \gamma_1}{r^5}\Bigg(8 m - \frac{9 q^2}{r} \Bigg)-\frac{16 q^3}{r^6}\gamma_2\,. \label{eq:E}
\end{align}

Using that a static metric has a Killing vector associated with the time symmetry $K^\mu = (1, 0, 0, 0)$, the total mass and electric charge are calculated by an integral at spatial infinity,
\begin{align}
    M_{\displaystyle \circ} &= \frac{4 \pi}{\kappa^2}\int_{\partial \Sigma} d^2 x \sqrt{\gamma^{(2)}} n^\mu \sigma^\nu \nabla_\mu K_\nu = \lim_{r\to\infty}\frac{4 \pi r^2 A^\prime(r)}{\kappa^2}\sqrt{\frac{B(r)}{A(r)}} = 4 \pi m \\
    Q_{\displaystyle \circ} &= - \int_{\partial \Sigma} d^2 x \sqrt{\gamma^{(2)}} n^\mu \sigma^\nu F_{\mu \nu} = \lim_{r\to\infty} 4 \pi r^2 \sqrt{\frac{B(r)}{A(r)}}E_r(r) = 4\pi q\;.
\end{align}
The $n^\mu$ vector  is normal to constant time slices and $n_\mu n^\mu = -1$. The  $\sigma^\mu$ vector is normal to the two-sphere and $\sigma_\mu \sigma^\mu = 1$. 

The inner $r_-$ and outer $r_+$ horizons, corrected to first order, take the form
\begin{equation}
    r_{\pm} = \frac{\kappa^2 m}{2} \pm \frac{\sqrt{2} \kappa |q|}{2}\sqrt{\frac{\kappa^2 m^2}{2 q^2} - 1 + \mathcal{O}(\alpha_i,\gamma_i)}\;.
\end{equation}
For extremal black holes, the inner and outer horizons coincide, $r_h = r_+ = r_-$. The extremal horizon is thus determined by the vanishing of   discriminant under the square root. This condition provides  the mass-to-charge ratio for extremal black holes in the EFT of gravity,
\begin{equation}
    \frac{\kappa}{\sqrt{2}}\frac{M_{\displaystyle \circ}}{|Q_{\displaystyle \circ}|} = 1 + \frac{44 \kappa^2}{63 r_h^4}\alpha_1 - \frac{16 \kappa^2 }{21 r_h^4}\alpha_2 - \frac{\kappa^2 }{21 r_h^4}\alpha_3 - \frac{2}{5 r_h^2}\gamma_1 - \frac{8}{5 \kappa^2 r_h^2}\gamma_2 + \mathcal{O}(\alpha_i^2,\gamma_i^2)\;.
    \label{eq:QM_extremal}
\end{equation}

The $\gamma_{1,2}$ and $\alpha_2$ terms are  consistent with those found respectively in \cite{Kats:2006xp} and \cite{DeLuca:2023mio}.

\subsubsection{Positivity Bound from Weak Gravity Conjecture}

There exist arguments that extremal black holes should always be able to decay, see \cite{Arkani-Hamed:2006emk} and the reviews \cite{vanBeest:2021lhn,Grana:2021zvf, Agmon:2022thq}. 
For an extremal  black hole to be able to decay even in the absence of light particles in the EFT of gravity, 
the mass-to-charge ratio  must be smaller than $1$, see \cite{Kats:2006xp},  
\be \frac{M}{Q}\bigg|_{\rm ext, EFT} < 1
\,. 
\label{eq:Q_M_condition_EFT}
\ee 
This  produces a positivity bound on the combination of coefficients in \eqref{eq:QM_extremal}, \be
\frac{2}{5 }\gamma_1 + \frac{8}{5  }\gamma_2 -\frac{44 \kappa^2}{63 r_h^2}\alpha_1 + \frac{16 \kappa^2 }{21 r_h^2}\alpha_2 + \frac{\kappa^2 }{21 r_h^2}\alpha_3   \quad > \quad  0 \,.
\label{eq:positivity}
\ee 
The interplay of this type of bound with constraints from  IR consistency of QFT has been discussed in \cite{Cheung:2014ega, 
Hamada:2018dde, Bellazzini:2019xts,
Chen:2019qvr, 
Arkani-Hamed:2021ajd, 
Bittar:2024xuc,
Knorr:2024yiu}. 
On the other hand, the linear combination in \eqref{eq:positivity} differs from the ones appearing in the Love numbers computed in the rest of the section. Hence there is no correlation between the Love numbers and the weak gravity conjecture.

\subsection{Vectorial Tidal Response \label{se:spin1}}

We compute the  deformability of the black hole under a vectorial tidal field living on the black hole background.\,\footnote{The parity-even and parity-odd degrees of freedom of the vector tidal field probe respectively the electric polarizability and magnetic susceptibility of the black hole. In $d=4$, these are related by electric-magnetic duality, such that the $k_\ell$ from each sector are necessarily equal.  On the other hand, the matching to the worldline EFT, given by $\lambda_\ell=N_\ell k_\ell$ (see \eqref{eq:matching_coef}),  differs for each sector. From  \cite{Hui:2020xxx},  the respective coefficients in $d=4$ are found to be
\be
N^{\rm el.}_\ell= (-1)^{\ell} \frac{\ell}{\ell+1}\frac{\sqrt{\pi}}{2^{\ell-4}}\Gamma\left(\frac{1}{2}-\ell\right)r_h^{2\ell+1}\,, \quad 
N^{\rm mag.}_\ell= (-1)^{\ell+1} \frac{\sqrt{\pi}}{2^{\ell-1}}\Gamma\left(\frac{1}{2}-\ell\right)r_h^{2\ell+1}\,. 
\ee
}

\subsubsection{ Equation of Motion}

Suppose a massless vector field $A_\mu$ propagating in the black hole background. 
% It could represent a perturbation to the electromagnetic field of a charged black hole.
The equation of motion is the Maxwell equation without sources,
\begin{align}
    \nabla^\nu F_{\mu \nu} = 
        \square A_\mu - \nabla^\nu \nabla_\mu A_\nu &= 0\,.
\label{eq:Maxwell_equation}
\end{align}
For a spherically symmetric spacetime, we can express $A_\mu$ as a linear combination of the vector spherical harmonics. A brief review of the vector spherical harmonics is provided in appendix \ref{sec:vector_spherical_harmonics}. 

Since angular momentum and parity are conserved in the background spacetime, we can treat different modes independently. We focus on the parity-odd degree of freedom,
\begin{align}
A_\mu^{\mathrm{odd}} \equiv \Psi_\ell(t, r)\begin{pmatrix}
0\\
0 \\
\epsilon\indices{_i^j}\nabla_j Y_{\ell m}
\end{pmatrix}\,.
\end{align}
By direct calculation, we find that $A_\mu^{\mathrm{odd}}$ is the transverse component of $A_\mu$, 
\begin{equation}
    \nabla^\mu A_\mu^{\mathrm{odd}} = 0\,.
\end{equation}

Commuting the covariant derivatives in \eqref{eq:Maxwell_equation}, the   equation of motion for the transverse component becomes
\be
    \square A_\mu^{\mathrm{odd}} - R\indices{_\mu ^\nu}A_{\nu}^{\mathrm{odd}} = 0\,. \ee 
    In terms of the $\Psi_\ell$ variable this is
\be
    \frac{\partial^2 \Psi_\ell}{\partial t^2} - A(r)B(r)\frac{\partial^2 \Psi_\ell}{\partial r^2} - \frac{1}{2}\frac{d}{dr}(A(r) B(r))\frac{\partial \Psi_\ell}{\partial r} + \frac{A(r)}{r^2}\ell(\ell + 1)\Psi_\ell  = 0\;.
\label{eq:vector_equation}
\ee
Hence the transverse component satisfies a second order linear differential equation of the form \eqref{eq:EOM}.
Our focus is static perturbations, $\partial_t \Psi_\ell=0$.

As discussed in section \ref{se:perturbation_theory}, we determine the solution to \eqref{eq:vector_equation} perturbatively using 
\begin{equation}
    \Psi_\ell = \Psi_\ell^{(0)} + \Psi_\ell^{(1)}+{\cal O}(\alpha_i^2,\gamma_i^2)\;.
\end{equation}
The EFT coefficients are included inside the perturbation, i.e.  $\Psi_\ell^{(1)}= \sum_i \alpha_i \Psi_{\ell,\alpha_i}^{(1)}+ \sum_i \gamma_i \Psi_{\ell,\gamma_i}^{(1)}$. 

The metric coefficients are given by
\begin{align}
    A(r) &= A^{(0)}(r) + A^{(1)}(r)+{\cal O}(\alpha_i^2,\gamma_i^2)\\
    B(r) &= B^{(0)}(r) + B^{(1)}(r)+{\cal O}(\alpha_i^2, \gamma_i^2)\;,
\end{align}
where the metric perturbations are given in \eqref{eq:A} and \eqref{eq:B}.
The  tidal equation of motion at order zero is 
\begin{align}
   {\cal D}^{(0)}\Psi_\ell^{(0)} & \equiv A^{(0)}B^{(0)}\frac{d^2 \Psi_\ell^{(0)}}{d r^2} + \frac{1}{2}\frac{d}{dr}(A^{(0)} B^{(0)})\frac{d \Psi_\ell^{(0)}}{d r} - \frac{A^{(0)}}{r^2}\ell(\ell + 1)\Psi_\ell^{(0)}  = 0 \,. 
   \label{eq:EOM_0_vector}
\end{align}
The equation of motion for the perturbation of the tidal field is then 
\be
   {\cal D}^{(0)}\Psi_\ell^{(1)} = -{\cal D}^{(1)}\Psi_\ell^{(0)}\,,
      \label{eq:EOM_1_vector}
\ee
which has the form of \eqref{eq:EOM_Psi1}.
The r.h.s  amounts to a source term that depends on the unperturbed solution.
The explicit  expression for the source is given in
App.~\ref{se:source}. 
Knowing the equation of motion of the  perturbation \eqref{eq:EOM_1_vector}, we can apply the analysis from section 
\ref{se:perturbation_theory}.

% \begin{align}
%      A^{(0)}B^{(0)}\frac{d^2 \Psi_\ell^{(0)}}{d r^2} + \frac{1}{2}\frac{d}{dr}(A^{(0)} B^{(0)})\frac{d \Psi_\ell^{(0)}}{d r} - \frac{A^{(0)}}{r^2}\ell(\ell + 1)\Psi_\ell^{(0)}  &= 0\\
%      A^{(0)}B^{(0)}\frac{d^2 \Psi_\ell^{(1)}}{dr^2}+ \frac{1}{2}\frac{d}{dr}(A^{(0)}B^{(0)})\frac{d \Psi_\ell^{(1)}}{dr} - \frac{A^{(0)}}{r^2}\ell(\ell + 1)\Psi_\ell^{(1)} &= -J[\Psi_\ell^{(0)}]\;.
% \end{align}
% The first order equation acquires a source $J[\Psi_\ell^{(0)}]$ that is function of the unperturbed solution. See appendix \ref{se:source} for the explicit form for the source.

% As expected, we have derived an EOM for the tidal vector field in the form of \eqref{eq:EOM_Psi1}.

\subsubsection{Neutral Black Hole}

\label{se:TLN_vec_Sch}

We compute the Love numbers of the EFT-corrected Schwarzschild black hole at $\partial^6$ order. 

% using the  method derived in  section 
% \ref{se:perturbation_theory}. 

We choose the solution regular on the horizon as
\begin{align}
\psi_{\ell}^{(0)} = S_\ell \frac{r^2}{r_h^2} \,_2F_1\left(1-\ell,2+\ell,3;\frac{r}{r_h}\right)   \underset{\ell\in\mathbb{N^*}}{=} S_\ell\sum^{\ell - 1}_{n=0} \frac{(1-\ell)_n (2+\ell)_n}{(3)_n \,n!} \frac{r^{n+2}}{r^{n+2}_h}
\end{align}
with \be
S_\ell= \frac{(1+\ell)\Gamma(\ell)\Gamma(2+\ell)}{4(-1)^{\ell+1} \Gamma(2\ell)}r^{\ell+1}_h\,, \quad (k)_n = \frac{\Gamma(k+n)}{\Gamma(n)}\,. 
\ee 
With this normalization, the  asymptotic behavior is $\psi_\ell^{(0)} \underset{ r\gg r_h}{\sim} r^{\ell+1}$, such that  $a^{(0)}_{\ell}=1$.
For the solution regular at infinity, 
we just have to specify the asymptotic behavior, $\psi_+^{(0)} \underset{ r_h=0}{\sim} r^{-\ell}$ such that $b^{(0)}_{+,\ell}=1$. This is sufficient to evaluate the Wronskian at larger $r$, from which we obtain the coefficient $N_{\psi,\psi_+}=2\ell+1$.

We then apply our general  formulae \eqref{eq:k1_full}, \eqref{eq:B_full}. It turns out that for $\ell\leq 2$, there is no running, hence we apply   \eqref{eq:k1_full}. We find 
\be
k_1=  - \frac{4 \alpha \kappa^2}{r_h^4}\,,\quad \quad k_2=  - \frac{219 \alpha \kappa^2}{20 r_h^4} \,. 
\ee
% \textbf{[In the notebook , need to include the overall factor $\kappa^2/r_h^{2\ell+1}$]}
This matches exactly the results found using $\alpha_2$ in \cite{DeLuca:2022tkm}. 

For $\ell\geq 3$  the Love numbers run. We compute the beta functions using \eqref{eq:B_full}. For the first harmonics we find 
\begin{center}
\begin{tabular}{ |c|c|c|c|c|c|c|c|c| } 
 \hline
 $\ell$ & 3 & 4 & 5 & 6 & 7 & 8  \\ 
 \hline
  \rule{0pt}{20pt}
 $\beta_{k_\ell}$ & $-\frac{704}{7}\frac{\alpha\kappa^2}{r_h^4}$  & 
 $-\frac{1525}{7}\frac{\alpha\kappa^2}{r_h^4}$ & 
  $-\frac{14288}{77}\frac{\alpha\kappa^2}{r_h^4}$ & 
   $-\frac{448105}{4719}\frac{\alpha\kappa^2}{r_h^4}$ & 
      $-\frac{5888}{169}\frac{\alpha\kappa^2}{r_h^4}$ & 
            $-\frac{1592703}{158015}\frac{\alpha\kappa^2}{r_h^4}$ 
 \\[10pt] 
 \hline
\end{tabular}
\end{center}
The $\ell=3$ result matches the coefficient of the logarithm found in \cite{DeLuca:2022tkm}. We have also checked  the first harmonics by direct computation. 

We find a closed-form expression for the beta function of any
running Love number of the neutral black hole. The trick is to notice that for a given $D^{(1)}$, only a fixed number  of terms from $\psi^{(0)}_\ell$ contributes from any $\ell$. We find
\be
\beta_{k_\ell} = \frac{ (\ell-2)(\ell-1)(\ell+2)(\ell+3)(40-96\ell(1+\ell)+23 \ell^2(1+\ell)^2) (1+\ell)!^4 ) }{120 (1+2\ell)(2\ell)!^2}\frac{\alpha \kappa^2}{r_h^4} \,,\quad \ell\in \mathbb{N}_{\geq 3}\,.
\ee
We conclude that all the Love number beta functions are negative. They are also  exponentially suppressed as  $\beta_{k_\ell}\sim 2^{-4\ell}$ at large $\ell$.

All the above results depend only on the physical  combination $\alpha$. We have checked this feature explicitly by separately computing the $\alpha_{1,2,3}$ contributions. 
The dependence in the physical  combination $\alpha$ arises nontrivially, only at the level of  the final expression given by  \eqref{eq:B_full}.

In contrast, for $\ell>2$, we checked explicitly that the constant terms that accompany the logarithm as in \eqref{eq:hatb_def} are \textit{not} proportional to $\alpha$.\,\footnote{For example, for $\ell = 3$ we identify the coefficient of $r^{-\ell}$ as
\be
 \left(\left(-\frac{436248}{1225}\alpha_1-\frac{184616}{1225}\alpha_2+\frac{70258}{1225}\alpha_3\right)+\frac{704}{7} \left(-3 \alpha_1+\alpha_2+\frac{\alpha_3}{2}\right) \log \left(\frac{r}{r_h}\right)\right)\frac{\kappa ^2}{r_h^4}\,.
\ee  The  $\alpha_2$ term  matches the one   from \cite{DeLuca:2022tkm}.}
This confirms our claim that these constant terms should not be considered as physical --- only the  beta function is. 

\subsubsection{Charged Black Hole}

\label{se:TLN_vec_RN}

We  compute the Love numbers of the EFT-corrected Reissner-Nordstr\"om black hole  at $\partial^4$ order.

In the case of the charged non-spinning black hole, the leading order solutions that are 
regular on the horizon are found via a power series.
The general leading order solution takes the polynomial form 
\begin{equation}
  \psi^{(0)}_\ell = \sum _{n=0}^{\ell+
    1} a^\ell_{n} r^{\ell + 1 -n}\,,
\end{equation}
where the normalized coefficients satisfy the  recurrence relation
\begin{align}
a^\ell_0&= 1 \nn  \\
a^\ell_1&=\frac{1-\ell^2}{2 \ell}m\kappa ^2 \nn  \\
a^\ell_{n+2} &= \frac{ (n+2-\ell) \left(2 m (n-\ell)a^\ell_{n+1}+q^2 (\ell+1-n) a^\ell_{n}\right)}{2 (n+2) (n+1-2 \ell)}\kappa ^2\,.
\end{align}
For the first values of $\ell$ we find
\begin{align}
\psi^{(0)}_1 &= r^2-\frac{1}{2}q^2 \nn  \\
\psi^{(0)}_2 &= r^3-\frac{3}{4}mr^2+\frac{1}{8}mq^2 \nn  \\
\psi^{(0)}_3 &= r^4 -\frac{4}{3}mr^3 +\frac{1}{60}(2m^2+q^2)(12r^2-q^2)  \nn  \\
\psi^{(0)}_4 &= r^5  -\frac{15}{8} m r^4 +\frac{5}{14}(3m^2+q^2)r^3-\frac{5}{56}m(2m^2+3q^2)r^2 +\frac{3}{224}mq^4+\frac{1}{112}m^3q^2\,, 
\end{align}
where we set $\kappa=1$ here for convenience.

We focus on the leading ${\cal O}(\partial^4)$ corrections from the EFT of gravity. The analysis follows the same steps as in section \ref{se:TLN_vec_Sch}.
We find that the renormalization flow appears for $\ell\geq 2$. 
The $\ell=1$  Love number is
\be
k_1= \frac{16}{9}\left(3 - \frac{r_-}{r_+}\right) \frac{r_-\, \gamma_1}{r_+^3} 
+\frac{32}{45} \left(5 - 3\frac{r_-}{r_+}\right) \frac{r^2_-\,  \gamma_2}{r_+^4\kappa^2}
\,.
\ee
The first beta functions are  given by
\begin{align}
   r^5_+ ~\beta_{k_2}& = \frac{54}{5}mq^2 \kappa^4 \gamma_1  \nn \\ 
   r^7_+~  \beta_{k_3}& =  \frac{32}{35}mq^2\left((13m^2\kappa^2+10q^2)\kappa^6 \gamma_1 + 4q^2 \kappa^4 \gamma_2 \right) \nn \\ 
   r^9_+ ~   \beta_{k_4}& = \frac{5}{882}mq^2 \left( (3m^2 \kappa^2+q^2 )( 335 m^2\kappa^2 + 723q^2 ) \kappa^8 \gamma_1  
   +28   q^2 (29m^2\kappa^2+12q^2) \kappa^6\gamma_2
   \right)\,,
   % \nn \\ 
   % r^{11}_+ ~   \beta_{k_5}& = \frac{192}{385}mq^2 (560 \kappa^6 m^6+1704 \kappa^4 m^4q^2+ 876\kappa^2m^2q^4 + 73q^6) \kappa^{10} \beta_1   
\end{align}
where we restored the gravitational constant. Notably, only the $R_{\mu\nu\rho\sigma} F^{\mu\nu}F^{\rho\sigma}$ operator contributes to the $\ell=2$ beta function.

\subsubsection*{EFT-corrected extremal black hole}

The particular case of the extremal black hole is defined by $r_+=r_-\equiv r_h$ in the charged solution. 
In that case we find
\be
k_1= \frac{32}{45}\left(
5\frac{\gamma_1}{r_h^2}+ 2 \frac{\gamma_2}{r_h^2\kappa^2}
\right)
\ee
and the first beta functions are 
\begin{center}
\begin{tabular}{ |c|c|c|c|c|c|c|c| } 
 \hline
 $\ell$ & 2 & 3 & 4 & 5 & 6 \\ 
 \hline
 \rule{0pt}{20pt}
 $\beta_{k_\ell}$
 & $ \frac{216}{5}\frac{ \gamma_1}{r_h^2}$
 & $ \frac{9216\gamma_1 }{35 r_h^2}+ \frac{1024\gamma_2 }{35 r_h^2 \kappa^2}$
 & $ \frac{7960\gamma_1 }{9 r_h^2}+ \frac{1600\gamma_2 }{9 r_h^2 \kappa^2}$
 
   & $ \frac{122112\gamma_1 }{55 r_h^2}+ \frac{32256\gamma_2 }{55 r_h^2 \kappa^2}$
   & $ \frac{60760\gamma_1 }{13 r_h^2}+ \frac{18816\gamma_2 }{13 r_h^2 \kappa^2}$
 \\[10pt] 
 \hline
\end{tabular}
\end{center}

\subsection{Tensorial Tidal Response}

\label{se:spin2}

We compute the  deformability of the black hole under a tensorial tidal field.\,\footnote{The matching to the worldline EFT is given by $\lambda_\ell= N_\ell k_\ell$  (see \eqref{eq:matching_coef}) with \cite{Hui:2020xxx}
\be
N_\ell= (-1)^{\ell}\frac{\ell-1}{\ell+2}\frac{\ell}{\ell+1}\frac{\sqrt{\pi}}{2^{\ell-2}}\Gamma\left(\frac{1}{2}-\ell\right)r_h^{2\ell+1}
\label{eq:Nl_tensor}
\,.
\ee
We remind that, even though the graviton propagation is modified by the EFT operators in the vicinity of the black hole, the perturbative treatment developed in section \ref{se:TLNs} implies that, by construction,  the solutions to the equation of motion at large $r$ are the unperturbed ones. Hence the tidal response has the same structure as those in \cite{Hui:2020xxx}, which is why  we can use the matching coefficient \eqref{eq:Nl_tensor}.   
}

\subsubsection{ Equation of Motion}

Consider a massless tensor field $h_{\mu \nu}$ that propagates in the black hole background $g_{\mu \nu}$. Such a field can be interpreted as a fluctuation of the spacetime metric,
\begin{equation}
    g_{\mu \nu} = \bar g_{\mu \nu} + h_{\mu \nu}, \; \; \; \; \; \; |h_{\mu \nu}|  \ll 1 \;.
\end{equation}
The full metric  $g_{\mu \nu}$ satifies the Einstein equations $G_{\mu \nu} = \kappa^2 T_{\mu \nu}$. At first order in $h_{\mu \nu}$ we have 
\begin{align}
    G_{\mu \nu} &= \bar G_{\mu \nu} + \delta G_{\mu \nu}, \; \; \; \; \; \; \; \; \; \;  T_{\mu \nu} = \bar T_{\mu \nu} + \delta T_{\mu \nu}\;.
\end{align}
The background spacetime satisfies the Einstein equations hence
 the equation of motion for the tensor fluctuation are
\begin{equation}
    \delta G_{\mu \nu} = \kappa^2 \delta T_{\mu \nu}\;.
\end{equation}
We mention that this equation can also be computed from the quadratic Lagrangian of the perturbation  $h_{\mu\nu}$. Here we obtain it more directly via the variation of the Einstein field equations.

For a spherically symmetric background it is convenient to write the components of $h_{\mu\nu}$ in function of the spherical harmonics. See appendix \ref{se:tensor_spherical_harmonics} for a brief review of tensorial spherical harmonics. We are interested in the parity-odd degrees of freedom,
\begin{align}
    h_{t i} &= h_0(t, r) \epsilon\indices{_i ^j} \nabla_j Y_{\ell m}\,, \\
    h_{r i} &= h_1(t, r) \epsilon\indices{_i ^j} \nabla_j Y_{\ell m} \,,\\
    h_{i j} &= h_2(t, r) \epsilon\indices{_( _i ^k} \nabla_{j)T} \nabla_k Y_{\ell m}\;. 
\end{align}
These harmonics exist for $\ell\geq 2$. 

The field equations are invariant under diffeomorphisms 
\begin{align}
    h_{\mu \nu} &\rightarrow h_{\mu \nu} + \nabla_\mu \xi_\nu + \nabla_\nu \xi_\mu\;.
\end{align}
We can use the gauge redundancy to simplify the equations. For parity-odd degrees of freedom, it is enough to consider the  gauge transformation
\begin{equation}
    \xi_t = 0, \; \; \; \; \; \; \; \xi_r = 0, \; \; \; \; \; \; \; \xi_i = \xi(t, r) \epsilon\indices{_i ^j} \nabla_j Y_{\ell m}\,,
\end{equation}
under which the  odd components transforms as
\begin{align}
    h_0 \rightarrow h_0 + \dot{\xi} \,,\quad\quad
    h_1 \rightarrow h_1 + \xi^\prime - \frac{2}{r}\xi \,,\quad\quad
    h_2 \rightarrow h_2 + 2 \xi\;.
    \label{eq:h_gauge}
\end{align}
We can choose $\xi(t, r)$ to eliminate $h_2$, which is the Regge-Wheeler gauge. 

Two degrees of freedom remain: $h_0$ and $h_1$.
We inspect their coupled equations of motion, raising one index for convenience. In particular, we obtain the components
\begin{align}
    \delta G\indices{^r _\phi} &= 0 \; \; \; \Rightarrow \; \; \;  A(r)(\ell(\ell+1) - 2) h_1 + r^2 \frac{d}{dt}\Bigg(\frac{2}{r}h_0 -h_0^\prime + \dot{h}_1 \Bigg) = 0\label{eq:condition_2}\\
    \delta G\indices{^\theta _\phi} &= 0 \; \; \; \Rightarrow \; \; \; \frac{d}{dr}\Big(A(r)B(r) \Big)h_1 + 2 A(r)B(r)h_1^\prime - 2 \dot{h}_0 = 0 \label{eq:condition_1}\;,
\end{align}
where $\dot{f} = \frac{\partial f}{\partial t}$, $f^\prime = \frac{\partial f}{\partial r}$. Conveniently, the $\delta T\indices{^r _\phi}$, $\delta T\indices{^\theta _\phi}$ perturbations vanish. 

Equation \eqref{eq:condition_2} can be written as
\begin{equation}
    h_1 = -\frac{r^2}{(\ell(\ell+1)-2)A(r)}\dot{\chi}\,, \quad \quad \quad \chi = \frac{2}{r}h_0 -h_0^\prime + \dot{h}_1\,,
     \label{eq:h1}
\end{equation}
where the combination $\chi$ is gauge invariant, as can be seen from \eqref{eq:h_gauge}. 
We then use \eqref{eq:condition_1} to  solve $h_0$ as a function of $\chi$. This step has to be done perturbatively. At zero order,
\begin{align}
    h_0^{(0)} &= -\frac{r A^{(0)}}{(\ell(\ell+1)-2)}(2 \chi^{(0)} + r \chi^{(0)\prime})\,.
     \label{eq:h00}
\end{align}
% \begin{align}
%     h_0^{(0)} &= -\frac{r A^{(0)}}{(\ell(\ell+1)-2)}(2 \chi^{(0)} + r \chi^{(0)\prime})
%      \label{eq:h00}
%     \\
%     h_0^{(1)} &= - \frac{r}{2(\ell(\ell+1)-2)A^{(0)}}\Bigg(\Big(r A^{(0) ^\prime}(A^{(1)} - B^{(1)}) + A^{(0)}(4 B^{(1)} - r A^{(1) ^\prime} + r B^{(1) ^\prime}) \Big)\chi^{(0)}  \nonumber \\ &+ 2 r A^{(0)} B^{(1)} \chi^{(0) ^\prime} + 4 A^{(0)}B^{(0)}\chi^{(1)} + 2 r A^{(0)} B^{(0)}\chi^{(1) ^\prime}\Bigg)\;. 
%     \label{eq:h01}
% \end{align}
All the degrees of freedom are thus expressed as a function of a single gauge invariant variable $\chi$. This variable, which arises naturally from our analysis, is  proportional to the conventional  Regge-Wheeler variable \begin{equation}
    \Psi = \frac{r}{\sqrt{2(\ell(\ell+1)-2)}}\chi\;. \label{eq:RW_def}
\end{equation}

Finally, the remaining component  $\delta G\indices{_t _\phi} = \kappa^2 \delta T_{t \phi}$, combined with  \eqref{eq:h1}, \eqref{eq:h00} and \eqref{eq:RW_def}, provides the equation of motion for $\Psi_\ell$. At zero order, we find 
\begin{align}
  {\cal D}^{(0)}\Psi_\ell^{(0)} \equiv   \frac{\partial^2 \Psi_\ell^{(0)}}{\partial t^2} - A^{(0)}B^{(0)}\frac{\partial^2 \Psi^{(0)}_\ell}{\partial r^2}& - \frac{1}{2}\frac{d}{dr}\Big(A^{(0)}B^{(0)}\Big)\frac{\partial \Psi_\ell^{(0)}}{\partial r} \label{eq:EOM_0_tensor}  \\ \nn  &+ \frac{A^{(0)}}{r^2}\Bigg(\ell(\ell+1) -2 + 2 A^{(0)} - r \frac{dA^{(0)}}{dr}  \Bigg)\Psi_\ell^{(0)} = 0\,. 
\end{align}
In the charged black hole case, the electromagnetic stress tensor contribute, but its contribution ends up canceling out so that \eqref{eq:EOM_0_tensor} holds. 
The equation for the perturbation takes the form 
\be    {\cal D}^{(0)}\Psi_\ell^{(1)} = -{\cal D}^{(1)}\Psi_\ell^{(0)}\,,
\label{eq:EOM_1_tensor} 
\;\ee 
where the explicit  source term is given in App.\,\ref{se:source}.  Both $\delta G$ and $\delta T$ contribute to the source.

\subsubsection{Neutral Black Hole}

\label{se:TLN_tensor_Sch}

We choose the solution regular on the horizon as
\begin{align}
\Psi_{\ell}^{(0)} = S_\ell \frac{r^3}{r_h^3} \,_2F_1\left(2-\ell,3+\ell,5;\frac{r}{r_h}\right)   \underset{\ell\in\mathbb{N^*}}{=} S_\ell\sum^{\ell - 1}_{n=0} \frac{(2-\ell)_n (3+\ell)_n}{(5)_n \,n!} \frac{r^{n+3}}{r^{n+3}_h}
\label{eq:NeutralTensorSolutions}\end{align}
with \be
S_\ell= \frac{(-1)^{l} (l+2)!^2}{24 (2 l)!}r^{\ell+1}_h\,.
\ee 
With this normalization, the  asymptotic behavior is $\psi_\ell^{(0)} \underset{ r\gg r_h}{\sim} r^{\ell+1}$, such that  $a^{(0)}_{\ell}=1$.
For the solution regular at infinity, 
we just have to specify the asymptotic behavior, $\psi_+^{(0)} \underset{ r_h=0}{\sim} r^{-\ell}$ such that $b^{(0)}_{+,\ell}=1$. This is sufficient to evaluate the Wronskian at larger $r$, from which we obtain the coefficient $N_{\psi,\psi_+}=2\ell+1$.

We then apply our general results \eqref{eq:k1_full}, \eqref{eq:B_full}. It turns out that for $\ell = 2$, there is no running, hence we apply   \eqref{eq:k1_full}. We find
\be
k_2=  - \frac{240 \alpha \kappa^2}{r_h^4} \,. 
\label{eq:TLNTensl2}
\ee
 This matches exactly the results found using $\alpha_2$ in \cite{DeLuca:2022tkm}. {This result also agrees with \cite{Cai:2019npx} after  translating the conventions, as noted in \cite{Cano:2025zyk}}.

For $\ell\geq 3$  the Love numbers run. We compute thus  the beta functions using \eqref{eq:B_full}. For the first harmonics we find 
\begin{center}
\begin{tabular}{ |c|c|c|c|c|c|c|c|c| } 
 \hline
 $\ell$ & 3 & 4 & 5 & 6 & 7 & 8  \\ 
 \hline
  \rule{0pt}{20pt}
 $\beta_{k_\ell}$ & $-\frac{8000}{7}\frac{\alpha\kappa^2}{r_h^4}$  & 
 $-\frac{12000}{7}\frac{\alpha\kappa^2}{r_h^4}$ & 
  $-\frac{12740}{11}\frac{\alpha\kappa^2}{r_h^4}$ & 
   $-\frac{2383360}{4719}\frac{\alpha\kappa^2}{r_h^4}$ & 
      $-\frac{259200}{1573}\frac{\alpha\kappa^2}{r_h^4}$ & 
            $-\frac{81000}{1859}\frac{\alpha\kappa^2}{r_h^4}$ 
 \\[10pt] 
 \hline
\end{tabular}
\end{center}
The $\ell=3$ result matches the coefficient of the logarithm found in \cite{DeLuca:2022tkm}. We have also checked  the first harmonics by direct computation.
We find a closed-form expression for the neutral black hole Love number beta functions for any harmonic,   
\be \beta_{k_\ell} = \frac{5 (\ell-2) (\ell+1)^4 (\ell+2)^4 (\ell+3) \left(4-\ell(\ell + 1)\right) \ell!^4 }{12 (1+ 2 \ell) (2 \ell)!^2} \frac{\alpha \kappa^2}{r_h^4} \,,\quad \ell\in \mathbb{N}_{\geq 3}\,. \ee 
As in the vectorial case we find that all the Love number beta functions are negative and   exponentially suppressed at large $\ell$. 
The  tensor-to-vector ratio of the beta functions is found to be 
\be \frac{\beta^{(2)}_{k_\ell}}{\beta^{(1)}_{k_\ell}} =\frac{50 (\ell+2)^3 \left(\ell^2+\ell-4\right)}{(\ell-1) (\ell (\ell+1) (23 \ell (\ell+1)-96)+40)} \xrightarrow{\ell\rightarrow\infty} \frac{50}{23}\,. \ee

All the above results depend only on the physical  combination $\alpha$, a nontrivial feature that we checked  by separately computing the $\alpha_{1,2,3}$ contributions. Again, we verified that  
 the constant terms that accompany the logarithms  are \textit{not} proportional to $\alpha$ and are thus unphysical.\,\footnote{For example, for $\ell = 3$ we identify the coefficient of $r^{-\ell}$ as
\be
 \left(\left(\frac{151525}{49}\alpha_1-\frac{48875 }{49}\alpha_2-\frac{101425}{196}\alpha_3\right)+\frac{8000}{7} \left(-3 \alpha_1+\alpha_2+\frac{\alpha_3}{2}\right) \log \left(\frac{r}{r_h}\right)\right)\frac{\kappa ^2}{r_h^4}\,.
\ee 
 The  $\alpha_2$ term  matches the one   from \cite{DeLuca:2022tkm}.
}

\subsubsection{Charged Black Hole}

We  compute the Love numbers of the EFT-corrected Reissner-Nordstr\"om black hole  at $\partial^4$ order.

 In contrast to the previous cases, the leading-order solution for charged non-rotating black holes that remain regular on the horizon
is no longer a polynomial but instead takes the form of an infinite  series in $r-r_h$. Starting from this form, knowing the asymptotic behavior at large $r$ is  challenging.

We circumvent this difficulty by computing the solution by applying the Frobenius method at $r=\infty$ as shown in section \ref{se:large_r}. This is useful because in the computation of beta functions, only a finite number of the highest monomials of the regular solution contributes. 
We find 
\begin{equation}
  \psi^{(0)}_\ell = \sum _{n=0}^{2\ell} a_{n} r^{\ell + 1 -n} + \dots \,,
\end{equation}
where the ellipses correspond to higher negative powers starting at $r^{-\ell}$. These terms feature a logarithm, but the powers are sufficiently negative  that they do not contribute in the  computation of the beta functions. 
The normalized coefficients satisfy the  recurrence relation
\begin{align}
a_0&= 1 \nn  \\
a_1&=\frac{4-\ell^2}{2 l}m\kappa ^2 \nn
% \\a^\ell_{n+2}&=\frac{ (n+\ell+4) \left(2 m (n+\ell)a^\ell_{n+1} -q^2  (n+\ell-1) a^\ell_{n}\right)}{2 (n+2) (2 \ell+n+3)}\kappa ^2
\\a_{n+2}&= \frac{(\ell-1-n) \left(q^2 (\ell+4-n) a_{n}-2 m (\ell+3-n) a_{n+1} \right)}{2 n (2 \ell+1 -n)}\kappa ^2 \,.
 \end{align}
For the first values of $\ell$ we find
% \begin{align}
% % \psi^{(0)}_1 &= r^2-\frac{1}{2}q^2 \nn  \\
% \psi^{(0)}_2 &= r^3-\frac{ q^2 r}{3} -\frac{2 m q^2}{9}  +\frac{9 q^4-16 m^2 q^2}{81 r} + \frac{81 m q^4-80 m^3 q^2}{450 r^2} -\frac{4320 m^4 q^2-6374 m^2 q^4+1125 q^6}{26730 r^3} + \dots \nn  \\
% \psi^{(0)}_3 &=  r^4  -\frac{ 5 m r^3}{6} +  \frac{5 m q^2 r}{36} +\frac{25 m^2 q^2}{176} + \frac{25 \left(27 m^3 q^2-11 m q^4\right)}{4752 r} + \frac{25 \left(189 m^4 q^2-158 m^2 q^4\right)}{33696 r^2} + \dots \nn  \\
% \psi^{(0)}_4 &= r^5 + 
% \end{align}
% where we set $\kappa=1$.
\begin{align}
\psi^{(0)}_2 &= r^3-\frac{ q^2 r}{3} -\frac{2 m q^2}{9}  +\frac{3 q^4-2 m^2 q^2}{12 r} +  \dots \nn  \\
\psi^{(0)}_3 &=  r^4  -\frac{ 5 m r^3}{6} +  \frac{5 m q^2 r}{36} +\frac{5 m^2 q^2}{108} + \frac{m^3 q^2 -3 mq^4}{72 r}-\frac{5m^2q^4}{324r^2} +  \dots \nn \\
\psi^{(0)}_4 &=  r^5 -\frac{3m r^4}{2} + \frac{(15m^2  +6 q^2) r^3}{28}-\frac{ (15m^2q^2+6q^4)r}{280 }-\frac{ 15m^3q^2+6mq^4}{1400 }
+\ldots
\end{align}
Setting $q=0$,  recovers the solutions for neutral black holes, a simplification arises due to a cancellation in the recurrence equation for the coefficients $a_n$, leading to the
polynomial solutions of \ref{eq:NeutralTensorSolutions}.

We focus on the leading ${\cal O}(\partial^4)$ corrections from the EFT of gravity. The running of the Love numbers appears for $\ell\geq 2$. 
The first beta functions are  given by
\begin{align}
   r^5_+ ~\beta_{k_2}& = \frac{128}{5}mq^2 \kappa^4 \gamma_1  \nn \\ 
   r^7_+~  \beta_{k_3}& = \frac{40}{63}m q^2   \left( \left(26 m^2\kappa^2 +40 q^2\right) \kappa ^6 \gamma_1 +27 q^2\kappa ^4 \gamma_2\right) \nn \\ 
   r^9_+ ~   \beta_{k_4}& = \frac{1}{11025}mq^2  \big(\left(62500  m^4\kappa ^2 +289540  m^2 q^2 + 130316 q^4\right)\kappa^8 \gamma_1  
   \nn \\ &\; \; \; \; +350 q^2 \left(445 m^2\kappa ^2 +248 q^2\kappa^6\right)\gamma_2
   \big)\,.
\end{align}
Notably, as observed in the spin-1 analysis, only the $R_{\mu\nu\rho\sigma} F^{\mu\nu}F^{\rho\sigma}$ operator contributes to the $\ell=2$ beta function. 

\subsubsection*{EFT-corrected extremal black hole }

The  beta functions of the first Love numbers of the  extremal black hole ($r_+=r_-\equiv r_h$) are given by
\begin{center}
\begin{tabular}{ |c|c|c|c|c|c|c|c| } 
 \hline
 $\ell$ & 2 & 3 & 4 & 5 
 % & 6 
 \\ 
 \hline
 \rule{0pt}{20pt}
 $\beta_{k_\ell}$
 & $ \frac{512
 }{5}\frac{ \gamma_1}{r_h^2}$
 & $ \frac{29440\gamma_1 }{63 r_h^2}+ \frac{960\gamma_2 }{7 \kappa^2r_h^2}$
 & $ \frac{15350336\gamma_1 }{11025 r_h^2}+ \frac{36416\gamma_2 }{63 \kappa^2r_h^2 }$
 
   & $\frac{16486542016\gamma_1 }{5011875 r_h^2}+ \frac{19491808\gamma_2 }{12375 \kappa^2r_h^2 }$
%%%%%%%%%%%%%%%%%%%%%%%
% L= 6 
%%%%%%%%%%%%%%%%%%%%%%%
    % & $ \frac{806504232448\gamma_1 }{120405285 r_h^2}+ \frac{46394845184\gamma_2 }{13378365 \kappa^2r_h^2} $
 \\[10pt] 
 \hline
\end{tabular}
\end{center}

\subsection{Extremal Black Holes, EFT Breakdown and the Schwinger Effect } \label{se:EFT_breakdown}

We can see from both vector and tensor tidal deformations that 
the contributions from the $F^4$ operators to the Love numbers are enhanced by a $\kappa^{-2}$ factor. To understand the implications of this phenomenon,  let us  evaluate some higher-order contributions to the Love numbers. It is enough to focus on the vector Love numbers of  the  extremal black hole. 

The  order-$\partial^6$ pure-curvature operators from \eqref{eq:EFT_lag1} contributions to the first Love numbers from the vector response are
\begin{align}
& k_1 \supset  -
\frac{  \kappa^2(452\alpha_1 + 92\alpha_2  - 79\alpha_3)}{60 r_h^4} 
 \,,\quad\quad  k_2 \supset - \frac{3  \kappa^2(2812\alpha_1 +612\alpha_2-489\alpha_3)}{400 r_h^4}  \\ \nn 
&
\beta_{k_3}\supset -
\frac{2  \kappa^2(1788\alpha_1 +676\alpha_2-293\alpha_3)}{7 r_h^4} 
 \,,\quad \ldots
\end{align}
They are thus subleading contributions due to the small $\frac{\kappa^2}{r_h^2}$ factor. 
Notice how the $\alpha$ combination does not appear anymore because space is not empty. 
We remind that  in the charged case this set of order-$\partial^6$ operators does not form a complete basis, there are also operators such as $F^6$, Riem\,$F^4$, etc.

We then consider   operators of the $F^6$ kind, focussing  on the $\gamma_{F^6}(F_\mu^\nu)^6$ operator. We find its contributions to the first  vector Love numbers are 
 \be
 k_1\supset\frac{64}{189}\frac{\gamma_{F^6} }{\kappa^4r_h^4}\,,\quad\quad 
 k_2\supset\frac{16}{525}\frac{\gamma_{F^6} }{\kappa^4r_h^4}\,,
 \label{eq:TLN_F6} 
 \ee
 while the contributions to the beta functions start at $\ell\geq5$.

We see that the $F^6$ contribution comes with a $\frac{1}{\kappa^2 r_h^2}$ factor compared to the $F^4$ contribution.  
Requiring that the  derivative expansion of the EFT be valid implies that the $F^6$ contribution be small with respect to the $F^4$ one. 
  Thinking in terms of loops of particles of mass $m$ as in section \ref{se:OneLoopEFT}, we have $\frac{\gamma_{F^6}}{\gamma_{F^4}}\sim \frac{q^2}{m^4} $. Requiring the $F^6$ contribution to be smaller than the $F^4$ one  leads to the condition
 \be
\frac{\kappa m^2 r_h}{|q| }> 1\,. \label{eq:val_EFT_EBH}
 \ee
 Repeating the same analysis for higher order operators $(F)^n$ leads again to the condition \eqref{eq:val_EFT_EBH}, hinting that it may be fundamental in some sense.

 The combination in \eqref{eq:val_EFT_EBH} is in fact precisely the one that controls the Schwinger effect of charged black holes, first derived in \cite{Hiscock_Weems}, see also \cite{Brown:2024ajk}. 
The Schwinger effect  causes  the loss of charge in charged black holes, 
due to the strong electric field splitting electron-positron pairs from the vacuum.  
For a near-extremal black hole with charge $Q_{\displaystyle \circ}$ the semi-classical emission rate is schematically
\be
\frac{d ({\rm pairs}) }{dt} \sim \exp\left(-\frac{Q_{\displaystyle \circ}}{Q_*}\right)\,,\quad \quad Q_*= \frac{|q| }{\pi\kappa^2 m^2 }\,.
\ee
The Schwinger effect is thus exponentially suppressed if $Q_{\displaystyle \circ}>Q_*$, which in term of radius occurs if $r_h> \frac{|q| }{\pi\kappa m^2 }$ where $m$ is here the electron mass. Rearranging, we see that, up to a $\pi$ factor, this is precisely the condition \eqref{eq:val_EFT_EBH}. 

The logics behind these conditions is that the charged particle must be heavy enough such that on-shell effects such as the Schwinger effect are exponentially suppressed.\,\footnote{This is analogous to the reason why an EFT with cutoff $\Lambda$ can be safely put at finite temperature $T<\Lambda$. Processes with nonlocal contributions from on-shell heavy states with $m>\Lambda$ exist, but are exponentially suppressed by a Boltzmann factor, hence their effects are  negligible with respect to the local effects encapsulated by the EFT \cite{Fichet:2019ugl}. 
}
If \eqref{eq:val_EFT_EBH} is satisfied, $m$ is heavy enough such that the EFT is valid and there is no  Schwinger effect essentially. This implies that the Love numbers we obtained in this section  apply specifically to  extremal black holes with $Q_{\displaystyle \circ}>Q_*$. Conversely, if \eqref{eq:val_EFT_EBH} is not satisfied, $m$ is too light, the charged particle cannot be integrated out, hence the full quantum loops have to be taken into account in the Love number computation.  This  regime is the focus of an upcoming work.

Finally, the limit of validity of the EFT also provides a typical maximal value for the Love numbers computed in the EFT framework, we find 
\be
k_\ell|_{\rm max} ={\cal O}(1)\,,\quad\quad  \beta_{k_\ell}|_{\rm max} ={\cal O}(1) \,. 
\ee

\section{Probing  Abelian Dark Sectors}
\label{se:bounds}

The overwhelming evidence for dark matter and dark energy suggests the existence of a light hidden sector. To avoid experimental bounds, the particles in this hidden sector should have suppressed interactions with visible matter; these sectors are broadly referred to as \textit{dark sectors}.

\subsection{On Generic Gravitational Searches for Light Dark  Particles}\label{se:generic_searches}

We could in principle search gravitationally for light particles  through a measurement of the GREFT coefficients. An assumption-independent approach 
could be to search for the GREFT-induced Love numbers of neutral back holes using gravitational waves data.

An example of target is  the lightest neutrino, which may be much lighter than the average neutrino mass $\sim 0.1$\,eV. Another candidate is the axion. 
There is also the logical possibility that many copies  of the SM coexist  \cite{Dvali:2007wp}, in which case  the GREFT coefficients are enhanced by the species number $N_s$.
 All these possibilities may  in principle be probed through a measurement of the GREFT coefficients.

However, current observations are not sensitive enough to put bounds on light particles using the tidal deformability of neutral black holes within the validity of the  GREFT.
For example, the predictions for black hole-related observables have a validity domain of $m \gtrsim \frac{1}{r_h}$. For $r_h\sim 10$~km, this condition requires $m\gtrsim 2\cdot 10^{-11} $\,eV. 
The current search methods produce bounds at much lower $m$,  that are thus outside the EFT validity domain. 

For example, a bound  from gravitational waves \cite{Chia:2023tle} applied to our predicted $\ell=2$ Love number for the neutral black hole \eqref{eq:TLNTensl2}
gives roughly $k_{2} \lesssim \O(1000)$, which implies  $m \gtrsim 10^{-54}$~eV.  
Bounds on the $(R_{\mu\nu}^{~~\,\rho\sigma})^3$ operator from causality \cite{Melville:2024zjq} would imply $m \gtrsim 10^{-50}$~eV. 
In fact, even if we assume the extreme hypothesis of $N_s\sim 10^{30}$ species, the EFT coefficients get enhanced by $N_s$, hence the above limits are enhanced by $\sqrt{N_s}\sim 10^{15}$, and thus remain outside of the EFT validity domain.
We conclude that searching for new light particles through the tidal deformation of neutral black holes would require immensely more sensitive probes. 

Some amplifying mechanism could however improve the situation. 
In the next subsection, we  exploit the fact that the  Love numbers of  charged black holes can be much larger than for the neutral ones.

\subsection{Towards a Search Through Dark-Charged Black Holes }
\label{se:extremal_BH_searches}

The dark sector may feature a dark Abelian gauge symmetry with gauge coupling $\tilde e$. Black holes can  be charged under this dark $U(1)$. 

The  time evolution of charged black holes in the semi-classical regime is well-known, at least qualitatively, see e.g. \cite{Hiscock_Weems, Brown:2024ajk, jess_BN}. 
  Aspects beyond  the semiclassical regime are still under scrutiny (see \cite{Brown:2024ajk}), but are irrelevant here. 
In short,  assuming that all the dark-charged particles are massive, the Hawking radiation of a sufficiently large black hole  drives it towards extremality, because the charged Hawking radiation is exponentially suppressed compared to the neutral one.
A competing Schwinger effect allows the charge to dissipate, but is exponentially suppressed if $Q_{\displaystyle \circ}>Q_*$ as explained in section \ref{se:EFT_breakdown}. Such a black hole
spends  a considerable amount of time near extremality. This is precisely the type of black hole for which our computation of the Love numbers is valid, as shown in section \ref{se:EFT_breakdown}. 

Given the time evolution described above, it is  possible that some black holes of the present-day Universe be near-extremal under a dark charge. This fact can in turn  be used to probe gravitationally  the dark sector. 

Let us focus on the lightest $U(1)$-charged particle of the dark sector with mass $m$ and charge $\tilde q>0$.  Consider a large enough black hole with $r_h\gtrsim \frac{\tilde q\tilde e}{\kappa m^2}$, 
so that the dark particle can be consistently integrated out (see \ref{se:EFT_breakdown}). This produces an Einstein-Maxwell EFT for the dark photon $X_\mu$. The EFT features the familiar $R_{\mu\nu\rho\sigma}X^{\mu\nu} X^{\rho\sigma}$ and $(X^{\mu\nu} X_{\mu\nu})^2$ operators, as well as higher-dimensional ones like  the $X^6$ ones.  

Based on our results, a rough  estimate of the $\ell=2$ Love number from the $X^6$ operators with $\gamma_{X^6} \propto \frac{\tilde q^6\tilde e^6}{m^8}$  gives
\be
k_2\sim  \frac{c}{16\pi^2}\frac{\tilde q^6\tilde e^6}{ \kappa^4 r_h^4 m^8}\,,
\ee
where $c\sim[0.01,1]$ depending on the dark particle spin and multiplicity. 
Let us assume a black hole radius near the EFT validity limit 
$
r_h\sim \frac{\tilde q\tilde e}{\kappa m^2} $. 
This implies   $k_2\sim \frac{\tilde q^2 \tilde e^2}{16\pi^2}$,  which can reach $k_2={\cal O}(1)$ if the dark $U(1)$ has strong coupling  $\tilde q \tilde e\sim 4\pi$. 
For black hole masses of roughly $10$--$100$ solar masses probed by LIGO/VIRGO, this case corresponds to a dark particle mass in the range 
\be
m\sim 0.1\textrm{--}1 ~{\rm GeV}\,. 
\ee
Interestingly, for these typical values of mass and coupling, the time evolution studied in \cite{jess_BN} implies that the dark-charged extremal black holes are typically long-lived enough to be primordial, i.e. have their origin in the Early Universe. 

In summary, we have shown that gravitational waves could be used to probe an Abelian dark sector 
in a scenario in which dark-charged extremal black holes exist at present times. While 
the effect is smaller by several order of magnitudes compared to current gravitational waves bounds,  one should take into account that our prediction is restricted to the EFT regime. In fact the largest effect occurs at the limit of validity of the EFT. This provides further motivation to  compute  the tidal properties of extremal black holes beyond  EFT,  a task we leave for future work.

\section{Summary and Outlook}

\label{se:Conclusion}

Gravity at  macroscopic distances is described by an EFT.   The gravitational effective Lagrangian
is structured as a derivative expansion whose leading-order term corresponds to GR, while the next-to-leading operators appear at order $\partial^6$ in the vacuum and  at order $\partial^4$ in the presence of matter. 

As a preliminary step, to  provide a concrete example of gravitational EFT and illustrate the EFT principles, we present a computation of the order-$\partial^6$  term  from loops of massive particles of spin $0$, $\frac{1}{2}$, $1$ via the heat kernel technique. Our result  confirms a finding from \cite{Goon:2016mil}. Together  with the order-$\partial^4$ Einstein-Maxwell from loops of charged particles \cite{Bastianelli:2008cu,Drummond80,Bittar:2024xuc},  this set of one-loop coefficients provides the gravitational EFT of the real world, that is dominated by neutrino and  electron loops. 

Our main focus  is the tidal deformability of black holes in the EFT of gravity. The corrections from the EFT contribute to the black hole tidal Love numbers --- which are known to vanish for GR in $d=4$ dimensions.

We have thoroughly analyzed the tidal deformation problem at perturbative level.  
The theory of differential equations with regular singular points, i.e. the Frobenius method, predicts that the tidal response may feature a logarithm. We make clear that this logarithm corresponds to a classical running of the corresponding Wilson coefficients in the worldline EFT. Consistency checks from our explicit computations confirm that the EFT-induced Love number  renormalization flow is a well-defined physical effect. 

A recurring  question in the literature is whether there is an ambiguity between the tidal response and subleading corrections to the source. 
Using the Frobenius method, we show that the corrections to the source do not experience running. 
Hence, whenever a Love number runs, no ambiguity is possible in its identification.  In the EFT of gravity, most of Love numbers run except the very first ones, hence the said ambiguity  essentially vanishes. 

The EFT-induced contributions to the Love numbers are computed using a  perturbation of the tidal solution that is regular on the horizon. 
We point out that this regular solution can be computed by introducing an appropriate tidal Green function. The Green function approach greatly simplifies the extraction of both constant and running Love numbers. We end up with a couple of   formulae, \eqref{eq:k1_full} and \eqref{eq:B_full}, that efficiently provide the Love numbers for any harmonics. 
Our formulae successfully reproduce all the available results from \cite{Charalambous:2022rre, DeLuca:2022tkm}. 

Using this new tool,  we investigate the deformations of non-spinning black holes from vectorial and tensorial tidal fields. 
Our treatment of the tidal field equation in  the vectorial case is fairly standard, but we present  simplifications to the computation of the tidal field equation  in the tensorial case. 

In the case of the neutral (i.e. EFT-corrected Schwarzschild) black hole, the running occurs for $\ell\geq 3$. We derive a closed-form formula for the Love beta functions with  arbitrary $\ell$,  for both vectorial and tensorial tidal deformations. 
The neutral back hole beta functions are exponentially suppressed at large $\ell$. 
In the charged (i.e. EFT-corrected Reisnner-Nordstr\"om) black hole case, the running occurs for $\ell\geq2$. We explicitly present   the results for the first harmonics. 
The higher harmonics  can be easily obtained from our general formula. A closed-form expression for the beta functions with any $\ell$ could be reached if one was able to obtain a closed form for the leading order regular solution, just  like in the neutral case.

We perform a consistency check  of our Love number calculations relying on the general EFT property that physical quantities must not depend on field redefinitions. To this end, 
we compute the Love numbers using a non-reduced basis of the order-$\partial^6$ Lagrangian.  We find that the constant Love numbers and all the beta functions are proportional to the  physical combination $\alpha$. In contrast, extra constant terms accompanying the logarithms are not proportional to $\alpha$, hence confirming that  these constant terms are unphysical.

One broad conclusion from our Love number results is that the electric charge greatly enhances the tidal deformability of the EFT-corrected black holes.  
We also find that the contributions from  the  $(F)^n$-type operators   to the Love numbers of the extremal black hole are so enhanced  that an extra condition is required for the EFT to remain valid. 
We show that this validity condition matches the one for the Schwinger effect to be suppressed, which is perfectly consistent from the EFT viewpoint. We conclude that the Love numbers  can reach up to ${\cal O}(1)$ values within the EFT validity regime.

We notice that the combination of operators in the Love number of extremal black holes  are independent of the one modifying the black hole charge-to-mass ratio. Therefore the Love numbers are not directly tied  to the Weak Gravity Conjecture, hence disproving a speculation about a possible correlation that had been proposed in the literature.

One may naturally ask whether the EFT-induced tidal deformability could be used to search for physics beyond the Standard Model at the purely gravitational level. 
Indeed, constraints on the $\ell=2$ Love number are available from gravitational waves data. 
While the prospects for a generic search through neutral black holes are rather gloomy, a more interesting possibility appears when considering  the hypothesis of a dark $U(1)$ gauge symmetry. 
We show that searching for a dark particle may be feasible in a scenario where dark-charged extremal black holes exist in the present-day Universe. The probed mass range is compatible with such extremal black holes being primordial. 
The effect is smaller than current gravitational waves bounds by several order of magnitudes. 
However, our estimate is  restricted to the EFT regime, while the largest effect occurs at the limit of validity of the EFT. This further motivates the investigation of the  extremal black holes tidal response beyond  EFT, that we will study in an upcoming work.

Finally, as a brief outlook, we believe it would be fruitful to further investigate the tidal deformability of spinning black holes in the EFT of gravity. The Love numbers of spinning black holes can be efficiently  computed using the technique we introduced. The extremal spinning case is especially interesting, because divergences in EFT-induced deformations of the near-horizon region of certain extremal black holes have been pointed out in \cite{Horowitz:2023xyl,Horowitz:2024dch}, see also \cite{Chen:2024sgx}. It would be certainly interesting to clarify the connection between these results and the standard Love numbers that are  computed via our technique.

\begin{acknowledgments}

We thank Jessica Santiago, Ricardo Sturani and Dmitri Vassilievich  for useful discussions. 
SF thanks the IPhT/CEA-Saclay for funding  a visit during which this work was initiated. 
%This work was supported in part by the S\~ao Paulo Research Foundation (FAPESP), grant 2021/10128-0. The work of LS was supported by grant 2023/11293-0 of FAPESP. 
%This study was financed in part by the Coordenação de Aperfeiçoamento de Pessoal de Nível Superior – Brasil (CAPES) – Finance Code 001.
%
This work has been supported by the following Brazilian research agencies: CAPES, FAPESP and CNPq. The work of SB was supported by grant 001 of CAPES, SF was supported by grant 2021/10128-0 of FAPESP, and LS was supported by grant 2023/11293-0 of FAPESP.
\end{acknowledgments}
\appendix

\section{Computing the Variation of ${\cal L}_{\rm eff}$  }
\label{app:Var_L4}

We collect standard variation formulae  and present some useful identities beyond standard textbooks that are needed to compute the variations of  order-$\partial^6$ operators in Section \ref{se:EFE}. We also include the variations of the order-$\partial^4$ operators.

\subsection{Conventions and Standard Formulae}

Throughout this work we use the conventions of Misner-Thorne-Wheeler  \cite{Misner:1973prb}, which include the mostly-plus  metric signature ${\rm sgn}(g_{\mu\nu})=(-,+,+,+)$ and positive scalar curvature for spheres. 
The commutator of covariant derivatives act on a generic rank two tensor as
% \be [\nabla_\rho, \nabla_\sigma]T_{\mu \nu} = R_{\mu_\lambda \rho \sigma}T\indices{^{\lambda}_\nu} + R_{\nu\lambda \rho \sigma}T\indices{_{\mu}^{\lambda}} \label{eq:commutator}
% \ee
\be [\nabla_\rho, \nabla_\sigma]T_{\mu \nu} = R_{ \rho \sigma \mu}^{~~~~\lambda}T_{\lambda\nu} + R_{ \rho \sigma \nu}^{~~~~\lambda}T_{\mu\lambda} \,. \label{eq:commutator}
\ee
The contraction of Bianchi's second identity $\nabla_{[\lambda}R_{\mu\nu]\rho\sigma}=0$ implies that
\be \nabla^\mu R_{\mu \nu }  = \dfrac{1}{2}\nabla_\nu R  \,. \label{eq:SecBi2}
\ee
The following  standard variational formulae  can be found in \cite{carroll2019spacetime}. 
\begin{align}
\delta g_{\mu\nu} &= -g_{\mu\rho}g_{\nu\sigma}\delta g^{\rho\sigma} \,
% \label{4.56_Carroll} 
\nonumber\\
\delta \sqrt{-g} &= -\dfrac{1}{2}\sqrt{-g}g_{\mu \nu}\delta g^{\mu \nu}
% \label{4.69_Carroll} 
\nonumber\\
\delta \Gamma^{\sigma}_{\mu \nu} &= -\dfrac{1}{2}\left[ g_{\lambda \mu}\nabla_{\nu}(\delta g^{\lambda \sigma}) + g_{\lambda \nu}\nabla_{\mu}(\delta g^{\lambda \sigma}) - g_{\mu \alpha}g_{\nu \beta}\nabla^{\sigma}(\delta g^{\alpha \beta}) \right] \,
% \label{4.64_Carroll} 
\nonumber\\
\nabla_\lambda (\delta \Gamma^{\rho}_{\nu \mu}) &=\partial_\lambda (\delta\Gamma^{\rho}_{\nu\mu}) + \Gamma^{\rho}_{\lambda \sigma}\delta\Gamma^{\sigma}_{\nu \mu} - \Gamma^{\sigma}_{\lambda \nu}\delta \Gamma^{\rho}_{\sigma \nu} - \Gamma^{\sigma}_{\lambda \mu} \delta \gamma^{\rho}_{\nu \sigma} \,
% \label{4.61_Carroll} 
\nonumber\\
\delta R^{\rho}_{~~\mu\lambda\nu} &= \nabla_\lambda (\delta \Gamma^{\rho}_{\nu \mu}) - \nabla_\nu (\delta \Gamma^{\rho}_{\lambda \mu}) \,
% \label{4.62_Carroll} 
\nonumber\\
\delta(\sqrt{-g}R) &= \sqrt{-g}\left(R_{\mu \nu} -\frac{R}{2}g_{\mu \nu} \right)\delta g^{\mu \nu} \,
\end{align}

\subsection{Supplementary Formulae}

We present more useful formulae that are not so easily found in standard references. 
In what follows $T$'s are generic tensors and $R$'s are curvature tensors.
\begin{align}
    T \delta R &= (T R_{\mu \nu} + g_{\mu \nu} \square T - \nabla_\mu \nabla_\nu T)\delta g^{\mu \nu} \nonumber\\
    T^{\mu \nu} \delta R_{\mu \nu} &= \frac{1}{2}(\square T_{\mu \nu} + g_{\mu \nu}\nabla_\rho \nabla_\sigma T^{\rho \sigma} - \nabla_\rho \nabla_\mu T\indices{^\rho _\nu} - \nabla_\rho \nabla_\mu T\indices{_\nu ^\rho})\delta g^{\mu \nu} \nonumber\\
    T\indices{_\mu ^\nu ^\rho ^\sigma}\delta R\indices{^\mu _\nu _\rho _\sigma} &= \frac{1}{2}\Big(\nabla_\rho \nabla_\sigma T\indices{_\mu ^\rho _\nu ^\sigma} + \nabla_\rho \nabla_\sigma T\indices{_\mu _\nu ^\rho ^\sigma} + \nabla_\rho \nabla_\sigma T\indices{^\rho _\mu ^\sigma _\nu}  \nonumber \\ & \quad - \nabla_\rho \nabla_\sigma T\indices{_\mu ^\rho ^\sigma _\nu} - \nabla_\rho \nabla_\sigma T\indices{_\mu _\nu ^\sigma ^\rho} - \nabla_\rho \nabla_\sigma T\indices{^\rho _\mu _\nu ^\sigma}\Big)\delta g^{\mu \nu} \nonumber\\
\end{align} 
\begin{align}
 \delta(\nabla_\delta R\indices{^\mu _\nu _\rho _\sigma}) &= \nabla_\lambda \nabla_\delta \delta R\indices{^\mu _\nu _\rho _\sigma} + (\delta \Gamma^\mu_{\delta \zeta})\nabla_\lambda R\indices{^\zeta _\nu _\rho _\sigma} - (\delta \Gamma^\zeta_{\delta \nu})\nabla_\lambda R\indices{^\mu _\zeta _\rho _\sigma} \nn \\ & \quad - (\delta \Gamma^\zeta_{\delta \rho})\nabla_\lambda R\indices{^\mu _\nu _\zeta _\sigma} \nonumber - (\delta \Gamma^\zeta_{\delta \sigma})\nabla_\lambda R\indices{^\mu _\nu _\rho _\zeta} - (\delta \Gamma^\zeta_{\lambda \delta})\nabla_\zeta R\indices{^\mu _\nu _\rho _\sigma} \\ & \quad + (\delta \Gamma^\mu_{\lambda \zeta})\nabla_\delta R\indices{^\zeta _\nu _\rho _\sigma} - (\delta \Gamma^\zeta_{\lambda \nu})\nabla_\delta R\indices{^\mu _\zeta _\rho _\sigma} \nonumber - (\delta \Gamma^\zeta_{\lambda \rho})\nabla_\delta R\indices{^\mu _\nu _\zeta _\sigma} \\& \quad - (\delta \Gamma^\zeta_{\lambda \sigma})\nabla_\delta R\indices{^\mu _\nu _\rho _\zeta} + \nabla_\lambda(\delta \Gamma^\mu_{\delta \zeta})R\indices{^\zeta _\nu _\rho _\sigma} - \nabla_\lambda(\delta \Gamma^\zeta_{\delta \nu})R\indices{^\mu _\zeta _\rho _\sigma} \nn \\
    & \quad - \nabla_\lambda(\delta \Gamma^\zeta_{\delta \rho})R\indices{^\mu _\nu _\zeta _\sigma} - \nabla_\lambda (\delta \Gamma^\zeta_{\delta_\sigma})R\indices{^\mu _\nu _\rho _\zeta} \nonumber \\
    (\delta \Gamma^\lambda_{\rho \sigma})T\indices{_\lambda ^\rho ^\sigma} &= \frac{1}{2}\nabla_\rho\Big( T\indices{_\mu _\nu ^\rho} +  T\indices{_\mu ^\rho _\nu} -  T\indices{^\rho _\mu _\nu}\Big)\delta g^{\mu \nu} \nonumber \\
    \delta(\nabla_\lambda T\indices{_\delta ^\mu _\nu _\rho _\sigma})&= \nabla_\lambda \delta T\indices{_\delta ^\mu _\nu _\rho _\sigma} - (\delta \Gamma^\zeta_{\lambda \delta})T\indices{_\zeta ^\mu _\nu _\rho _\sigma} + (\delta \Gamma^\mu_{\lambda \zeta})T\indices{_\delta ^\zeta _\nu _\rho _\sigma}\nonumber \\  & \quad - (\delta \Gamma^\zeta_{\lambda \nu})T\indices{_\delta ^\mu _\zeta _\rho _\sigma} - (\delta \Gamma^\zeta_{\lambda \rho})T\indices{_\delta ^\mu _\nu _\zeta _\sigma} - (\delta \Gamma^\zeta_{\lambda \sigma})T\indices{_\delta ^\mu _\nu _\rho _\zeta} \nonumber\\
    (\nabla_\mu \delta \Gamma^\nu_{\rho \sigma})T\indices{^\mu _\nu ^\rho ^\sigma} &= \frac{1}{2}\nabla_\rho \nabla_\sigma\Big( T\indices{^\sigma ^\rho _\mu _\nu} -  T\indices{^\sigma _\mu _\nu ^\rho} -  T\indices{^\sigma _\mu ^\rho _\nu} \Big)\delta g^{\mu \nu} \nonumber\\
    (\nabla_\lambda \nabla_\delta \delta R\indices{^\mu _\nu _\rho _\sigma})T\indices{^\lambda ^\delta _\mu ^\nu ^\rho ^\sigma} & = \frac{1}{2}\nabla_\rho \nabla_\sigma \nabla_\delta \nabla_\lambda\Big(T\indices{^\lambda ^\delta _\mu ^\rho _\nu ^\sigma} + T\indices{^\lambda ^\delta _\mu _\nu ^\rho ^\sigma} +  T\indices{^\lambda ^\delta ^\rho _\mu ^\sigma _\nu} \nonumber\\ & \quad - T\indices{^\lambda ^\delta _\mu ^\rho ^\sigma _\nu} - T\indices{^\lambda ^\delta _\mu _\nu ^\sigma ^\rho} - T\indices{^\lambda ^\delta ^\rho _\mu _\nu ^\sigma}\Big)\delta g^{\mu \nu}
\end{align}

\subsection{Curvature-squared Operators}

The Riem$^2$ contributions to the  effective Lagrangian, denoted ${\cal L}_4$ in \eqref{eq:L_eff_gen},  are
\be
{\cal L}_{4} = \frac{1}{2}\hat R + \beta_1 R^2
+\beta_2  R_{\mu\nu}R^{\mu\nu} +\beta_3 R_{\mu \nu\rho \sigma}R^{\mu \nu \rho \sigma}\,.
\ee
Analogously to Section \ref{se:EFE}, the  ${\cal L}_4$ Lagrangian corrects the field equations as 
\be
\hat R_{\mu\nu}-\frac{1}{2}\hat R g_{\mu\nu}= -\frac{2}{\sqrt{-g}}\frac{\delta(\sqrt{-g}{\cal L}_4)}{\delta g^{\mu\nu}} + \O(\partial^{6}) \,.
\ee
By computing the variations using the formulae presented in this appendix, we find that our equations reproduce the one from \cite{Kats:2006xp}, upon adjusting the conventions. %(we are working with $\kappa^2 = 8 \pi G = 1$)
The variations of the order-$\partial^4$ operators are as follows, 
    \begin{align}
        -\frac{2}{\sqrt{-g}}\delta(\sqrt{-g}R^2) &=  \left(R^2 g_{\mu \nu} -4 R R_{\mu \nu}  -4 g_{\mu \nu}\square R +4 \nabla_\mu \nabla_\nu R
        \right)\delta g^{\mu \nu} \nn \\ 
        -\frac{2}{\sqrt{-g}}\delta(\sqrt{-g}R_{\mu \nu}R^{\mu \nu}) &= \left. \big(  g_{\mu \nu}R_{\rho \sigma}R^{\rho \sigma} - g_{\mu \nu}\square R -2 \square R_{\mu \nu} \right. \nn \\ 
        & \left. \quad +2 \nabla_\mu \nabla_\nu R +4 R\indices{^\rho _\mu _\nu _\sigma}R\indices{^\sigma _\rho}\right)\delta g^{\mu \nu} \nn \\
  -\frac{2}{\sqrt{-g}}\delta (\sqrt{-g} R_{\delta \lambda\rho \sigma}R^{\delta \lambda \rho \sigma}) &= \left. \big( g_{\mu \nu} R_{\delta \lambda \rho \sigma}R^{\delta \lambda \rho \sigma} -4 R\indices{_\rho _\sigma _\lambda _\mu} R\indices{^\rho ^\sigma ^\lambda _\nu} +8 R_{\rho \mu \nu \delta}R^{\rho \delta } \nonumber \right.\\ 
  & \left. \quad -8 \square R_{\mu \nu} +4 \nabla_\mu \nabla_\nu R +8 R_{\rho \mu}R\indices{^\rho _\nu}  \right.\big)
        \delta  g^{\mu \nu} \,.
    \end{align}

\section{The Heat Kernel Coefficients}
\label{app:HK}

The general expressions for the coefficients appearing in \eqref{eq:Gam1_b} and \eqref{eq:Leff_oneloop},
that can be obtained by translating  the coefficients of  \cite{Vassilevich:2003xt,Gilkey_original} to  the conventions of this paper, are as follows.
\begin{align}
b_0&=I \nn \\
b_2&=\frac{1 }{6}RI-X \nn \\ \nn 
b_4&=\frac{1}{360}\Big(
12 \square R+5 R^2-2R_{\mu\nu}R^{\mu\nu}+2R_{\mu\nu\rho\sigma}R^{\mu\nu\rho\sigma} \Big)I \label{eq:b4}
\\ 
 & \quad - \frac{1}{6} \square X - \frac{1}{6} R X + \frac{1}{2} X^2 + \frac{1}{12} \Omega_{\mu\nu}\Omega^{\mu\nu}
\end{align}
\begin{align}
b_6 &=  \frac{1}{360}\bigg(
8 D_{\rho}\Omega_{\mu\nu}D^{\rho}\Omega^{\mu\nu}
+2 D^{\mu}\Omega_{\mu\nu} D_{\rho}\Omega^{\rho\nu}
+12 \Omega_{\mu\nu}\square \Omega^{\mu\nu}
-12 \Omega_{\mu\nu}\Omega^{\nu\rho}\Omega^{~~\mu}_{\rho}  \nn  \\ \nn 
& \quad +6 R_{\mu\nu\rho\sigma}\Omega^{\mu\nu}\Omega^{\rho\sigma}
-4 R_{\mu}^{~\nu}\Omega^{\mu\rho}\Omega_{\nu\rho} + 5 R\Omega_{\mu\nu}\Omega^{\mu\nu} \\ &\quad \nn
- 6 \square^2 X +60 X\square X+ 30 D_\mu X D^\mu X - 60 X^3
\\ &\quad \nn
- 30 X \Omega_{\mu\nu}\Omega^{\mu\nu} - 10 R \square X - 4 R_{\mu\nu} D^\nu  D^\mu X - 12 D_\mu R D^\mu X + 30 XX R \\ &\quad \nn 
- 12 X \square R  - 5 X R^2 + 2 X R_{\mu\nu}R^{\mu\nu} - 2 X R_{\mu\nu\rho\sigma}R^{\mu\nu\rho\sigma}
\bigg)   
\\ &\quad \nn
+\frac{1}{7!}\bigg(
18 \square^2  R + 17 D_\mu R D^\mu R  - 2D_\rho R_{\mu\nu} D^\rho R^{\mu\nu}
- 4 D_\rho R_{\mu\nu} D^\mu R^{\rho\nu}  \\ &\quad \nn
+9 D_\rho R_{\mu\nu\sigma\lambda} D^\rho R^{\mu\nu\sigma\lambda}  + 28 R \square R - 8 R_{\mu\nu} \square R^{\mu\nu} 
 \\ &\quad \nn
+ 24 R_{\mu\nu} D_\rho D^\nu  R^{\mu\rho} + 12 R_{\mu\nu\sigma\lambda} \square R^{\mu\nu\sigma\lambda} +\frac{35}{9} R^3 
\\ &\quad \nn
-\frac{14}{3} R R_{\mu\nu} R^{\mu\nu} 
+ \frac{14}{3} R R_{\mu\nu\rho\sigma}R^{\mu\nu\rho\sigma} -\frac{208}{9} R_{\mu\nu} R^{\mu\rho} R_{~~\rho}^{\nu} 
\\ &\quad \nn
+ \frac{64}{3} R_{\mu\nu}R_{\rho\sigma} R^{\mu\rho\nu\sigma}  
-\frac{16}{3} R^{\mu}_{~\nu } R_{\mu\rho\sigma\lambda}R^{\nu\rho\sigma\lambda}
\\ &\quad 
+ \frac{44}{9} R^{\mu\nu}_{~~\alpha\beta} R_{\mu\nu\rho\sigma} R^{\rho\sigma \alpha\beta }   + \frac{80}{9} R_{\mu~~\rho~~}^{~~\nu~~\sigma} R^{\mu\alpha\rho \beta } R_{\nu\alpha \sigma \beta} 
\bigg) I 
\label{eq:b6full}
\end{align}
Here $I$ is the identity matrix for internal indexes. 

\section{Computing the Tidal Green Function}

\label{app:Green}

We want to solve the tidal equation of motion in the presence of a source ${\cal J}(r)$, 
\be
{\cal D}_r^{(0)} \Psi^{\cal J}(r) = f(r)  {\cal J}(r)\,. \label{eq:EOM_generic}
\ee
To this end we introduce the Green function of the ${\cal D}_r^{(0)}$  differential operator, defined in \eqref{eq:Green_def} and reproduced here, 
\be {\cal D}_r^{(0)} G(r,r') = {f(r)} \delta(r-r') = \delta(r_\star-r_\star')
\,.\ee
The particular solution to \eqref{eq:EOM_generic} is then obtained via the convolution
\be
\Psi^{\cal J}(r) = \int^\infty_{r_h} dr' G(r,r') {\cal J}(r')\,. \label{eq:conv_generic}
\ee
Below we compute the Green function, following \cite{Fichet:2019owx}.
We define generic regularity conditions at two points $r=a$ and $r=b>a$ such that   
\begin{align}
{\cal B}_a \Psi =0\,,\quad \quad {\cal B}_b\Psi =0\,. 
\end{align}
The homogeneous solutions are denoted 
\be
\Psi(r)= A \psi_1(r)+B\psi_2(r)\,,
\label{eq:sols_hom}
\ee
where $A$, $B$ are constants. The associated Wronskian is defined as
\be
W(\psi_1, \psi_2)= \psi_1\psi_2' - \psi'_1\psi_2\,. 
\ee
Taking its derivative and using the homogeneous equation of motion leads to 
\be
W(r)= \frac{N_{\psi_1, \psi_2}} {f(r)}\,,
\ee
where $N_{\psi_1, \psi_2}$ is a nonzero constant. The $r$-dependence of $W(r)$ is thus fixed, only $N_{\psi_1, \psi_2}$ depends on the choice of solutions \eqref{eq:sols_hom}. 

The solution to the sourced equation of motion takes the form
\be
\Psi(r)= A(r) \psi_1(r)+B(r)\psi_2(r)
\label{eq:sols_part}\,.
\ee
In order to apply standard  ODE solving methods it is convenient to introduce ${\cal D}_r^{(0)}\equiv f^2(r) \hat{\cal D}_r^{(0)}$ such that 
 $\hat{\cal D}_r^{(0)}=\partial_{r}^2+\ldots$ 
  and $\hat {\cal D}_r^{(0)} \Psi^{\cal J}(r) = \frac{1}{f(r)}  {\cal J}(r)$. 
Following a standard method we set $A'(r)\psi_1(r)+B'(r)\psi_2(r)=0$ and obtain
\begin{align}
A'(r)&= -\frac{\psi_2'(r)}{W(r)} \frac{1}{f(r)} {\cal J}(r)= 
-\frac{\psi_2'(r)}{N_{\psi_1, \psi_2}} {\cal J}(r)
\,\\
B'(r)&= \frac{\psi_1'(r)}{W(r)} \frac{1}{f(r)}  {\cal J}(r)
= \frac{\psi_1'(r)}{N_{\psi_1, \psi_2}}  {\cal J}(r)
\,. 
\end{align}
The boundary conditions  take the form
\be
{\cal B}_a \Psi = A_a\psi_{1,a}+ B_a\psi_{2,a} =0
\ee
\be
{\cal B}_b \Psi = A_b \psi_{1,b}+ B_b\psi_{2,b} =0
\ee
with the definitions $\psi_{i,a}\equiv \psi_i(a)$, $\psi_{i,b}\equiv \psi_i(b)$, 
$A_a\equiv A(a)$, $A_b\equiv A(b)$ and similarly for $B$.

In order to compute $A(r)$ and $B(r)$, we integrate appropriate linear combinations of $A'(r)$, $B'(r)$ that give
\begin{align}
 \int^r_{r_a}\left( \psi_{1,a}A'(r) + \psi_{2,a} B'(r)\right) & \nn =  
\psi_{1,a}A(r)+ \psi_{2,a}B (r) 
\\ &=\int^r_{r_a}\left(
 \psi_{2,a}\psi_{1}(r')-\psi_{1,a}\psi_{2}  (r') 
\right) 
\frac{{\cal J}(r')}{N_{\psi_1, \psi_2}} \,,
\end{align}
\begin{align}
 \int_r^{b}\left( \psi_{1,b}A'(r) + \psi_{2,b} B'(r)\right) &\ nn =
-\psi_{1,b}A(r)- \psi_{2,b}B (r) \\ &=\int_r^{b}\left(  \psi_{2,b}\psi_{1}(r')- \psi_{1,b}\psi_{2}  (r')
\right) 
\frac{{\cal J}(r')}{N_{\psi_1, \psi_2}} \,.
\end{align}
This can be put in the matrix form
\be
\begin{pmatrix}
    \psi_{1,a} & \psi_{2,a} \\
        \psi_{1,b} & \psi_{2,b}
\end{pmatrix}
\begin{pmatrix}
    A(r) \\B(r)
\end{pmatrix} = 
\begin{pmatrix}
\int^r_{r_a}\left(
 \psi_{2,a}\psi_{1}(r')-\psi_{1,a}\psi_{2}  (r') 
\right) 
\frac{{\cal J}(r')}{N_{\psi_1, \psi_2}}
\\
    \int_r^{b}\left(   \psi_{1,b}\psi_{2} (r')- \psi_{2,b}\psi_{1}(r')
\right) 
\frac{{\cal J}(r')}{N_{\psi_1, \psi_2}}
\end{pmatrix}\,.
\ee
Inverting the matrix and plugging $A(r)$, $B(r)$ into \eqref{eq:sols_hom} gives the particular solution
\begin{align}
& \Psi^{\cal J}(r)= \frac{1}{\psi_{1,a}\psi_{2,b}-\psi_{1,b}\psi_{2,a}} \times \\ \nn & 
\Bigg[
\int_a^rdr'
\left(
\psi_{1,b}\psi_{2} (r) -  \psi_{2,b}\psi_{1}(r)
\right) 
\left(
\psi_{1,a}\psi_{2}  (r') - \psi_{2,a}\psi_{1}(r')
\right) 
\frac{{\cal J}(r')}{N_{\psi_1, \psi_2}}~+ 
\\ \nn &
\int_r^bdr'
\left(
\psi_{1,a}\psi_{2}  (r)- \psi_{2,a}\psi_{1}(r) 
\right) 
\left(
\psi_{1,b}\psi_{2}  (r') - \psi_{2,b}\psi_{1}(r')
\right) 
\frac{{\cal J}(r')}{N_{\psi_1, \psi_2}}
\Bigg]\,.
\end{align}

Finally observe that the solution can be rewritten as 
\be
\Psi^{\cal J}(r)= \int_a^b dr' G(r,r'){\cal J}(r')
\ee
with 
\be
G(r,r')=\frac{1}{N_{\psi_1, \psi_2}}\frac{
\left(
\psi_{1,a}\psi_{2}  (r_<)- \psi_{2,a}\psi_{1}(r_<) 
\right) 
\left(
\psi_{1,b}\psi_{2}  (r_>) - \psi_{2,b}\psi_{1}(r_>)
\right) 
}
{\psi_{1,a}\psi_{2,b}-\psi_{1,b}\psi_{2,a}}\,
\label{eq:Green_general}
\ee
which is the general Green function with arbitrary boundary conditions at $r=a,b$.

To obtain the tidal Green function in the black hole background we take the limits $a\to r_h$, $b\to\infty$. Identifying the $\psi_1\equiv \psi^{(0)}$ as the solution regular on the horizon and $\psi_2\equiv \psi^{(0)}_+$ as the solution regular at infinity implies that $\psi_{2,a},\psi_{1,b}\to \infty$ in \eqref{eq:Green_general}. This proves \eqref{eq:Green}.

\section{Computing Non-Spinning Geometries \label{se:bh_calculation}}

From the ansatz \eqref{eq:metric_perturbated}-\eqref{eq:electric_perturbated} for a static and spherically symmetric solution, we need to determine the three unknown functions $A^{(1)}(r)$, $B^{(1)}(r)$ and $E_r^{(1)}(r)$.  Three linearly independent equations are thus needed. We choose
\begin{align}
    G\indices{^t _t} &= \kappa^2 T\indices{^t _t}\\
    G\indices{^t _t} - G\indices{^r _r} &= \kappa^2(T\indices{^t _t} - T\indices{^r _r})\\
    \nabla_\nu F\indices{_t ^\nu} &= \Tilde{J}_t \;.
\end{align}
Remember that $T_{\mu \nu} = T_{\mu \nu}^{{\rm e.m.}} + \Tilde{T}_{\mu \nu}$, where $\Tilde{T}_{\mu \nu}$ are the corrections due to the higher-order terms in the effective theory. At zeroth order in $\alpha_i$ and $\gamma_i$, the equations are automatically satisfied. At first order we obtain,
\begin{align}
    \frac{d}{dr}\Big(r B^{(1)}(r) \Big) &=  \frac{\kappa^2 q^2}{2 r^2}C^{(1)}(r) + \kappa^2 r^2 \Tilde{T}\indices{^t _t}\label{eq:first_equation}\\
    \frac{d}{dr}D^{(1)}(r) &= \frac{\kappa^2 r}{1 - \frac{\kappa^2 m}{r} + \frac{\kappa^2 q^2}{2 r^2}}\Big(\Tilde{T}\indices{^t _t} - \Tilde{T}\indices{^r _r} \Big)\label{eq:second_equation}\\
    \frac{d}{dr}C^{(1)}(r) &= -\frac{2 r^2}{q}\Tilde{J}^t \label{eq:third_equation}\;,
\end{align}
where
\begin{align}
    C^{(1)}(r) &= \frac{A^{(1)}(r) - B^{(1)}(r)}{1 - \frac{\kappa^2 m}{r} + \frac{\kappa^2 q^2}{2 r^2}} - \frac{2 r^2 E_r^{(1)}(r)}{q}\ \\
    D^{(1)}(r) &= \frac{B^{(1)}(r) - A^{(1)}(r)}{1 - \frac{\kappa^2 m}{r} + \frac{\kappa^2 q^2}{2 r^2}} \;.
\end{align}
Imposing an asymptotically flat spacetime, $A^{(1)}(\infty) = B^{(1)}(\infty) = C^{(1)}(\infty) = 0$, we obtain the solution via the integrals
\begin{align}
    C^{(1)}(r) &= \frac{2}{q}\int_r^\infty dr r^2 \Tilde{J}^t\\
    B^{(1)}(r) &= -\frac{\kappa^2}{r}\Bigg(\frac{q^2}{2}\int_r^\infty dr \frac{C^{(1)}(r)}{r^2} + \int_r^\infty dr r^2 \Tilde{T}\indices{^t _t} \Bigg)\\
    A^{(1)}(r) &= B^{(1)}(r) + \kappa^2 \Bigg(1 - \frac{\kappa^2 m}{r} + \frac{\kappa^2 q^2}{2 r^2} \Bigg)\int_r^\infty dr \frac{r}{\Big(1 - \frac{\kappa^2 m}{r} + \frac{\kappa^2 q^2}{2 r^2} \Big)}\Big(\Tilde{T}\indices{^t _t} - \Tilde{T}\indices{^r _r} \Big)\;.
\end{align}
This provides the results  \eqref{eq:A}, \eqref{eq:B}, \eqref{eq:E}.

\section{Decomposition into Spherical Harmonics \label{sec:spherical_harmonics}}

The spherical harmonics $Y_{\ell m}:S^2\to \mathbb{R}$  are governed by the representation theory of $SO(3)$ \cite{Tung:1985na}. They correspond to a   subset of the irreps of $SO(3)$ restricted to the quotient $SO(3)/SO(2) \equiv S^2 $. 
 Due to its group-theoretical nature, the set is orthogonal and complete, hence the spherical harmonics provide an orthogonal basis for the  space of square-integrable functions.  
 Any  square integrable function $f(t, r, \theta, \phi)$ on a manifold with $SO(3)$ symmetry can thus be  expressed as a linear combination of spherical harmonics 
\begin{equation}
    f(t, r, \theta, \phi) = \sum_{\ell=0}^\infty \sum_{m=-\ell}^\ell
    f_{\ell m}(t, r) Y_{\ell m}(\theta, \phi)\;.
\end{equation}
The spherical harmonics satisfy the orthogonality theorem 
\begin{equation}
    \int Y_{\ell m}(\theta, \phi) Y^{*}_{\ell^\prime m^\prime}(\theta, \phi) d\Omega = \delta_{\ell \ell^\prime} \delta_{m m^\prime}\;
\end{equation}
and the corresponding completeness theorem. 

Spherical harmonics are eigenfunctions of the squared angular momentum operator $L^2$ (i.e. the Casimir operator in function space),
\begin{equation}
     L^2 Y_{\ell m}(\theta, \phi) = \ell(\ell+1)Y_{\ell m}(\theta, \phi), \; \; \; \; \; \; \;  L^2 = -\Bigg[\frac{1}{\sin^2{\theta}}\frac{\partial^2}{\partial \phi^2} + \frac{1}{\sin{\theta}}\frac{\partial}{\partial \theta}\bigg(\sin{\theta}\frac{\partial}{\partial \theta} \bigg) \Bigg]\;.
\end{equation}
In particular for the spherically symmetric spacetime defined in \eqref{eq:metric_perturbated}, the Laplacian satisfies
\begin{equation}
    \square Y_{\ell m}(\theta, \phi) 
    =  -\frac{1}{r^2}L^2 Y_{\ell m}(\theta, \phi) = - \frac{\ell(\ell+1)}{r^2}Y_{\ell m}(\theta, \phi)\;.
\end{equation}
For this reason, spherical harmonics are ideally suited for describing fields propagating in a spherically symmetric background. 
 Since angular momentum is conserved, the equations of motion for free fields in such spacetime decompose into an infinite set of decoupled terms with fixed $\ell$.

The spherical harmonics are eigenvectors of the parity operator,
\begin{equation}
    \mathcal{P} Y_{\ell m}(\theta, \phi) = Y_{\ell m}(\pi - \theta, \pi + \phi) = (-1)^l Y_{\ell m}(\theta, \phi)\;,
\label{eq:parity_Y}
\end{equation}
hence the decomposition can be split into parity even and odd pieces. 
Degrees of freedom  with definite angular momentum and parity eigenvalue are kinetically decoupled and  can thus be treated  independently.

% In addition to angular momentum conservation, spherically symmetric spacetimes are parity invariant. 

\subsection{Vector \label{sec:vector_spherical_harmonics}}

Vector spherical harmonics are the   extension of spherical harmonics to  vector fields. 

Let us consider first three-dimensional flat space with spherical coordinates, i.e. the $S^2$ slicing of $\mathbb{R}^3$.
For a given spherical harmonic, we can define three orthogonal vectors that form a basis of the vector field space \cite{Hui:2020xxx}.
First define a vector in the radial direction
\begin{equation}
     Y_{\ell m} \mathbf{\hat{e}_r}\;.
\end{equation}
Since $Y_{\ell m}(\theta, \phi)$ depends only on the angles, its gradient  is orthogonal to the radial direction,
\begin{equation}
     \nabla Y_{\ell m}\;.
\end{equation}
Finally, since the cross product between two vectors is orthogonal to both, the third basis vector   can be chosen as
\begin{equation}
     \mathbf{\hat{e}_r}\times \nabla Y_{\ell m}\;.
\end{equation}

Any vector field of $\mathbb{R}^3$ can be expressed as a linear combination of the vector spherical harmonics
\begin{align}
    \mathbf{V} = \sum_{\ell,m} \begin{pmatrix}
v^{(1)}_{\ell m}(r)Y_{\ell m}(\theta, \phi) \\
v^{(2)}_{\ell m}(r) \gamma^{i j} \partial_j Y_{\ell m}(\theta, \phi) + v^{(3)}_{\ell m}(r) \epsilon^{i j} \partial_j Y_{\ell m}(\theta, \phi)
\end{pmatrix}\;,
\label{eq:vector_harmonics1}
\end{align}
where the Latin indices run over the sphere coordinates $ (\theta, \phi)$, $\gamma_{i j}$ is the two-sphere metric 
\begin{equation}
    ds^2 = d\theta^2 + \sin^2{\theta}d\phi^2\;,
\label{eq:two_sphere_metric}\end{equation}
and $\epsilon_{i j} = \sqrt{|\gamma|}\tilde{\epsilon}_{i j} = \sin{\theta}\tilde{\epsilon}_{ij}$ is the  Levi-Civita tensor, being $\tilde{\epsilon}_{i j}$ the Levi-Civita symbol.

The form given in \eqref{eq:vector_harmonics1} is easily generalized to four-vectors in spherically symmetric curved spacetime. One simply appends an orthogonal vector  in the time direction, $Y_{\ell m}\mathbf{\hat{e}}_t$. Therefore, any four-vector field in spherically symmetric spacetime can be expressed as
\begin{align}
    V_\mu = \sum_{\ell,m} \begin{pmatrix}
v^{(0)}_{\ell m}(t, r)Y_{\ell m}(\theta, \phi)\\
v^{(1)}_{\ell m}(t, r)Y_{\ell m}(\theta, \phi) \\
v^{(2)}_{\ell m}(t, r) \nabla_i Y_{\ell m}(\theta, \phi) + v^{(3)}_{\ell m}(t, r) \epsilon\indices{_i ^j} \nabla_j Y_{\ell m}(\theta, \phi)
\end{pmatrix}\;,
\label{eq:vector_harmonics}
\end{align}
where $\nabla_i$ is the covariant derivative with respect to the two-sphere metric \eqref{eq:two_sphere_metric}. 

A vector in curved space generally transforms according to
\begin{equation}
    V_{\mu^\prime} = \frac{\partial x^\mu}{\partial x^{\mu^\prime}}V_{\mu}\;.
    \label{eq:transfo_vector}
\end{equation}
Hence, from the viewpoint of $S^2$, the  $V_t$, $V_r$ components transform as scalars, while $V_i$ transform as a vector of $S^2$. This can also be seen from \eqref{eq:vector_harmonics}.

Combining \eqref{eq:transfo_vector}  with  the parity property \eqref{eq:parity_Y}, we can deduce the parity of the components of \eqref{eq:vector_harmonics}. 
 The time and radial components transforms as scalars under parity with eigenvalue $(-1)^{\ell}$. The gradient of a scalar transforms as a vector,
\begin{equation}
    \mathcal{P}\nabla_i Y_{\ell m}(\theta, \phi) = (-1)^{\ell+1} \nabla_i Y_{\ell m}(\theta, \phi)\;.
\end{equation}
The last component transforms as a pseudo-vector,
\begin{equation}
    \mathcal{P}\epsilon^j_i \nabla_j Y_{\ell m}(\theta, \phi) = (-1)^{\ell} \epsilon\indices{_i ^j} \nabla_j Y_{\ell m}(\theta, \phi)\;.
\end{equation}
It follows that the $v^{(0,1,2)}$ are parity-even while $v^{(3)}$ is parity-odd. 
 Therefore, the pseudo-vector term decouples from the other three.

% If \eqref{eq:vector_harmonics} describes a free vector field propagating in a spherically symmetric   spacetime, parity is conserved, and terms with different transformation under parity do not mix. Therefore, the pseudo-vector term, the odd-mode, decouples from the other three, the even-modes.

 This choice of basis is convenient to separate the vector field degrees of freedom such that the angular dependence factorizes in the equation of motion. This is used in section \ref{se:spin1}. 

\subsection{Tensor \label{se:tensor_spherical_harmonics}}

We can extend the spherical harmonic decomposition  to tensors, distinguishing the components according to their behavior under transformation of the $S^2$ slices \cite{Regge:1957td, DeFelice:2011ka}. 

A tensor in curved space transforms according to
\begin{equation}
    T_{\mu^\prime \nu^\prime} = \frac{\partial x^\mu}{\partial x^{\mu^\prime}}\frac{\partial x^\nu}{\partial x^{\nu^\prime}}T_{\mu \nu}\;.
\end{equation}
We assume  a symmetric tensor. 
From the viewpoint of the $S^2$ slices, 
the components $T_{tt}$, $T_{tr}$ and $T_{rr}$ transform as scalars, $T_{t i}$ and $T_{r i}$ transform as vectors, while $T_{i j}$ transforms as a tensor of $S^2$.

For a symmetric tensor $h_{ij}$, given a spherical harmonic,  three independent tensors are needed to form a basis. First, we define a tensor proportional to the two-sphere metric,
\begin{equation}
    \gamma_{i j}Y_{\ell m}\;.
\label{eq:tensor1}
\end{equation}
We can construct a rank-two tensor by applying two covariant derivatives to $Y_{\ell m}$, resulting in two independent tensors: one symmetrized and the other antisymmetrized. To guarantee orthogonality to \eqref{eq:tensor1}, we subtract the trace,
\begin{align}
    &\nabla_{(i} \nabla_{j)T} Y_{\ell m}\\
    &\epsilon\indices{_{(i} ^k}\nabla_{j)T}\nabla_{k}Y_{\ell m}\;.
\end{align}

Finally, similar to our approach with vectors, we can separate the contributions based on parity. The parity-even components are
\begin{equation}
T_{\mu \nu}^{\mathrm{even}} =  \sum_{\ell,m}
\begin{pmatrix}
\begin{array}{cc|c}
-A(r)H_0(t, r)Y_{\ell m} & H_1(t, r)Y_{\ell m} & \mathcal{H}_0(t, r)\nabla_i Y_{\ell m}\\
\mathrm{Sym} & \frac{1}{B(r)}H_2(t, r)Y_{\ell m} & \mathcal{H}_1(t, r)\nabla_i Y_{\ell m} \\
\hline
\mathrm{Sym} & \; \; \mathrm{Sym} \; \; & \; \; \mathcal{K}(t,r)\gamma_{ij}Y_{\ell m} + G(t, r)\nabla_{(i} \nabla_{j)T} Y_{\ell m}\;\;
\end{array}
\end{pmatrix}\;.
\end{equation}
The parity-odd components are
\begin{equation}
T_{\mu \nu}^{\mathrm{odd}} =  \sum_{\ell,m}
\begin{pmatrix}
\begin{array}{cc|c}
0 & 0 & h_0(t, r)\epsilon\indices{_i ^j}\nabla_j Y_{\ell m} \\
0 & 0 & h_1(t, r)\epsilon\indices{_i ^j}\nabla_j Y_{\ell m} \\
\hline
\mathrm{Sym} & \; \; \mathrm{Sym} \; \; & \; \; h_2(t, r) \epsilon\indices{_( _i ^k} \nabla_{j)T} \nabla_k Y_{\ell m}\;\;
\end{array}
\end{pmatrix}\;.
\end{equation}
Note that the odd components are those involving the Levi-Civita tensor, which contribute with an additional sign change under the parity transformation. This is used in section \ref{se:spin2}.

\section{Source of the Equations of Motion \label{se:source}}

The source for the equation of motion   of the vector tidal field  perturbation in \eqref{eq:EOM_1_vector} is 
\begin{align}
    {\cal D}^{(1)}\Psi_\ell^{(0)} = &\,\,(A^{(0)}B^{(1)} + A^{(1)}B^{(0)})\frac{d^2 \Psi_\ell^{(0)}}{dr^2} + \frac{1}{2}\frac{d}{dr}(A^{(0)}B^{(1)} + A^{(1)}B^{(0)})\frac{d\Psi_\ell^{(0)}}{dr}\nn \\   & - \frac{A^{(1)}}{r^2}\ell(\ell+1)\Psi_\ell^{(0)}\;. 
    \label{eq:source_vector}
\end{align}

The source for the tensor tidal field perturbation \eqref{eq:EOM_1_tensor}  for a neutral black hole is     
\begin{align}
    &{\cal D}^{(1)}\Psi_\ell^{(0)} =\frac{\kappa^2}{2 r^{10} r_h^3}\bigg(  12 \alpha _1 \left(-42 \left(j^2-32\right) r^2 r_h^5+\left(41 j^2-2721\right) r r_h^6+1380 r_h^7+\left(j^2-3\right) r^7\right) \nn \\ 
    &-20 \alpha _2 \left(-18 j^2 r^2 r_h^5+\left(17 j^2+63\right) r r_h^6-60 r_h^7+\left(j^2-3\right) r^7\right)\nn \\
    & -\alpha _3 \left(-72 \left(j^2-42\right) r^2 r_h^5+\left(71 j^2-6201\right) r r_h^6+3180 r_h^7+\left(j^2-3\right) r^7\right)\bigg)\Psi_\ell^{(0)}\nn \\
    &  +\frac{(r-r_h)^2\kappa^2}{2 r^9 r_h^3}\bigg(\left( 12 \alpha _1-20 \alpha _2-\alpha _3 \right)(120 r_h^5 -2 r^4 r_h-3 r^3 r_h^2-4 r^2 r_h^3-5 r r_h^4-r^5)\bigg) \frac{\partial\Psi_\ell^{(0)}}{\partial r}\,.
    \label{eq:source_neutral}
\end{align}
Notice that the physical combination $\alpha$ does not appear in the source term. The $\alpha$ combination emerges only in the final expressions given by \eqref{eq:k1_full} and \eqref{eq:B_full}. 
The source term for $\alpha_2$ in \eqref{eq:source_neutral} is equivalent to that in \cite{DeLuca:2022tkm}, although we made different reductions by using the equation of motion  of $\Psi_\ell^{(0)}$, \eqref{eq:EOM_0_tensor}.

The source for the tensor tidal field perturbation \eqref{eq:EOM_1_tensor} for a charged black hole is 
% \begin{align}
%     &{\cal D}^{(1)}\Psi_\ell^{(0)} = \frac{2 r_+ r_- (r - r_-)(r - r_+)}{5(j^2-2) \kappa^{2} r^{12}}\Bigg( \kappa^2 \gamma_1 \Big(960 r^4 (\ell(\ell +1)-2) + 1440 r_+^2 r_-^2 \nn \\   &- 5 r^3 (r_+ + r_-)(201j^2 -322) - 1840 r r_+ r_-(r_- + r_+) +16 r^2 (25 r_+^2 + 18 r_+ r_- \nn \\ &+ 61 j^2 + 25 r_-^2)\Big) - 96 \kappa^2 \gamma_2 r_+ r_-\Big( r^2(4+\ell)(\ell-3) - 10 r_+ r_- + 10 r (r_+ + r_-)\Big)\Bigg)\Psi_l^{(0)}\nn \\
%     &-\frac{2 r_+ r_- (r - r_-)(r - r_+)}{5(j^2-2) \kappa^{2} r^{11}} \Bigg(\kappa^2 \gamma_1 \Big(260 r^4 (\ell(\ell +1)-2) - 1440 r_+^2 r_-^2 - 5 r^3 (r_+ + r_-)(61j^2 -202) \nn \\ &+ 1840 r r_+ r_-(r_- + r_+) - 4 r^2 (100 r_+^2 + 734 r_+ r_- - 87  j^2 r_+ r_- + 100 r_-^2)\Big) \nn \\ &- 48 \kappa^2 \gamma_2 r_+ r_-\Big( r^2 (j^2 +18) + 20 r_+ r_- - 20 r (r_+ + r_-)\Big) \Bigg)\frac{\partial \Psi_l^{(0}}{\partial r} \nn \\ &+\frac{2 r_+ r_- (r - r_-)(r - r_+)}{5 \kappa^2 r^8}\Bigg( \kappa^2 \gamma_1\Big(20 r^2 + 28 r_+ r_- - 25 r (r_+ + r_-) \Big) - 8 \gamma_2 r_+ r_-\Bigg)\frac{\partial^2 \Psi_l^{(0)}}{\partial r^2} \nn \\ 
%     &+ \mathcal{O}(\alpha)\,,
% \end{align}
\begin{align}
    &{\cal D}^{(1)}\Psi_\ell^{(0)} = \bigg( \gamma _1  r^4 \Big(\kappa ^2 \left(14 q^2-25 m r\right)+20 r^2\Big) -4 \gamma _2  q^2 r^4 \bigg) \frac{\kappa ^2 q^2 \left(\kappa ^2 \left(q^2-2 m r\right)+2 r^2\right)}{10 r^{12}}  \frac{\partial^2 \Psi_l^{(0)}}{\partial r^2}\nn \\
    & + \bigg(\gamma _1  \Big( 40 r \kappa ^4 \left(9 q^2-5 m r\right) \left(q^2-2 m r\right) -260 \left(j^2-2\right) r^5\nn\\ 
    &  \quad\quad\quad\quad  -\kappa ^2 r^3 \left(5 \left(202-61 j^2\right) m r +6 \left(29 j^2-178\right) q^2\right)\Big) \nn \\
    &+ 24 \gamma _2 q^2 r \Big(\left(j^2+18\right) r^2+10 \kappa ^2 \left(q^2-2 m r\right)\Big)  \bigg) \frac{\kappa ^2 q^2 \left(\kappa ^2 \left(q^2-2 m r\right)+2 r^2\right)}{10 \left(j^2-2\right) r^{12}} \frac{\partial \Psi_l^{(0)}}{\partial r} \nn\\
    & + \bigg( \gamma _1 \Big(\kappa ^2 r^2 \left(5 \left(322-201 j^2\right) m r+8 \left(61 j^2-32\right) q^2\right)+960 \left(j^2-2\right) r^4 \nn \\
    & \quad\quad\quad\quad +40 \kappa ^4 \left(q^2-2 m r\right) \left(9 q^2-5 m r\right)\Big) \nn \\
    & + 48 \gamma _2 q^2 \Big(5 \kappa ^2 \left(q^2-2 m r\right)-\left(j^2-12\right) r^2\Big)  \bigg) \frac{\kappa ^2 q^2 \left(\kappa ^2 \left(q^2-2 m r\right)+2 r^2\right)}{10 \left(j^2-2\right) r^{12}} \Psi_l^{(0)} \nn \\ 
    &+ \mathcal{O}(\alpha_i)\,.
    \label{eq:source_charged}
\end{align}
We used $j^2 = \ell (\ell + 1)$ and $\partial_t \Psi_l = 0$ in both \eqref{eq:source_neutral} and \eqref{eq:source_charged}.

\bibliographystyle{JHEP}
\bibliography{biblio}

\end{document}